\pdfoutput=1
\documentclass[11pt,twoside,a4paper,cmspaper,final,collab]{cms-tdr}
\def\svnVersion{b9bd56c}\def\svnDate{2022/06/06}\def\cmsCernNoTag{CERN-EP-2022-008}\def\cmsCernDate{\today}\def\cmsMessage{Published in Physical Review D as \href{http://dx.doi.org/10.1103/PhysRevD.105.112007}{\doi{10.1103/PhysRevD.105.112007}.}}
\begin{document}\cmsNoteHeader{EXO-21-002}

\ifthenelse{\boolean{cms@external}}{\newcommand{\cmsLeft}{upper\xspace}}{\newcommand{\cmsLeft}{left\xspace}}
\ifthenelse{\boolean{cms@external}}{\newcommand{\cmsRight}{lower\xspace}}{\newcommand{\cmsRight}{right\xspace}} 
\newcommand{\cmsTable}[1]{\resizebox{\textwidth}{!}{#1}}
\newcommand{\MT}{\ensuremath{M_{\mathrm{T}}}\xspace}
\newcommand{\LT}{\ensuremath{L_{\mathrm{T}}}\xspace}
\newcommand{\ST}{\ensuremath{S_{\mathrm{T}}}\xspace}
\newcommand{\LTmet}{\LT$+$\ptmiss}
\newcommand{\mossf}{\ensuremath{M_{\text{OSSF}}}\xspace}
\newcommand{\mmin}{\ensuremath{M_{\text{min}}}\xspace}
\newcommand{\ql}{\ensuremath{Q_{\Pell}}\xspace}
\newcommand{\ml}{\ensuremath{M_{\Pell}}\xspace}
\newcommand{\dR}{\ensuremath{\Delta{R}}\xspace}
\newcommand{\dRmin}{\ensuremath{\Delta {R}_{\text{min}}}\xspace}
\newcommand{\ttZ}{\ensuremath{\ttbar\PZ}}
\newcommand{\ttV}{\ensuremath{\ttbar\PV}}
\newcommand{\VVV}{\ensuremath{\PV\PV\PV}}
\newcommand{\WZ}{\ensuremath{\PW\PZ}}
\newcommand{\ZZ}{\ensuremath{\PZ\PZ}}
\newcommand{\ZG}{\ensuremath{\PZ\PGg}}
\newcommand{\vltau}{\ensuremath{\PGt^{\prime}}\Xspace}
\newcommand{\vlnu}{\ensuremath{\PGn^{\prime}}\Xspace}
\DeclareRobustCommand{\PS}{{\HepParticle{S}{}{}}\Xspace}
\newcommand{\nj}{\ensuremath{N_{\mathrm{j}}}\xspace}
\newcommand{\nbj}{\ensuremath{N_{\PQb}}\xspace}

\cmsNoteHeader{EXO-21-002}
\title{Inclusive nonresonant multilepton probes of new phenomena at \texorpdfstring{$\sqrt{s} = 13\TeV$}{sqrt(s) = 13 TeV}}

\date{\today}

\abstract{
  An inclusive search for nonresonant signatures of beyond the standard model (SM) phenomena in events with three or more charged leptons, including hadronically decaying \PGt leptons, is presented.  
  The analysis is based on a data sample corresponding to an integrated luminosity of 138\fbinv of proton-proton collisions at $\sqrt{s}= 13\TeV$, collected by the CMS experiment at the LHC in 2016--2018.
  Events are categorized based on the lepton and \PQb-tagged jet multiplicities and various kinematic variables.
 Three scenarios of physics beyond the SM are probed, and signal-specific boosted decision trees are used for enhancing sensitivity.
  No significant deviations from the background expectations are observed.
  Lower limits are set at 95\% confidence level on the mass of type-III seesaw heavy fermions in the range 845--1065\GeV for various decay branching fraction combinations to SM leptons.
  Doublet and singlet vector-like \PGt lepton extensions of the SM are excluded for masses below 1045\GeV and in the mass range 125--150\GeV, respectively. 
  Scalar leptoquarks decaying exclusively to a top quark and a lepton are excluded below 1.12--1.42\TeV, depending on the lepton flavor. For the type-III seesaw as well as the vector-like doublet model, these constraints are the most stringent to date. For the vector-like singlet model, these are the first constraints from the LHC experiments. Detailed results are also presented to facilitate alternative theoretical interpretations.
}

\hypersetup{
pdfauthor={CMS Collaboration},
pdftitle={Inclusive nonresonant multilepton probes of new phenomena at 13 TeV},
pdfsubject={CMS},
pdfkeywords={CMS, multileptons, type-III seesaw, heavy fermions, vector-like leptons, leptoquarks}}

\maketitle 

\section{Introduction}\label{sec:intro}
The standard model (SM) of particle physics describes the known fundamental particles and their interactions, and has been extensively tested by experiments. 
There are strong indications, however, that the SM is incomplete, and beyond-the-SM (BSM) models are required to answer the open questions such as the origin of neutrino masses, the particle nature of dark matter, and the observed baryon asymmetry in the universe. 
A multitude of compelling BSM theories have been proposed with characteristic signatures that would modify the production of SM particles in proton-proton ($\Pp\Pp$) collisions. 
In particular, new BSM particles decaying via the weak interaction could produce the distinctive signature of an excess of events containing multiple final-state leptons above the SM expectations. 

In this paper, we describe a search for anomalous production of events with at least three charged leptons (electrons, muons, and hadronically decaying \PGt leptons) using $\Pp\Pp$ collision data at $\sqrt{s} = 13\TeV$ collected by the CMS experiment at the CERN LHC during 2016 to 2018, corresponding to an integrated luminosity of 138\fbinv. The final states analysed in this result include production of up to four light leptons, and up to three hadronically decaying \PGt leptons.
The search is carried out in an inclusive fashion, encompassing a number of final states with numerous kinematic properties, which makes it sensitive to a broad range of BSM scenarios. 
Collision events are classified by the number of reconstructed objects, such as charged leptons and \PQb-tagged jets (identified from \PQb quark hadronization); kinematic properties, such as the momenta of individual objects; 
combined properties, such as the invariant mass of lepton pairs; and properties of the entire event, such as missing transverse momentum (\ptvecmiss) or total hadronic energy. 
A set of model-independent signal regions (SRs) are defined without reference to any specific signature or model, but rather to minimize the SM background contributions. 
Results are presented in the form of detailed tables of observed and predicted background yields for these mutually exclusive SRs.

We consider three specific BSM models that address shortcomings of the SM and predict complementary nonresonant multilepton signatures. These BSM models are type-III seesaw heavy fermions~\cite{Minkowski:1977sc,Mohapatra:1979ia,Magg:1980ut,Mohapatra:1980yp,Schechter:1980gr,Schechter:1981cv,Mohapatra:1986aw,Mohapatra:1986bd,Foot:1988aq}, doublet and singlet vector-like extensions of the third-generation of leptons~\cite{delAguila:1982fs,Fishbane:1985gu,Fishbane:1987tx,Montvay:1988av,delAguila:1989rq,Fujikawa:1994we,delAguila:2008pw}, and scalar leptoquarks coupled to a top quark and an SM lepton of any flavor~\cite{Pati:1973uk,Buchmuller:1986zs,Davidson:2011zn,Diaz:2017lit}. 
For the first time, we carry out dedicated analyses for these models using a multivariate approach based on boosted decision trees (BDTs). In addition, the model-independent SRs are also used to set constraints on these models.

This paper is organized as follows. 
We describe the three BSM models in Section~\ref{sec:signals}.
Sections~\ref{sec:detector} and \ref{sec:samples} describe the CMS detector and the data and simulation samples used in this search, respectively. 
Section~\ref{sec:objects} describes the reconstruction and identification of leptons, jets, and \ptvecmiss.
In Section~\ref{sec:eventselec}, we outline the broad event selection, and in Section~\ref{sec:backgrounds}, we describe the background estimation techniques.
Section~\ref{sec:sr} describes the model-independent search categories that span the multilepton phase space, as well as the model-specific event selections using BDTs. 
Section~\ref{sec:systematics} describes the systematic uncertainties in the predictions. 
Section~\ref{sec:results} presents the results of this search, and also discusses the procedure for future interpretations using the model-independent SRs and supporting information made available in a \textsc{HEPData} record~\cite{hepdata}.

\section{Signal models}\label{sec:signals}
\subsection{Type-III seesaw fermions}

The observed nonzero neutrino masses and mixing among lepton flavors can be explained by a seesaw mechanism, which introduces new heavy particles coupled to the SM leptons~\cite{Minkowski:1977sc,Mohapatra:1979ia,Magg:1980ut,Mohapatra:1980yp,Schechter:1980gr,Schechter:1981cv,Mohapatra:1986aw,Mohapatra:1986bd,Foot:1988aq}. 
In these models, the neutrino is a Majorana particle, and the neutrino mass arises via mixing with new massive fermions.
We consider the type-III seesaw model~\cite{Biggio:2011ja} in this paper, which introduces an SU(2) triplet of heavy leptons, including Dirac charged leptons ($\Sigma^\pm$) and a Majorana neutral lepton ($\Sigma^0$).

\begin{figure}[hbt!]
\centering
\includegraphics[width=0.4\textwidth]{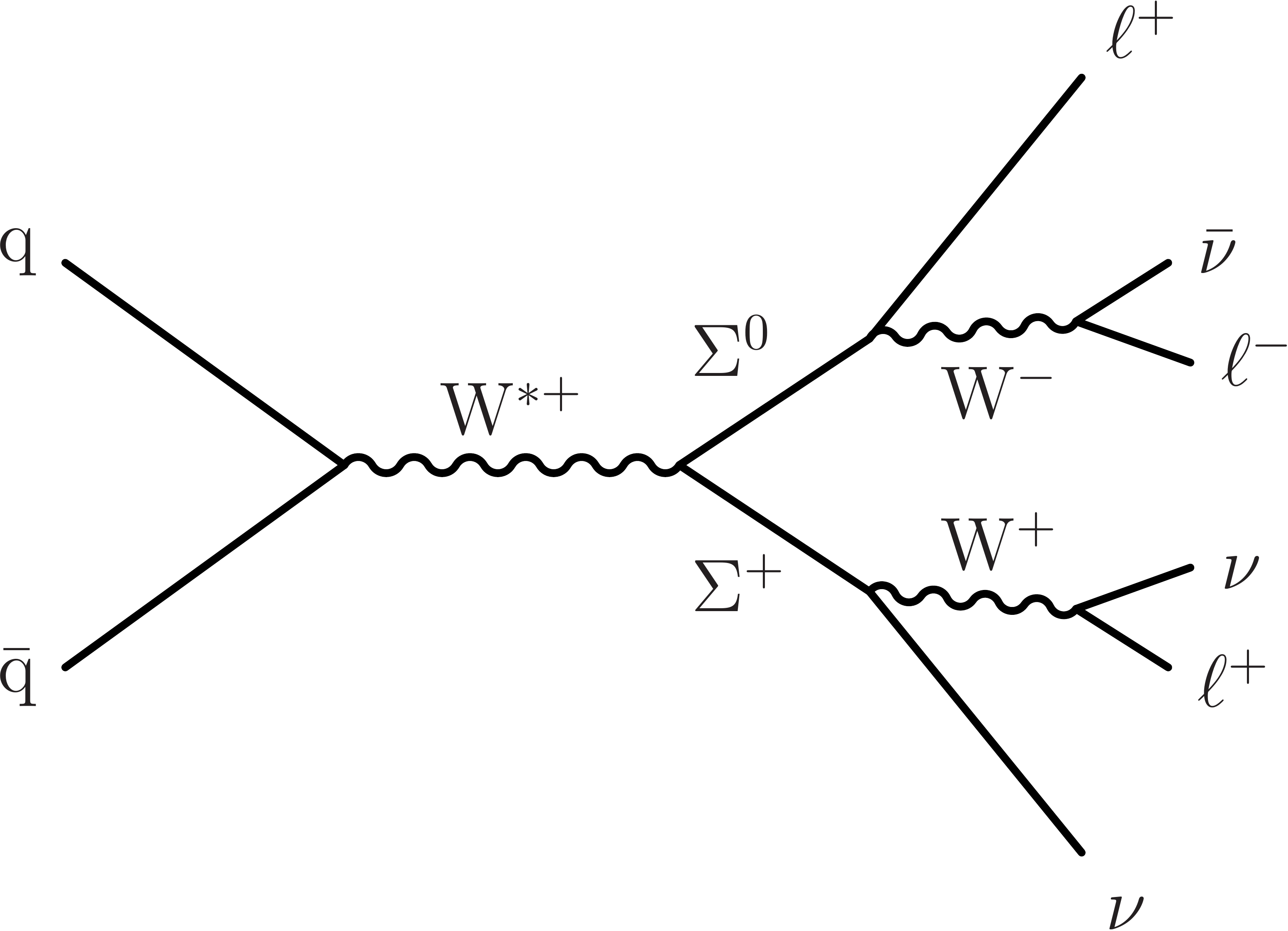} \hspace{.025\textwidth}
\includegraphics[width=0.4\textwidth]{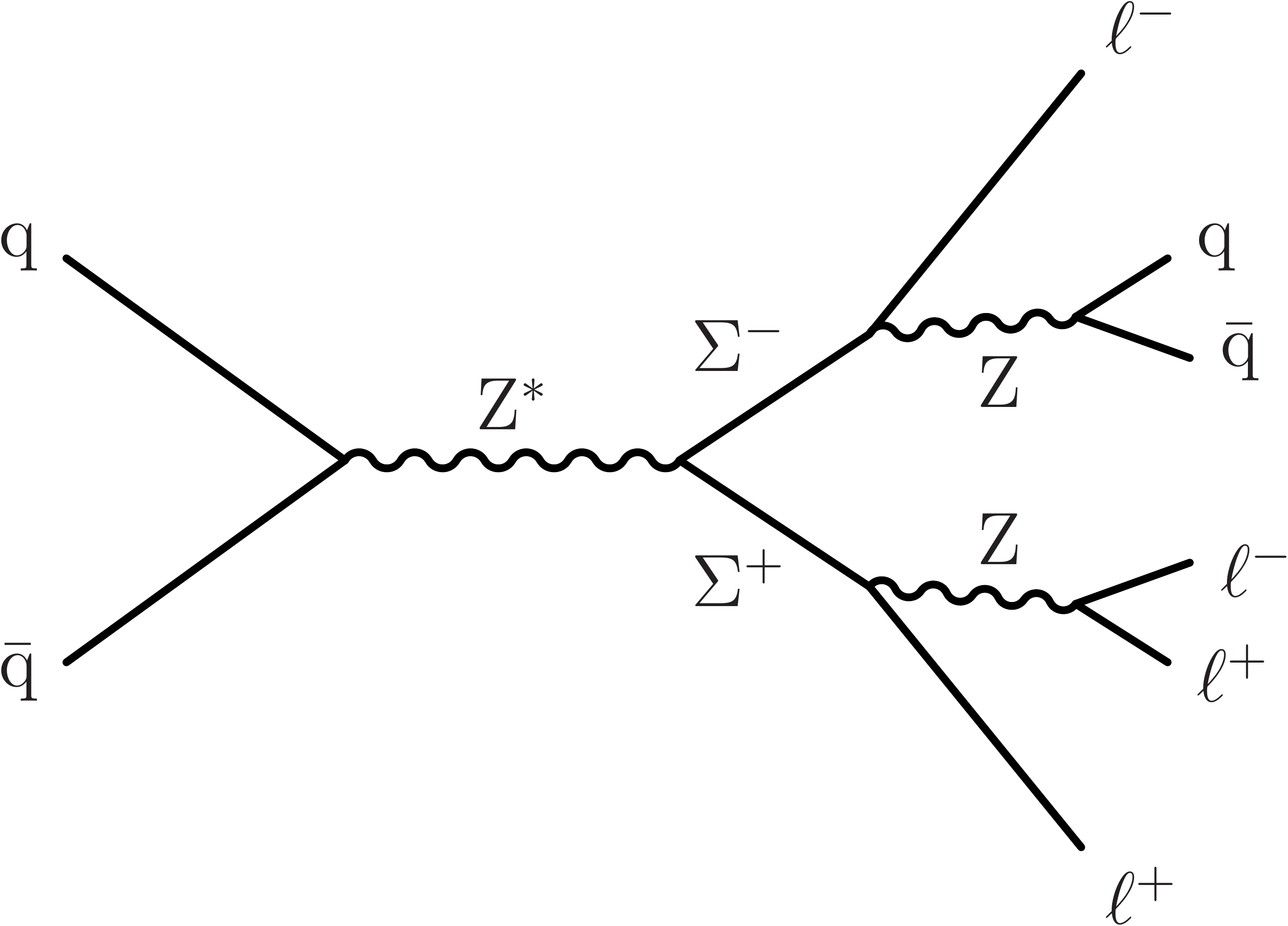}

\caption{\label{fig:SeesawFeynman}
Example processes illustrating production and decay of type-III seesaw heavy fermion pairs at the LHC that result in multilepton final states.
}
\end{figure}

At the LHC, these heavy fermions may be pair-produced through electroweak interactions in both charged-charged $(\Sigma^\pm\Sigma^\mp)$ and charged-neutral $(\Sigma^\pm\Sigma^0)$ modes. 
The seesaw fermions are assumed to mix with SM leptons, and decay to a \PW, \PZ, or Higgs boson (\PH) and an SM lepton (\PGn, or \Pell = \Pe,\PGm,\PGt).
The three production modes, combined with the nine possible combinations of boson-SM lepton decay 
yield 27 distinct signal production and decay modes.
An example of the complete decay chain is $\Sigma^\pm \Sigma^0\to \PW^\pm \PGn \PW^\mp \Pell^\pm \to \Pell^\pm \PGn \PGn \Pell^\mp \PGn \Pell^\pm$. 
Two diagrams exemplifying the production and decay of $\Sigma$ pairs that result in multilepton final states are shown in Fig.~\ref{fig:SeesawFeynman}.
Electroweak and low-energy precision measurements enforce an upper limit on the mixing angles of $10^{\mathrm{-4}}$ across all lepton flavors~\cite{Biggio:2019eeo,Das:2020uer}.
This bound allows for prompt decays of heavy fermions in the mass ranges accessible to collider experiments~\cite{Abada:2007ux,Abada:2008ea,Franceschini:2008pz,Cai:2017mow,Ashanujjaman:2021jhi,Ashanujjaman:2021zrh}.

In this analysis, the $\Sigma^{\mathrm{\pm,0}}$ are assumed to be degenerate in mass and their decays are assumed to be prompt. 
The effects of the radiative mass splitting between the neutral and charged heavy fermions are negligible. 
The $\Sigma$ decay branching fractions to different bosons are determined solely by their masses.
The free parameters are the $\Sigma$ mass, $m_\Sigma$, and the $\Sigma$ decay branching fractions to the SM lepton flavors, $\mathcal{B}_{\Pe}$, $\mathcal{B}_{\PGm}$, and $\mathcal{B}_{\PGt}$, where $\mathcal{B}_{\Pe}+\mathcal{B}_{\PGm}+\mathcal{B}_{\PGt}=1$.

The most stringent limits on the type-III seesaw model come from a search conducted by the ATLAS Collaboration using the combined LHC data set from
2016--2018 at $\sqrt{s}=13\TeV$ in multilepton final states with up to four electrons and muons~\cite{ATLAS:2022yhd}.
The search excluded at 95\% confidence level (\CL) type-III seesaw fermions with masses below 910\GeV in the lepton-flavor-democratic scenario.
Previous constraints in the same scenario by the CMS collaboration from a cut-based search using a comparable data set and in similar final states
excluded type-III seesaw fermions with masses below 880\GeV at 95\% \CL~\cite{Sirunyan:2019bgz}.

\subsection{Vector-like leptons}
Vector-like fermions are hypothetical particles whose left- and right-handed components transform under conjugate representations of the SM gauge symmetries~\cite{delAguila:1982fs,Fishbane:1985gu,Fishbane:1987tx,Montvay:1988av,Fujikawa:1994we}, and hence their masses are independent of the SM Higgs mechanism and are not constrained by electroweak precision measurements~\cite{delAguila:1989rq,delAguila:2008pw}.
Vector-like fermions arise in a wide variety of BSM scenarios, including, but not limited to, supersymmetric models~\cite{Martin:2009bg,Graham:2009gy,Endo:2011mc,Zheng:2019kqu}, models with extra spatial dimensions~\cite{Kong:2010qd,Huang:2012kz}, and grand unified theories~\cite{Nevzorov:2012hs,Dorsner:2014wva,Joglekar:2016yap}.
Extensions of the SM with one or more vector-like fermion families may provide a dark matter candidate~\cite{Schwaller:2013hqa,Halverson:2014nwa,Bahrami:2016has,Bhattacharya:2018fus}, and account for the mass hierarchy between the different generations of particles in the SM via their mixings with the SM fermions~\cite{Agashe:2008fe,Redi:2013pga,Falkowski:2013jya}.
Furthermore, vector-like leptons are also among the proposed solutions to the observed tensions between the experimental measurements and the SM prediction of the anomalous magnetic moment of the muon~\cite{Endo:2011mc,Dermisek:2013gta, Megias:2017dzd,Kawamura:2019rth,Hiller:2020fbu, Muong-2:2006rrc,Muong-2:2021ojo}.

\begin{figure}[hbt!]
\centering
\includegraphics[width=0.4\textwidth]{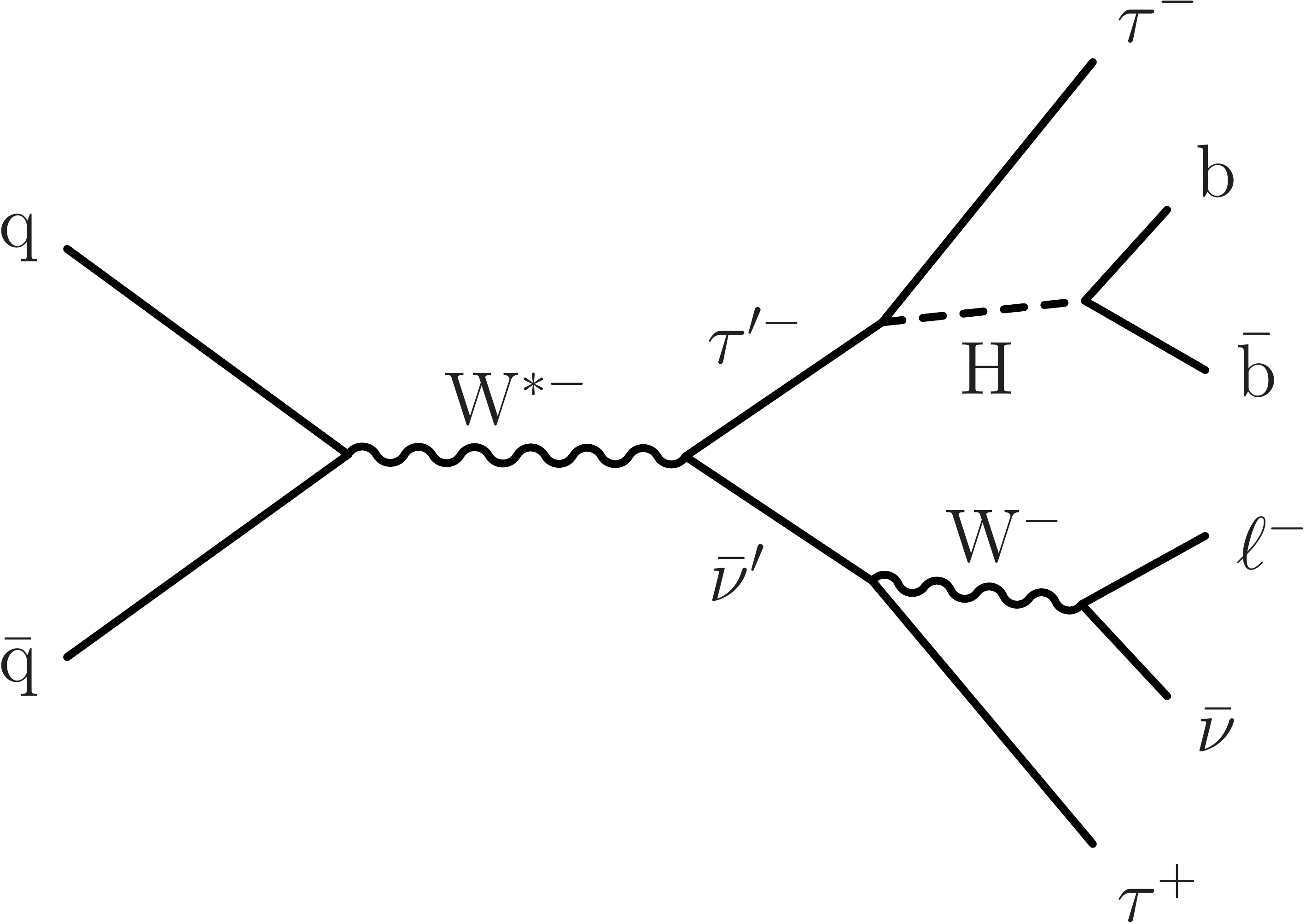} \hspace{.025\textwidth}
\includegraphics[width=0.4\textwidth]{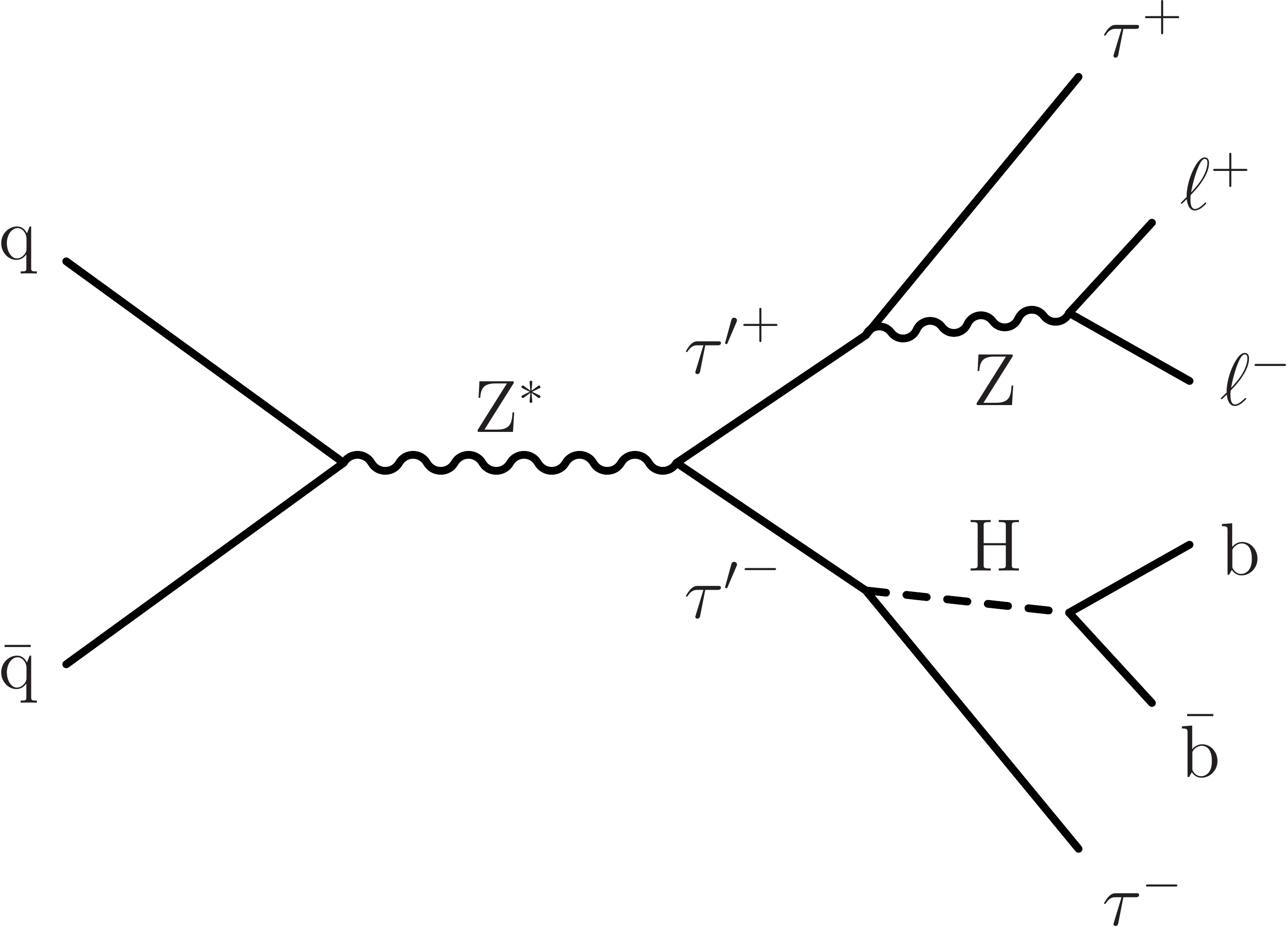}
\caption{\label{fig:VLLFeynman}
Example processes illustrating production and decay of doublet vector-like \PGt lepton pairs  at the LHC that result in multilepton final states. The right diagram also illustrates the singlet scenario.
}
\end{figure}

In this paper, we consider two distinct models in which the vector-like leptons couple to the SM \PGt lepton~\cite{Kumar:2015tna,Bhattiprolu:2019vdu}. 
The vector-like doublet model contains an SU(2) doublet (\vltau,\vlnu), where the \vltau and \vlnu are mass-degenerate at tree level and can be produced in pairs ($\Pp\Pp\to \PGt^{\prime+}\PGt^{\prime-} /~ \vlnu\overline{\vlnu}$) or in association ($\Pp\Pp\to \vltau\vlnu$). 
The decay modes are $\vltau\to \PZ \Pgt$ or $\PH\Pgt$, and $\vlnu\to \PW\Pgt$, with the branching fractions of the \vltau dependent on the mass $m_{\vltau}$.
An example of the complete decay chain for the associated production would be $\vlnu\PGt^{\prime\pm}\to \PW^\pm\Pgt^\mp\PH\Pgt^\pm \to \Pell^\pm\Pgn\Pgt^\mp b\overline{b}\Pgt^\pm$ and for the pair production would be $\vlnu\overline{\vlnu}\to \PW^\pm\Pgt^\mp\PW^\pm\Pgt^\mp \to \Pell^\pm\Pgn\Pgt^\mp \Pell^\pm\Pgn\Pgt^\mp$.
In the vector-like singlet model, only a charged lepton (\vltau) is present. 
The \vltau can decay to either $\PZ \Pgt$ or $\PH\Pgt$, or $\PW\Pgn$, with the branching fractions similarly governed by $m_{\vltau}$. 
Figure~\ref{fig:VLLFeynman} shows two processes from the doublet and singlet models, which exemplify the production and decay of vector-like \PGt lepton pairs that result in multilepton final states.

Electroweak precision data constrain the mixing angle between vector-like leptons and SM leptons to be less than about $10^{\mathrm{-2}}$, permitting prompt decays for mass values that are close to the electroweak scale~\cite{Dermisek:2014cia,Dermisek:2014qca}.
We assume prompt decays of vector-like \PGt leptons; aside from this assumption, the analysis is insensitive to the precise values of the mixing angles. 

The most stringent constraints on models with vector-like \PGt lepton doublets are from a search conducted by the CMS Collaboration~\cite{Sirunyan:2019ofn} with 77\fbinv of data collected in 2016--2017, which excludes them in the mass range of 120--790\GeV. The search is performed with multilepton final states consisting of up to four electrons and muons, and also an additional final state with two light leptons along with one hadronically decaying \PGt lepton.
There are, so far, no direct constraints on the vector-like \PGt lepton singlet model from any of the LHC experiments.
The L3 Collaboration at the LEP placed a lower bound of $\sim$100\GeV on the mass of additional heavy leptons~\cite{Achard:2001qw}.

\subsection{Leptoquarks}
Leptoquarks are color-triplet scalar or vector bosons that carry nonzero baryon and lepton quantum numbers and fractional electric charge~\cite{Buchmuller:1986zs}. 
Such particles commonly emerge in grand unified theories, \eg, based on $SU(4)$~\cite{Pati:1974yy}, $SU(5)$~\cite{Georgi:1974sy}, or $SO(10)$~\cite{Fritzsch:1974nn} schemes, models with compositeness~\cite{Gripaios:2014tna,DaRold:2018moy}, and \textit{R}-parity violating supersymmetry models~\cite{Weinberg:1981wj,Barbier:2004ez}. 

In $\Pp\Pp$ collisions at the LHC, scalar leptoquarks (\PS) could be pair-produced via strong interactions, with the production cross section depending only on the leptoquark mass, $m_S$, but not on the unknown Yukawa coupling.
Depending on the nature of the Yukawa coupling, such leptoquarks are expected to decay either to an up-type quark and a charged lepton or to a down-type quark and a neutrino, with branching fractions $\beta$ and $1-\beta$, respectively. 
We assume that the Yukawa couplings involve only one generation of quarks or leptons. 
The simultaneous coupling of leptoquarks to more than one generation of quarks or leptons that are not aligned with the SM Yukawa couplings may lead to quark or lepton flavor violation~\cite{Mandal:2019gff,Diaz:2017lit}. 

\begin{figure}[hbt!]
\centering
\includegraphics[width=0.4\textwidth]{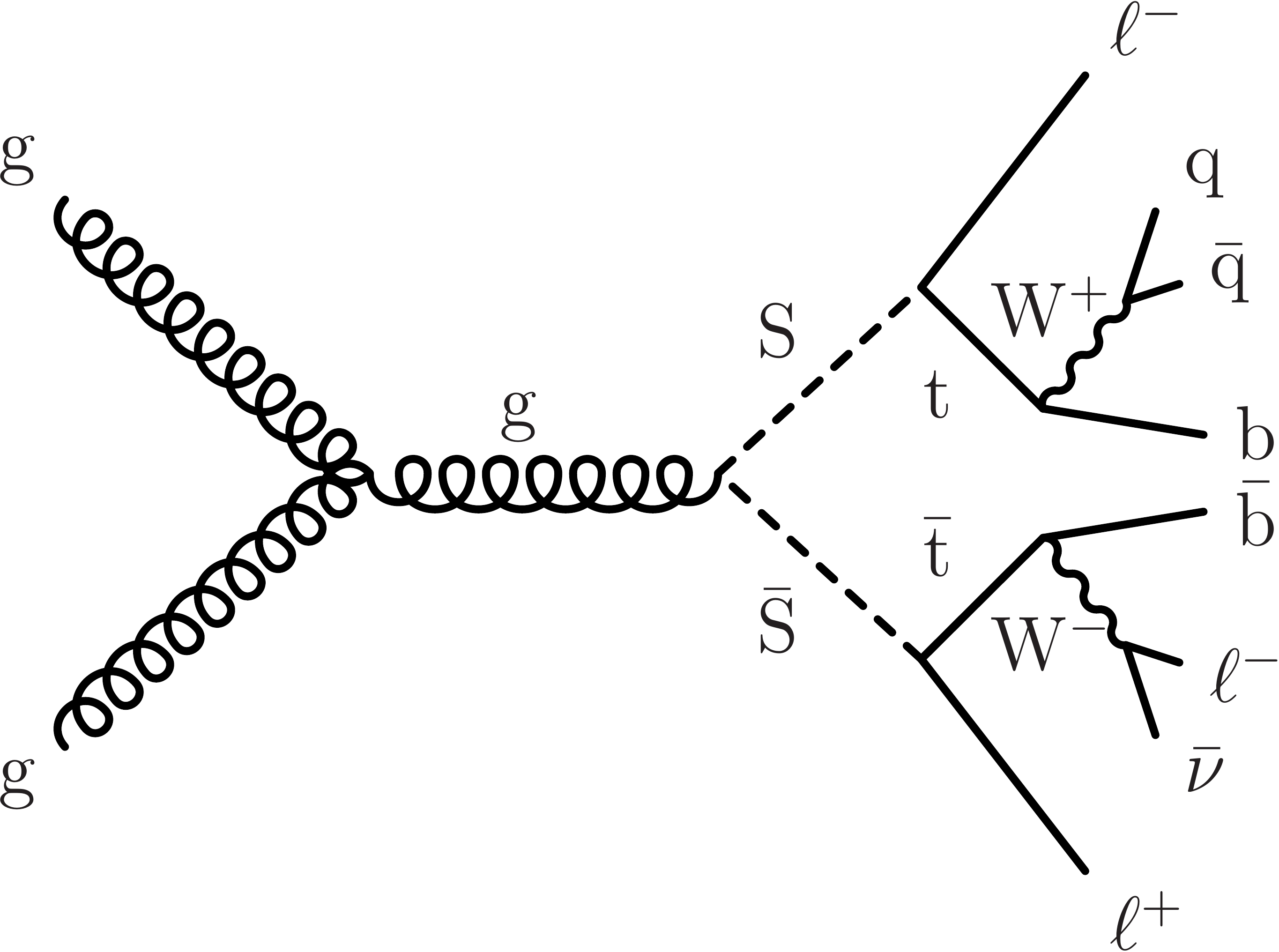} \hspace{.025\textwidth}
\includegraphics[width=0.4\textwidth]{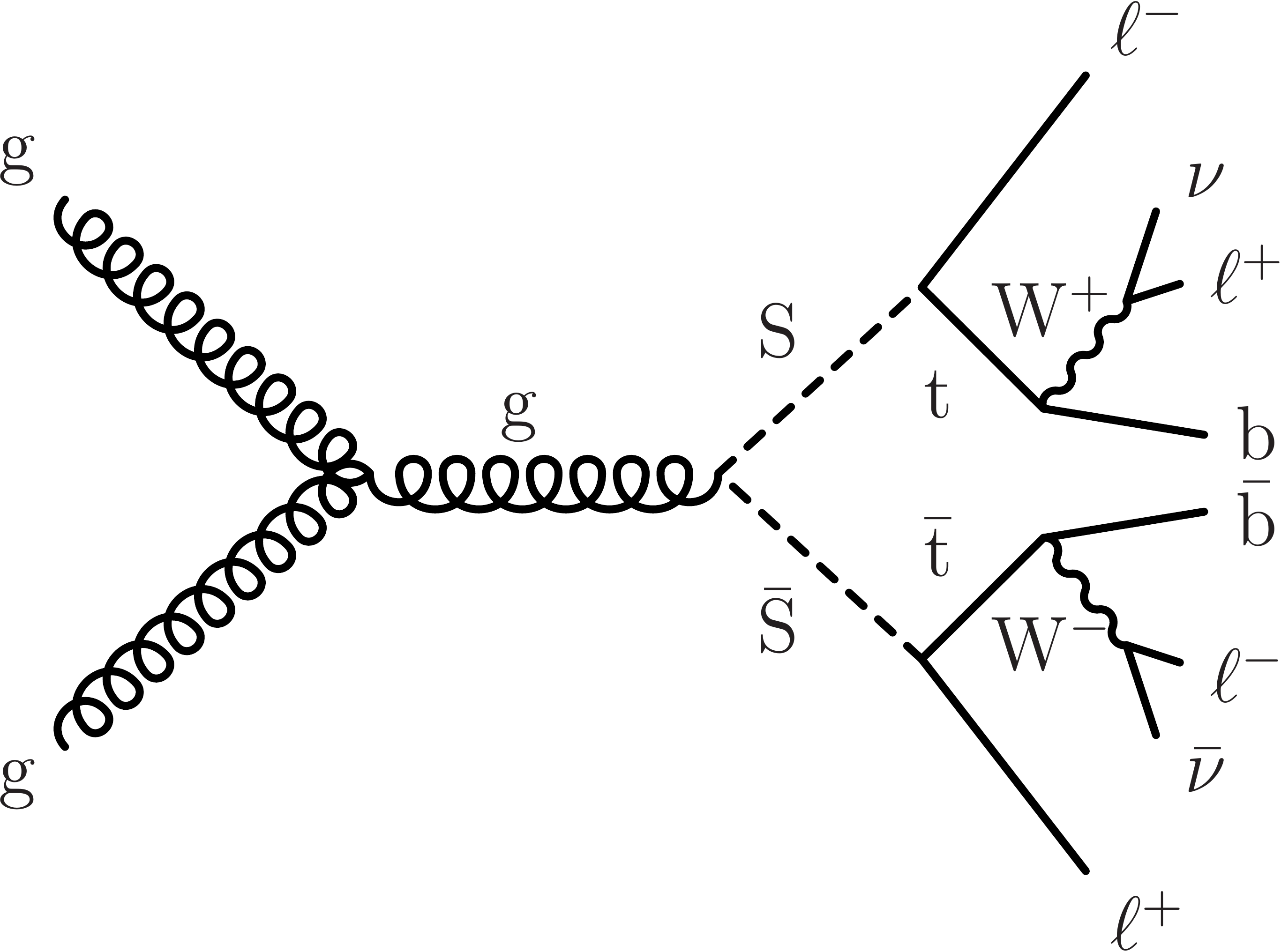}
\caption{\label{fig:LQFeynman}
Example processes illustrating the production and decay of scalar leptoquark pairs in $\Pp\Pp$ collisions at the LHC that result in multilepton final states.
}
\end{figure}

In this analysis, we consider scalar leptoquarks~\cite{Davidson:2011zn} with electric charge of $-1/3\abs{e}$, and a nonzero Yukawa coupling to the top quark and a single flavor of SM charged lepton.
In a supersymmetric theory, these leptoquarks are right handed down-type squarks that couple to the top quark and charged leptons through leptonic-hadronic \textit{R} parity violating interactions, where the down-type squarks are the scalar partners of the SM down-type quarks.
We assume that only one flavor of charged lepton coupling dominates at a time, and hence consider leptoquark branching fractions $\mathcal{B}_{\Pe}=1$, $\mathcal{B}_{\PGm}=1$, or $\mathcal{B}_{\PGt}=1$, for leptoquarks decaying into a top quark and a charged lepton of the first-, second-, or third-generation, respectively.
We target the mass range from just above the top quark mass up to the \TeV scale.
Furthermore, the leptoquark decays are assumed to be prompt, and the coupling is assumed to satisfy $\lesssim$0.1, within the bounds on such Yukawa couplings from leptonic \PZ boson decays~\cite{Mizukoshi:1994zy,Davidson:2011zn}.
As with the type-III seesaw and vector-like lepton models, the analysis is independent of the magnitude of the leptoquark Yukawa couplings aside from the assumption of prompt decays.
Figure~\ref{fig:LQFeynman} shows two processes exemplifying the production and decay of leptoquark pairs that result in multilepton final states.

Leptoquarks with preferential couplings to third-generation fermions have been suggested among the possible extensions of the SM~\cite{Alvarez:2018gxs,Angelescu:2018tyl,Crivellin:2019dwb,Saad:2020ucl,Haisch:2020xjd} motivated by a series of anomalies recently observed in charged- and neutral-current B meson decays, $\Pb\to \mathrm{c} \Pell\PGn$~\cite{BaBar:2013mob,Belle:2019rba,LHCb:2017vlu,LHCb:2017smo,LHCb:2021trn} and $\Pb\to \mathrm{s} \Pell\Pell$~\cite{Belle:2016fev,LHCb:2017avl,LHCb:2019hip}, respectively.
The ATLAS and CMS Collaborations have conducted a number of searches for leptoquarks with flavor-diagonal and cross-generational couplings involving third-generation fermions~\cite{Aad:2020jmj,Aad:2021rrh,ATLAS:2019qpq,Sirunyan:2018ruf,CMS:2018svy,Sirunyan:2020zbk,CMS:2018qqq,CMS:2018iye,CMS:2018txo}.
The most stringent constraints on scalar leptoquarks with 100\% branching fraction to a top quark and first-, second-, or third-generation lepton are set by ATLAS, excluding such particles with masses below 1.48, 1.47\TeV~\cite{Aad:2020jmj} and 1.43\TeV~\cite{Aad:2021rrh}, respectively.
Similarly, CMS has excluded scalar leptoquarks decaying to a top quark and a \PGt lepton or a bottom quark and a neutrino with equal branching fractions ($\beta=0.5$) with masses below 950\GeV~\cite{Sirunyan:2020zbk}. The final states include hadronically decaying top quark and \PGt lepton, \PQb-tagged jet, and significant missing energy.

\section{The CMS detector} \label{sec:detector}

The central feature of the CMS apparatus is a superconducting solenoid of 6\unit{m} internal diameter, providing a magnetic field of 3.8\unit{T}. 
Within the solenoid volume are a silicon pixel and strip tracker, a lead tungstate crystal electromagnetic calorimeter (ECAL), and a brass and scintillator hadron calorimeter (HCAL), each composed of a barrel and two endcap sections. 
Forward calorimeters extend the pseudorapidity ($\eta$) coverage provided by the barrel and endcap detectors. 
Muons are detected in gas-ionization chambers embedded in the steel flux-return yoke outside the solenoid. 
A more detailed description of the CMS detector, together with a definition of the coordinate system used and the relevant kinematic variables, can be found in Ref.~\cite{Chatrchyan:2008zzk}.

Events of interest are selected using a two-tiered trigger system. 
The first level, composed of custom hardware processors, uses information from the calorimeters and muon detectors to select events at a rate of around 100\unit{kHz} within a fixed latency of about 4\mus~\cite{Sirunyan:2020zal}. 
The second level, known as the high-level trigger, consists of a farm of processors running a version of the full event reconstruction software optimized for fast processing, and reduces the event rate to around 1\unit{kHz} before data storage~\cite{Khachatryan:2016bia}. 

\section{Data samples and event simulation} \label{sec:samples}

The total integrated luminosity recorded by CMS in $\Pp\Pp$ collisions at $\sqrt{s} = 13\TeV$ corresponds to 138\fbinv, with 36.3, 41.5, and 59.8\fbinv recorded in the years 2016, 2017, and 2018, respectively. 
The data presented here are collected using a combination of isolated single-electron (-muon) triggers with corresponding transverse momentum (\pt) thresholds of
27 (24)\GeV in 2016, and 32 (27)\GeV in 2017, and 32 (24)\GeV in 2018. 
The rates of signal and SM background processes that gives rise to isolated and nondisplaced leptons are estimated from Monte Carlo (MC) simulations, which incorporate detailed detector and $\Pp\Pp$ collision properties. 

The $\ZG$, $\WZ$, $\ttV$, and triboson ($\VVV$) backgrounds, where \PV denotes a \PW or \PZ boson, are generated using \MGvATNLO (versions 2.2.2 for 2016 data and 2.4.2 for 2017 and 2018 data)~\cite{Alwall:2014hca} at next-to-leading order (NLO) precision in perturbative quantum chromodynamics (QCD). 
The top quark mass used in all simulations is 172.5\GeV. 
The $\ZG$ background includes all diagrams contributing to $\Pp\Pp\to \Pell\Pell\PGg$, with photons from both initial-state radiation (ISR) and final-state radiation (FSR), and with an invariant mass cut of $m(\Pell^+\Pell^-)>10\GeV$. 
The $\ZZ$ background contribution from quark-antiquark annihilation production is generated using
\POWHEG 2.0~\cite{Nason:2004rx,Frixione:2007vw,Alioli:2010xd} at NLO, whereas the contribution from gluon-gluon fusion production is generated at leading order (LO) using \MCFM 7.0.1~\cite{Campbell:2010ff}. 
The SM processes involving Higgs boson production are generated using \POWHEG, \MGvATNLO and \textsc{JHUGen} 7.0.11~\cite{Gao:2010qx,Bolognesi:2012mm,Anderson:2013afp,Gritsan:2016hjl} at NLO, for a Higgs boson mass of 125\GeV. 
Processes with a single top quark and a \PZ boson or with four top quarks are simulated using \MGvATNLO at NLO in QCD.
Other small contributions from processes involving a single top quark and an electroweak or Higgs boson, two top quarks and two bosons, or three top quarks are simulated using \MGvATNLO at LO in QCD. 
Simulated event samples for the Drell--Yan (DY) and $\ttbar$ processes, which are used for systematic uncertainty studies and in the BDT training process, are generated at NLO with \MGvATNLO and \POWHEG, respectively.

All signal samples are simulated at LO precision. 
The type-III seesaw and vector-like lepton samples are generated with \MGvATNLO~2.6.1, whereas the leptoquark samples are generated with \PYTHIA 8.212 (8.230) in 2016 (2017 and 2018)~\cite{Sjostrand:2014zea}.
The production cross sections for the type-III seesaw signal model are calculated at NLO plus next-to-leading
logarithmic precision, assuming that the heavy leptons are SU(2) triplet fermions~\cite{Fuks:2012qx,Fuks:2013vua}.
Similarly, vector-like lepton and leptoquark cross sections are calculated at NLO precision~\cite{Bhattiprolu:2019vdu,Blumlein:1996qp,Kramer:2004df}. 
In this paper, these higher-order cross sections are used in the analysis of these BSM models.

The  NNPDF3.0 LO or NLO parton distribution function (PDF) sets~\cite{Ball:2014uwa} are used for all background and signal samples for 2016 data, with order matching that of the matrix element calculations. 
The NNPDF3.1 next-to-NLO order (NNLO) PDF set~\cite{Ball:2017nwa} is used for all 2017 and 2018 samples.
To perform the parton showering, fragmentation, and hadronization of the matrix-level events in all samples, \PYTHIA 8.212 is used with the event tune CUETP8M1~\cite{Khachatryan:2015pea} for 2016, and \PYTHIA 8.230 is used with the event tune CP5~\cite{CMS:2019csb} for 2017 and 2018. The MLM~\cite{Hoeche:2006ps} or FxFx~\cite{Frederix:2012ps} jet matching schemes are used for \MGvATNLO samples at LO or NLO, respectively. 
The simulation of the response of the CMS detector to incoming particles is performed using the $\GEANTfour$ toolkit~\cite{Agostinelli:2002hh}. 
Additional inelastic $\Pp\Pp$ interactions from the same or nearby bunch crossings (pileup) are simulated and incorporated in the MC samples.

\section{Event reconstruction and particle identification} \label{sec:objects}

In each event, the candidate vertex with the largest total physics-object $\pt^2$ is taken to be the primary $\Pp\Pp$ interaction vertex (PV). 
The physics objects are the jets, clustered using the anti-\kt algorithm~\cite{Cacciari:2008gp,Cacciari:2011ma} with the tracks assigned to candidate vertices as inputs, and the associated \ptvecmiss, which is the negative vector \pt sum of those jets.

The reconstruction and identification of individual particles in an event is based on the particle-flow (PF) algorithm~\cite{Sirunyan:2017ulk}, with an optimized combination of information from the various elements of the CMS detector. 
The energy of photons is obtained from the ECAL measurement. 
The energy of electrons is determined from the electron momentum at the PV as determined by the tracker, the energy of the corresponding ECAL cluster, and the energy sum of all bremsstrahlung photons spatially compatible with originating from the electron track. 
The momentum of muons is determined from the curvature of the corresponding track, and the energy is obtained from the momentum.
The energy of charged hadrons is determined from a combination of their momentum measured in the tracker and the matching ECAL and HCAL energy deposits, corrected for the response function of the calorimeters to hadronic showers. 
Finally, the energy of neutral hadrons is obtained from the corresponding corrected ECAL and HCAL energies.

Electrons are reconstructed by geometrically matching charged-particle tracks from the tracking system with energy clusters deposited in the ECAL~\cite{CMS:2020uim}.
The electron momentum is estimated by combining the energy measurement in the ECAL with the momentum measurement in the tracker.
The momentum resolution for electrons with $\pt \approx 45\GeV$ from $\PZ \to \Pe \Pe$ decays ranges from 1.7 to 4.5\%.
It is generally better in the barrel region than in the endcaps, and also depends on the bremsstrahlung energy emitted by the electron as it traverses the material in front of the ECAL.
To suppress undesired electrons originating from photon conversions in detector material, as well as the misidentification of hadrons, the electron candidates are required to satisfy shower shape and track quality requirements, using the medium cut-based criteria described in Ref.~\cite{CMS:2020uim}.
Electrons used in this analysis are also required to satisfy $\pt>10\GeV$ and $\abs{\eta}<2.4$.

Muons are reconstructed from compatible tracks in the inner tracker and the muon detectors~\cite{Sirunyan:2018fpa}.
Additional track fit and matching quality criteria suppress the misidentification of hadronic showers that punch through the calorimeters and reach the muon system.
Matching tracks measured in the inner tracker and the muon detectors results in a relative \pt resolution, for muons with \pt up to 100\GeV, of 1\% in the barrel and 3\% in the endcaps, and of better than 7\% in the barrel for muons with \pt up to 1\TeV~\cite{Sirunyan:2018fpa}.
Muons used in this analysis must lie within the tracking system acceptance, $\abs{\eta}<2.4$, and are required to have $\pt>10\GeV$.

Hadronically decaying \Pgt lepton candidates (\tauh) are reconstructed from jets, using the hadrons-plus-strips algorithm~\cite{Sirunyan:2018pgf}, which combines one or three tracks with energy deposits in the calorimeters, to identify the corresponding one- or three-prong \PGt lepton decay modes. 
Neutral pions from \PGt lepton decay are reconstructed as strips with variable size in $\eta$-$\phi$ from reconstructed electrons and photons, where the $\phi$ is azimuthal angle in radians and the strip size varies as a function of the \pt of the electron or photon candidate. 
The reconstructed \tauh candidate must satisfy $\abs{\eta}<2.3$ and $\pt>20\GeV$.

Jets are clustered using the anti-\kt algorithm~\cite{Cacciari:2008gp} with a distance parameter of 0.4, as implemented in the \textsc{FastJet} package~\cite{Cacciari:2011ma}. 
The minimum \pt threshold for the jets selected in this analysis is 30\GeV and the central axis of the jet is also required to be inside the tracking acceptance, $\abs{\eta}<2.4$. 
Jets are composite objects made up of several particles, hence the momentum is determined as the vectorial sum of all particle momenta, and is found from simulation to be, on average, within 5--10\% of the true momentum over the whole \pt spectrum and detector acceptance. 
Additional $\Pp\Pp$ interactions within the same or nearby bunch crossings can contribute additional tracks and calorimetric energy depositions, increasing the apparent jet momentum.
To mitigate the effect of the charged-particle contribution from pileup on reconstructed jets, a charged hadron subtraction technique is employed, which removes the energy of charged hadrons not originating from the PV~\cite{Sirunyan:2017ulk}.
In addition, the impact of neutral pileup particles in jets is mitigated by an event-by-event jet-area-based correction of the jet four-momenta~\cite{Cacciari:2008ca,Cacciari:2008ps,Sirunyan:2017jes}. 
Aside from pileup contamination removal, additional quality criteria are applied to each jet to remove those potentially mismeasured because of instrumental effects or reconstruction failures~\cite{CMS:2017jme}. 
Finally, the qualifying jets must lie outside a cone of $\Delta R \equiv \sqrt{\smash[b]{(\Delta\eta)^2+(\Delta\phi)^2}}=0.4$ around a selected muon, electron, or \tauh candidate, where $\Delta \phi$ is the $\phi$ angle between the jet and lepton.

Jet energy corrections are derived from simulation studies so that the average measured energy of jets matches that of particle level jets. 
In situ measurements of the \pt balance in dijet, photon+jet, leptonically decaying $\PZ$+jet, and multijet events are used to determine any residual differences between the jet energy scale in data and in simulation, and appropriate corrections are made to the jet \pt~\cite{Sirunyan:2017jes}. 

The reconstructed jets originating from \PQb hadrons are identified using the medium working point of the \textsc{DeepCSV} {\cPqb} tagging algorithm~\cite{Sirunyan:2017ezt}. 
This working point has an identification efficiency of 60--75\% for {\cPqb} quark jets, depending on jet \pt and $\eta$, and a misidentification probability of about 10\% for {\cPqc} quark jets and about 1\% for light-quark and gluon jets.

The vector \ptvecmiss is defined as the negative vector \pt sum of all the PF candidates in an event, and its magnitude is denoted as \ptmiss~\cite{Sirunyan:2019kia}. 
The pileup-per-particle identification algorithm~\cite{Bertolini:2014bba} is applied to reduce the pileup dependence of the \ptvecmiss observable.
The \ptvecmiss is computed from the PF candidates weighted by their probability to originate from the PV, and is modified to account for corrections to the energy scale of the reconstructed jets in the event. 

The leptons that are produced from the decays of the SM bosons \PW, \PZ, \PH (either directly, or via an intermediate \PGt lepton) are referred to as prompt leptons, and are often indistinguishable in momentum and isolation from those produced in signal events. 
Thus, the SM processes giving rise to three or more isolated leptons, such as $\WZ$, $\ZZ$, $\ttV$, $\VVV$, and Higgs boson production, constitute the irreducible backgrounds in this analysis.
On the other hand, reducible backgrounds come from SM processes in which the jets are misidentified as leptons, or where the leptons originate from heavy-quark decays. 
Some examples of such backgrounds are $\PZ$+jets or $\ttbar$+jets production, in which the prompt leptons are accompanied by leptons that are within or near jets, 
hadrons that traverse the HCAL and reach the muon detectors, or hadronic showers with large electromagnetic energy fractions. 
Leptons from such sources are referred to as misidentified leptons, and SM background processes with such misidentified leptons are collectively labeled as ``MisID'' backgrounds in the subsequent discussion.

The reducible backgrounds are significantly suppressed by applying stringent requirements on the lepton isolation and displacement. 
For electron and muon candidates, the relative isolation is defined as the scalar \pt sum, normalized to the lepton \pt, of photon and hadron PF objects within a cone of radius $\Delta R $ around the lepton.
For electrons, the relative isolation is required to be less than $0.0478+0.506\GeV/\pt$ in the barrel ($\abs{\eta}<1.479$) and less than $0.0658+0.963\GeV/\pt$ in the endcap ($\abs{\eta}>1.479$), with $\Delta R=0.3$.
The relative isolation for muons is required to be less than 0.15 with $\Delta R=0.4$.
The isolation quantities are also corrected for contributions from particles originating from pileup vertices.  
In addition to the isolation requirement, electrons in the barrel must satisfy $\abs{d_{\mathrm{z}}}<0.1\cm$ and $\abs{d_{\mathrm{xy}}}<0.05\cm$,
and in the endcap $\abs{d_{\mathrm{z}}}<0.2\cm$ and $\abs{d_{\mathrm{xy}}}<0.1\cm$, 
where $d_{\mathrm{z}}$ and $d_{\mathrm{xy}}$ are the longitudinal and transverse impact parameters of electrons with respect to the PV, respectively. 
Similarly, muons must satisfy $\abs{d_{\mathrm{z}}}<0.1\cm$ and $\abs{d_{\mathrm{xy}}}<0.05\cm$.
For both electrons and muons, the three-dimensional impact parameter significance, the impact parameter value divided by its uncertainty, must be less than 10, 12, and 9 in 2016, 2017, and 2018 data, respectively.
All selected electrons within a cone of $\Delta R<0.05$ of a selected muon are discarded in order to reduce bremsstrahlung contributions from muons. 

For \PGt leptons, the \textsc{DeepTau}~\cite{TAU-20-001} algorithm is used to distinguish genuine hadronic tau lepton decays from jets originating from the hadronization of quarks or gluons, as well as from electrons or muons. 
Information from all individual reconstructed particles near the \tauh axis is combined with properties of the \tauh candidate and the event. In addition to this multivariate requirement, \tauh candidates must satisfy $\abs{d_{\mathrm{z}}}<0.2\cm$. 
All selected \tauh candidates within a cone of $\Delta R<0.5$ of a selected electron or muon are also discarded to suppress misidentified tau leptons.

Additionally, to suppress misidentified leptons originating from heavy-flavor decays, leptons are discarded if a \PQb-tagged jet with $\pt>10\GeV$ and $\abs{\eta}<2.5$ is found within a cone of radius $\Delta R<0.4$ around the lepton. 

These lepton reconstruction and selection requirements result in typical efficiencies of 40--85\%, 65--90\%, and 20--50\% for electrons, muons, and \tauh leptons, respectively, depending on lepton \pt and $\eta$.

\section{Event selection} \label{sec:eventselec}

We consider seven distinct final states (channels) based on the number of light leptons and \tauh candidates.
These seven channels are orthogonal, and are defined as:
\begin{itemize}
\item $\ge$4 light leptons and any number of \tauh candidates (4L),
\item exactly 3 light leptons and $\ge$1 \tauh candidates (3L1T),
\item exactly 3 light leptons and no \tauh candidates (3L), 
\item exactly 2 light leptons and $\ge$2 \tauh candidates (2L2T),
\item exactly 2 light leptons and exactly 1 \tauh candidates (2L1T),
\item exactly 1 light lepton and $\ge$3 \tauh candidates (1L3T), and
\item exactly 1 light lepton and exactly 2 \tauh candidates (1L2T).
\end{itemize}
In the 4L channel, only the leading four light leptons in \pt are used in the subsequent analysis. 
Likewise, in the 3L1T, 2L2T, and 1L3T channels, only the leading 1, 2, and 3 \tauh candidates are used, respectively.
In each channel,
at least one muon with \pt $>$ 26 (29)\GeV in 2016 and 2018 (2017) or at least one electron with \pt $>$ 30 (35)\GeV in 2016 (2017 and 2018) is required, with the thresholds set in order to be consistent with the triggers used.

The events in these seven channels are further classified based on several event properties. 
This classification is used to enhance sensitivity to particular signal decay chains, or to define dedicated selections to help constrain the SM backgrounds. 
The quantities are defined below.
\begin{itemize}
\item Scalar momentum sums:  We define \LT as the scalar \pt sum of all charged leptons that constitute the channel. 
For example, in the 4L channel, \LT is calculated from the leading four light leptons in \pt, while for the 3L1T channel, it is calculated from the three light leptons and the leading \tauh. 
We define $\HT$ as the scalar \pt sum of all jets.
Additionally, the scalar sum of \LT, $\HT$, and \ptmiss is defined as \ST. 
The quantity $\LTmet$ is also of interest. 
For the signal models considered in this analysis, high signal mass hypotheses give rise to events with high \LT, $\HT$, \ptmiss, and \ST.
\item Charge and flavor combinations: We count the number OSSF$n$ as distinct opposite-sign (electric charge) same-flavor lepton pairs in an event.
Specific lepton pairs are labeled as OSSF (opposite-sign, same-flavor) and
OSDF (opposite-sign, different-flavor). 
We define \ql as the sum of charges of all leptons in the event.
\item Invariant and transverse masses:
 We define \ml as the invariant mass of all leptons in the event, and \mmin as the minimum invariant mass of all dilepton pairs in the event, irrespective of charge or flavor.
 Additionally, the invariant mass of leptons $i$ and $j$ is defined as $M^{\mathrm{ij}}$.
 The transverse mass for a single lepton $i$ is defined as $\MT^{\mathrm{i}} =(2\ptmiss \pt^{\mathrm{i}}[1-\text{cos}(\vec{p}_{\mathrm{T}}^{\text{miss}},\ptvec^{\mathrm{i}})])^{\mathrm{1/2}}$, where $\pt^{\mathrm{i}}$ is the \pt of lepton $i$.
 Similarly, $\MT^{\mathrm{ij}}$ is defined as the transverse mass calculated with the \ptmiss and the resultant 4-momentum sum of lepton $i$ and $j$. 
 The lepton indices run over up to 4 leptons, in descending \pt order.

We define the \mossf variable in a given event as the OSSF dielectron or dimuon mass closest to the \PZ boson mass at $91\GeV$~\cite{Zyla:2020zbs}, subject to some additional constraints, 
and label events with \mossf within 15\GeV of the \PZ boson mass (76--106\GeV mass window) as OnZ. 
Throughout this paper, OSSF1 or OSSF2 events that are not OnZ are labeled as OffZ. In the 3L event with a distinct OSSF pair (such as in $\Pe^+\Pe^-\PGm^+$, $\PGm^+\PGm^-\Pe^-$), the event is classified as BelowZ, OnZ, or AboveZ if the $\mossf<76\GeV$, within 76--106\GeV, or $>$106\GeV, respectively. In the 3L events with two nondistinct OSSF pairs (such as in $\Pe^+\Pe^-\Pe^+$), the events are classified as OnZ if either pair satisfies \mossf within 76--106\GeV, as BelowZ if both pairs have masses $<$76\GeV, or as AboveZ if both pairs have masses $>$106\GeV. In cases where one pair has mass $<$76\GeV and the other pair has mass $>$106\GeV, the event is classified as MixedZ.

In 3L OSSF1 events, the \MT variable is defined as $\MT^{\mathrm{i}}$, where the lepton $i$ is not part of the \mossf pair. 
In events with three electrons or three muons, the \mossf and \MT variables are chosen simultaneously so that the event is OnZ, and \MT is in the range 50--150\GeV, where this is kinematically possible. 
Similarly, in 4L OSSF2 events with four electrons or muons, \mossf is chosen to give the maximum number of nonoverlapping OSSF pairs with masses within the \PZ boson mass window.
Such events are labeled as Single- or Double-OnZ, respectively, depending on whether they have one or two nonoverlapping OnZ OSSF pairs.
Additionally, the \pt of the \mossf lepton pair is defined as $\pt^{\text{OSSF}}$.

The signal models and the SM backgrounds can have multiple \PW and \PZ bosons in the decay chains. 
The invariant and transverse mass quantities aid in
defining regions to isolate these specific decays. 
The \mossf and \MT variables primarily isolate events with $\PZ\to \Pell \Pell$ and $\PW\to \Pell \PGn$ decays, respectively, while $\MT^{\mathrm{ij}}$ is useful in describing signal events with two visible leptons and \ptmiss, such as $\vlnu\to \PW\PGt\to \Pell\PGt\PGn$ in the vector-like lepton doublet model.

\item Angular quantities: We define \dRmin as the minimum \dR between all the dilepton pairings in an event, irrespective of charge or flavor. 
Similarly, $\dRmin^{\tauh}$ is defined as the minimum \dR between any dilepton pair, where at least one of the leptons is a \tauh candidate.
The quantities $\Delta \phi^{\mathrm{ij}}$ and $\Delta\eta^{\mathrm{ij}}$ are defined as the azimuthal angle or pseudorapidity difference between the $i^{\mathrm{th}}$ and $j^{\mathrm{th}}$ lepton, whereas $\Delta \phi^{\mathrm{i}}$ is defined to denote the opening azimuthal angle between lepton $i$ and \ptvecmiss. 
These are quantities that help to characterize the topology of the signal and background events.

\item Counts: We define \nj as the multiplicity of jets and \nbj as the multiplicity of \PQb-tagged jets satisfying the selection criteria defined earlier.
\end{itemize}

Finally, all events with $\mmin<12\GeV$, $\dRmin<0.2$, or $\dRmin^{\tauh}<0.5$ are vetoed in order to suppress contributions due to low-mass resonances (\PJGy, \PGU) and low-\dR FSR photons.

\section{Background estimation}\label{sec:backgrounds}

A set of control regions (CRs) dominated by the primary background processes is used for the purpose of SM background determination. 
A summary of all the CR definitions is provided in Table~\ref{tab:CRdef}.
These CRs are utilized to develop and verify a mixture of methods based on both MC simulations and data: normalizing simulated samples for the dominant irreducible SM  processes, deriving any residual corrections to the simulated samples, and developing the background estimates based on data for the reducible background contributions. 
The CRs consist of the following selections: 4L events with two OSSF OnZ pairs and no \PQb~jets (4L $\ZZ$ CR); 
3L events with an OSSF OnZ pair, $\MT<150\GeV$, and $\ptmiss<125\GeV$ (3L OnZ CR); 
3L OffZ events with a trilepton mass OnZ and no \PQb~jets (3L $\ZG$ CR); 
and 2L1T events with an OSSF OnZ pair and $\ptmiss<100\GeV$ (2L1T MisID \PGt CR). 
The 3L OnZ CR is further split into three subregions with the following criteria: 
$\ptmiss<100\GeV$, $\MT<50\GeV$, and $\nbj =0$ (3L MisID $\Pe/\PGm$ CR); 
$50<\MT<150\GeV$, minimum lepton $\pt>20\GeV$, and $\nbj=0$ (3L $\WZ$ CR); 
and $\ptmiss<125\GeV$, $\MT<150\GeV$, minimum lepton $\pt>20\GeV$, $\nbj>0$, $\nj>2$, and $\ST>350\GeV$ (3L $\ttZ$ CR).
Events satisfying any of the CR selection criteria are not considered in the SRs.

\begin{table*}[hbt!]
  \centering
    \topcaption{A summary of control regions for the irreducible SM processes $\ZG$, $\WZ$, $\ttZ$, and $\ZZ$, and for the misidentified lepton backgrounds.
      The \ptmiss, \MT, the minimum 3L lepton \pt $(\pt^{\mathrm{3}})$, and \ST are in units of \GeV.
      The 3L OnZ CR is further split into 3L MisID $\Pe/\PGm$ CR, 3L $\WZ$ CR, and 3L $\ttZ$ CR.
    }
    \cmsTable{
      \begin{scotch}{l l l r r r r l}
                                         CR name           & OSSF$n$ & \mossf   & \nbj & \ptmiss & \MT   & $\pt^{\mathrm{3}}$ &  Other selections \\
        \hline
        2L1T MisID \PGt                 & OSSF1   & OnZ        & \NA     & $<$100    & \NA      & \NA    & \NA \\[1.5ex]
        3L $\ZG$           & OSSF1   & BelowZ     & 0      & \NA        & \NA      & \NA    & Trilepton mass OnZ \\[1.5ex]
        3L OnZ                                   & OSSF1   & OnZ        & \NA     & $<$125    & $<$150  & \NA    & \NA \\
        \multicolumn{1}{l}{3L MisID $\Pe/\PGm$}  & OSSF1   & OnZ        & 0      & $<$100    & $<$50   & \NA    & \NA \\
        \multicolumn{1}{l}{3L $\WZ$}          & OSSF1   & OnZ        & 0      & $<$125    & 50--150 & $>$20 & \NA \\
        \multicolumn{1}{l}{3L $\ttZ$}       & OSSF1   & OnZ        & $\ge$1 & $<$125    & $<$150  & $>$20 & $\nj>2$, $\ST>350$ \\[1.5ex]
        4L $\ZZ$                                 & OSSF2   & Double-OnZ & 0      & \NA        & \NA      & \NA    & \NA \\
      \end{scotch}
    }
    \label{tab:CRdef}
\end{table*}

\subsection{Irreducible backgrounds}

The irreducible backgrounds, as described earlier, arise from processes in which all reconstructed leptons originate from decays of SM bosons. 
These contributions are estimated using simulated event samples, after normalization and validation in dedicated CRs 
in data for the major $\WZ$, $\ZZ$, $\ZG$, and $\ttZ$ processes.  
The normalization correction factors and associated uncertainties, which include both statistical and systematic contributions, take into account the contamination of events from other processes, and are applied to the corresponding background estimates in the SRs.
The measurements for the diboson processes are largely independent because of the high purity of the corresponding CRs.
Since these backgrounds make significant contributions to the $\ttZ$-enriched CR, the normalization correction for this
process is measured after the corresponding corrections have been obtained for the other backgrounds.

The $\ZZ\to 4\Pell$ process is the primary background component in the channels with four leptons. 
The fraction of $\ZZ$ events in the 4L $\ZZ$ CR is greater than 99\%.
We consider the $\PQq\PQq\to \ZZ\to 4\Pell$ and $\Pg\Pg\to \ZZ\to 4\Pell$ processes collectively,
and observe relative normalization uncertainties of 4--5\% in each of the three data-taking periods. 
These uncertainties are dominated by the statistical uncertainties, as the contamination from background processes other than $\ZZ$ is negligible in this CR. 

The $\WZ\to 3\Pell\PGn$ process is the primary irreducible background source for channels with three leptons.
This background is normalized to data in the 3L $\WZ$ CR, where the minimum lepton \pt threshold is raised to 20\GeV to suppress the misidentified lepton contributions, and the selection yields a set of events $>75\%$ pure in $\WZ$. 
We observe relative normalization uncertainties in the range 3--6\% across the three data-taking periods for the $\WZ\to 3\Pell\PGn$ process, which  include the statistical and systematic components due to subtraction of background processes other than $\WZ$. 

The $\ZZ$ and $\WZ$ simulation samples are reweighted as functions of the jet multiplicity as well as the visible diboson \pt to match the simulated distributions to those of the data in these CRs, where the visible diboson \pt is defined as the vector \pt sum of the charged leptons in the event. 
This accounts for missing higher-order QCD and electroweak corrections, and yields an improved description of leptonic and hadronic quantities of interest in this analysis.

Production of $\ttZ$ is a major irreducible SM background process for all channels with $\nbj>0$.
This background is normalized to data in the 3L $\ttZ$ CR selection, which is orthogonal to the 3L $\WZ$ CR and the misidentified lepton CR selections.
Similarly to the selection in the 3L $\WZ$ CR, the minimum lepton \pt threshold is raised to 20\GeV to suppress the misidentified lepton contributions, and the selection yields a set of events $\sim$60\% pure in $\ttZ$.
Including the statistical and systematic uncertainties due to subtraction of other background processes, we measure relative normalization uncertainties in the range of 20--30\% across the three data-taking periods for the $\ttZ$ process.

A smaller background contribution arises from ISR or FSR photons that convert asymmetrically such that only one of the resultant leptons is reconstructed in the detector, or from misidentifying on-shell photons as electrons.
The dominant source of such backgrounds, collectively referred to as the conversion background, is DY events with an additional photon.The cross section of this process is normalized in a dedicated 3L $\ZG$ CR, 
consisting of BelowZ trilepton events with $\nbj=0$, where the mass of the trilepton system is within 15\GeV of the \PZ boson mass. 
The CR targets $\PZ\to\Pell\Pell+\PGg$ events, where, for example, the photon converts in the detector and one of the four leptons is too soft to satisfy our lepton selection criteria. 
This selection yields a set of events $>70\%$ pure in $\ZG$.
We obtain relative normalization uncertainties of about 10\% across the three data-taking periods, where the quoted value includes the statistical and systematic components due to subtraction of background processes other than $\ZG$, as well as a flavor-dependent component due to varying fractions of internal and external conversions as a function of electron multiplicity in the events.

Other irreducible processes that are not normalized in a dedicated CR in data are estimated from simulation samples and normalized to their theoretical cross sections.

In the following figures, diboson backgrounds from $\WZ$ and $\ZZ$ processes are denoted as \mbox{``$\PV\PV$''}, whereas the $\ttZ$ and $\ttbar\PW$ contributions are labeled as ``$\ttV$''.
Background processes involving a lepton conversion, particularly the $\ZG$ process, are labeled as ``Conv.''
Other irreducible backgrounds estimated using simulation consist of triboson, Higgs boson, and other rare SM contributions, and are collectively referred to as ``Rare'' backgrounds.

\subsection{Misidentified lepton backgrounds}

The misidentified lepton backgrounds are estimated via a three- or four-dimensional implementation of a matrix method~\cite{Khachatryan:2015bsa}, based on the lepton multiplicity in the targeted signal selections in data. 
The matrix method defines a set of sideband regions for each SR based on the isolation properties of the selected lepton objects in each event.
Leptons in the SR selections satisfy the tight lepton definitions given in Section~\ref{sec:objects}, whereas the sideband selections are defined by loose criteria with relaxed isolation requirements ($<$1.0 relative isolation for electrons and muons, and a relaxed working point of the \textsc{DeepTau} algorithm for \tauh), but are otherwise identical to the tight lepton criteria. 
Therefore, for a 3 (4) lepton event in an SR, the matrix method uses an additional 7 (15) nonoverlapping sideband regions with at least one lepton failing the tight isolation criteria. 
The sideband regions are therefore orthogonal to the SRs by construction.
The probabilities with which prompt- and misidentified lepton candidates pass the tight selection, given that they satisfy the loose selection, are denoted as prompt and misidentification rates, respectively. 
These are measured as a function of various kinematic features of leptons and hadronic properties of events that impact the lepton isolation.
These rates are used in the extrapolation from the sideband regions to the SR. 
This extrapolation is performed for each event, 
where the contamination due to prompt leptons that fail the tight lepton selection criteria is also corrected for. 
Because of the isolation requirements used in the single-lepton triggers, background contributions with up to 2 (3) simultaneous misidentified leptons in 3 (4) lepton events can be predicted by this implementation of the matrix method. 
The fraction of signal events where all lepton candidates are misidentified is found to be negligible in simulation based studies. 

The prompt rates are measured using a ``tag-and-probe'' method~\cite{CMS:2011aa} in various dilepton event selections. 
In data, prompt rates for electrons and muons are studied in a DY-enriched OnZ OSSF $\Pe\Pe$ and $\PGm\PGm$ events, respectively.
Similarly, prompt rates for \tauh candidates are studied in a DY-enriched set of opposite-sign $\Pe\tauh$ and  $\PGm\tauh$ events. 
In simulation, the prompt rates are measured in DY and $\ttbar$ MC samples, using reconstructed leptons kinematically matched to generator-level prompt leptons ($\dR<0.2$).
The measured prompt rates are primarily parametrized as a function of the lepton \pt and $\abs{\eta}$.
Prompt rates for electrons and muons vary from about $65\%$ at $\pt\sim10\GeV$ to about $95\%$ at $40\GeV$ and above.
For one- (three-) prong \tauh candidates, the prompt rates are about 50--70\% (30--70\%) in the \pt range of 20--50\GeV.
The final prompt rates for all lepton flavors are based on the DY-enriched data measurements, and the differences between rates derived from DY and $\ttbar$ MC simulations are taken as an estimate of the associated systematic uncertainty, accounting for the dependence of prompt rates on hadronic activity.
Prompt rate uncertainties are found to be unimportant in the matrix method, and the corresponding impact on the misidentified lepton background estimate is negligible.

\begin{figure*}[hbt!]
\centering
\includegraphics[width=0.49\textwidth]{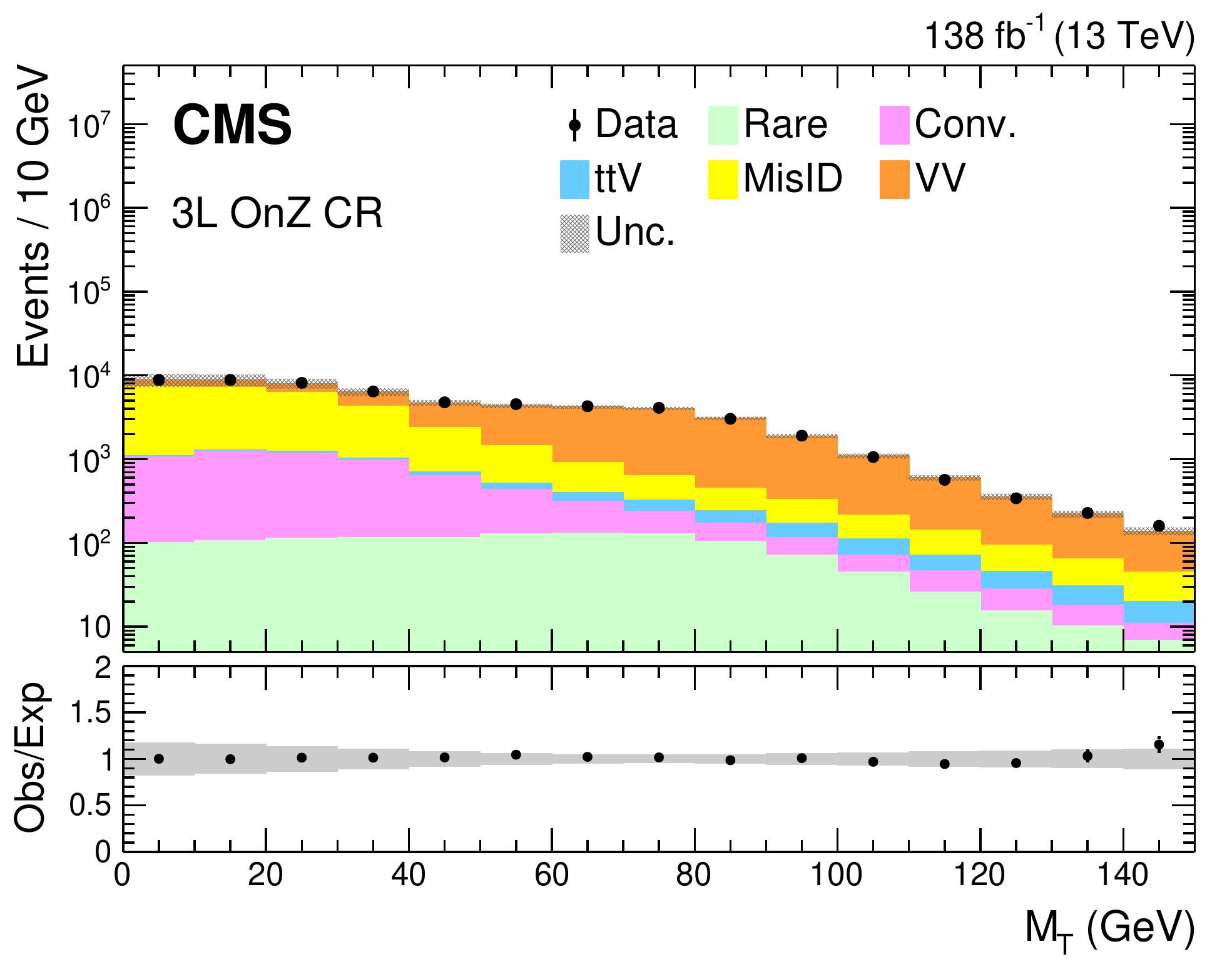}
\includegraphics[width=0.49\textwidth]{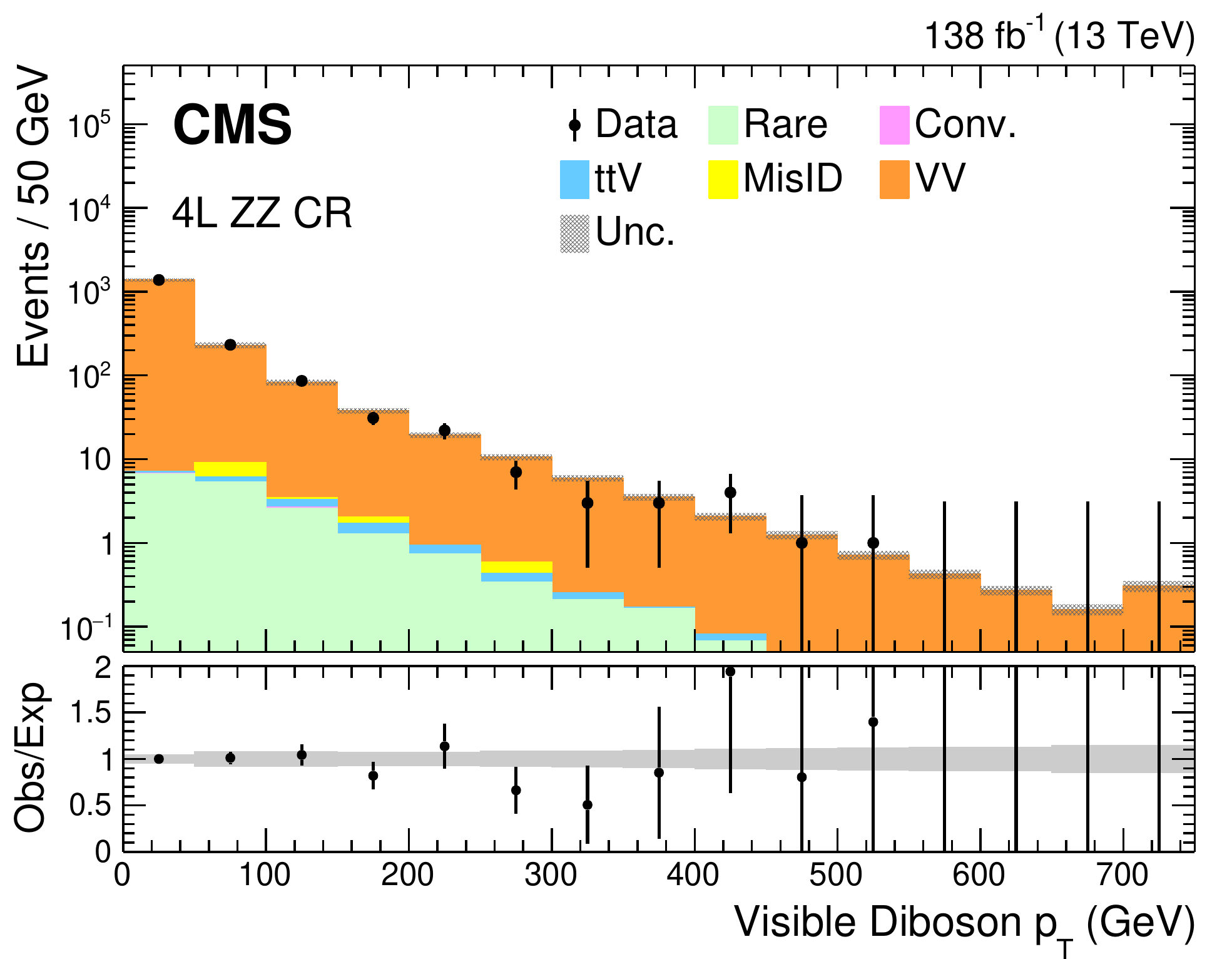}
\includegraphics[width=0.49\textwidth]{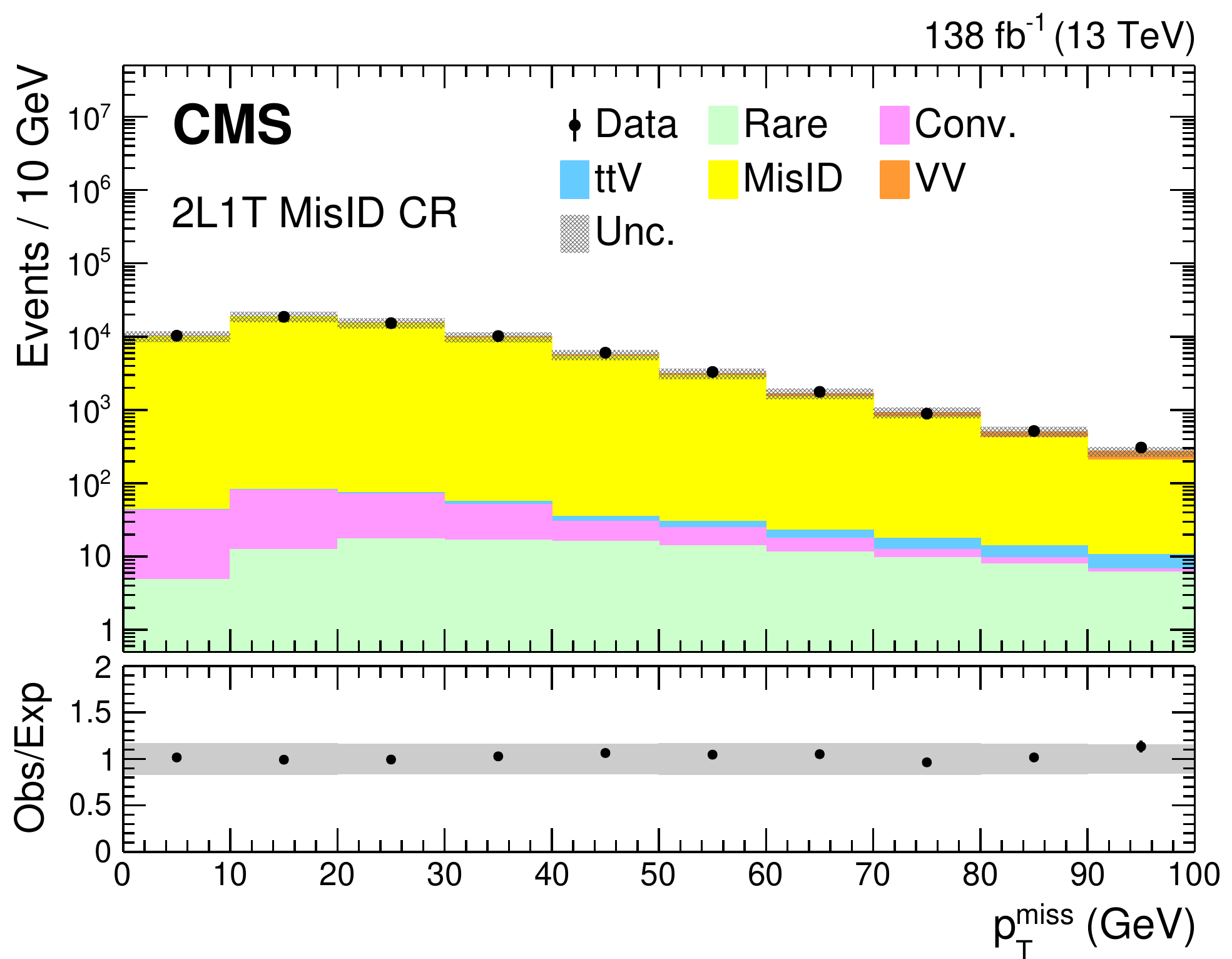}
\includegraphics[width=0.49\textwidth]{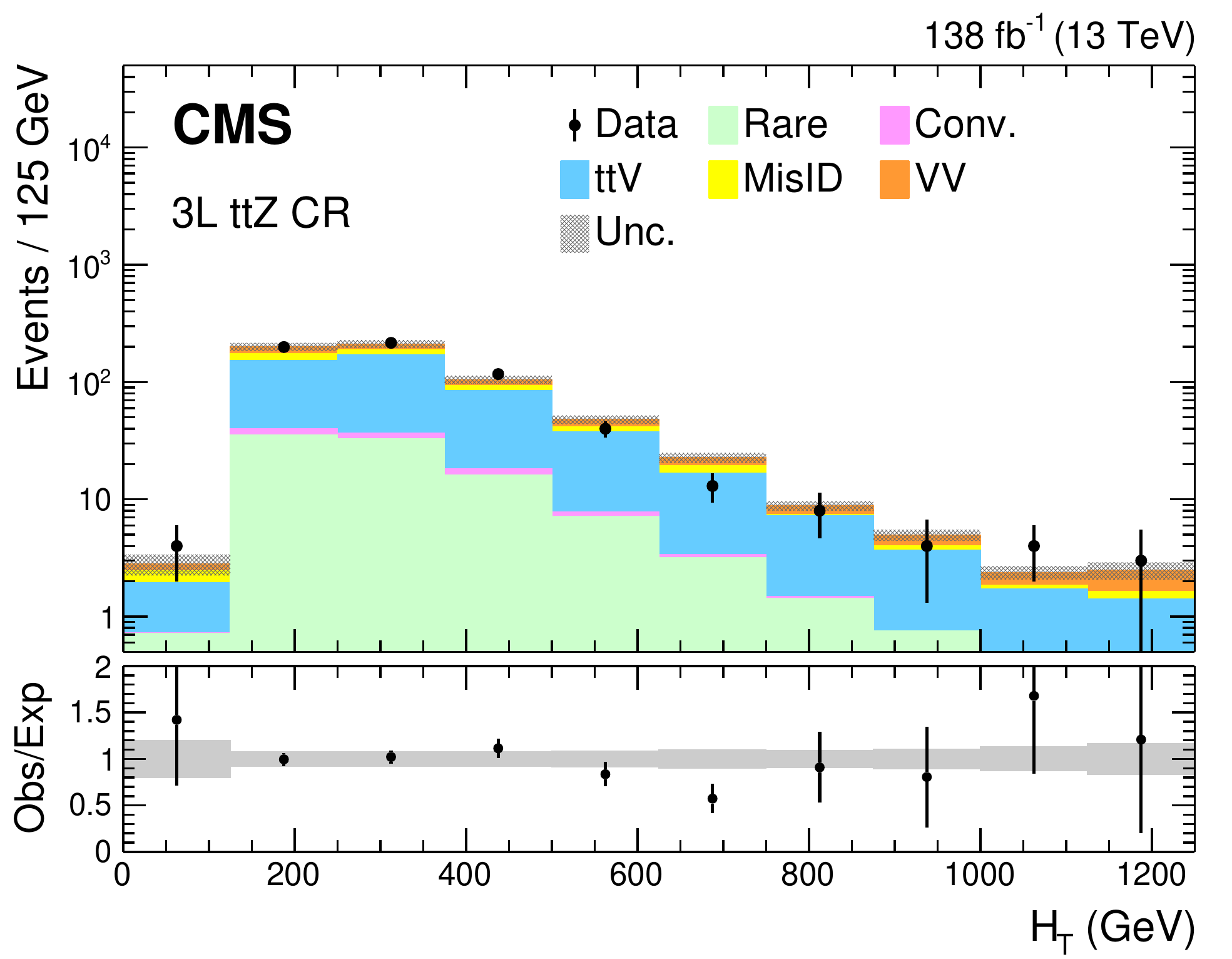}
\caption{\label{fig:Control} The distributions of \MT in 3L OnZ CR (upper left), visible diboson \pt in 4L $\ZZ$ CR (upper right), \ptmiss in 2L1T MisID CR (lower left), and $\HT$ in 3L $\ttZ$ CR (lower right) events. The rightmost bin contains the overflow events in each distribution. The lower panel shows the ratio of observed events to the total expected background prediction. The gray band on the ratio represents the sum of statistical and systematic uncertainties in the SM background prediction. 
}
\end{figure*}

The DY+jets and $\ttbar$+jets processes are the dominant SM contributions to the total misidentified lepton background in multilepton events. 
However, different gluon, light quark, and heavy quark compositions, as well as different event kinematic properties of these two processes, yield
misidentification rates that may differ by up to $50\%$ from each other for a given lepton flavor. 
Therefore, dedicated data and MC measurements are performed for both processes. 
A variant of the tag-and-probe method is used for the measurement of the misidentification rates.
In both 3L and 2L1T MisID CRs, the OnZ leptons are taken as the tag leptons, and the additional lepton is taken as the misidentified lepton probe, \eg, $\Pe\Pe\Pgm$ and $\Pgm\Pgm\Pgm$ events are used to measure muon misidentification rates, 
while $\Pe\Pe\tauh$ and $\Pgm\Pgm\tauh$ events are used to measure the \tauh misidentification rates. 
In measurements conducted in data, contributions due to prompt probe leptons are estimated and subtracted using MC simulation. 
Misidentification rates obtained in simulated $\ttbar$+jets samples are verified in dedicated data CRs enriched in such contributions, 
where one lepton is required to fail the three-dimensional impact parameter significance requirement or the {\cPqb} tag veto described in Section~\ref{sec:objects}.

The lepton misidentification rates are also parametrized as functions of the lepton \pt and $\abs{\eta}$.
The \tauh rates are further split for one- and three-prong objects. 
The central value of the misidentification rates for each lepton flavor is corrected for the recoil of the event, where the recoil is defined as a function of the vector sum of the \pt of all other leptons, jets, and \ptmiss in the event. 
These recoil-based corrections improve the modeling of misidentified lepton backgrounds in DY+jets events, 
in which the misidentified lepton often originates from a jet recoiling against the leptonically decaying \PZ boson system.
Similarly, the misidentification rates are corrected as a function of the multiplicity of tracks originating from the PV and the jet multiplicity.

The final misidentification rates for all lepton flavors are obtained by a weighted average of the DY- and $\ttbar$-based measurements.
These are evaluated according to the expected DY-$\ttbar$ composition of the MisID background, as obtained from simulated samples in each SR category and for each \PQb-tagged jet multiplicity. 
These DY and $\ttbar$ MC samples use normalization factors measured in dedicated dilepton control regions.
Half of the difference between rates derived from DY- and $\ttbar$-based measurements is assigned as a systematic uncertainty to allow for inaccurate modeling of the expected background composition.
Typical electron and muon misidentification rates, relative to the loose selection, are in the range of 5--30\%, whereas those of \tauh objects are found to be in the range of 1--15\%.

Figure~\ref{fig:Control} shows a selection of kinematic distributions in the various control regions, with the sum of statistical and systematic uncertainties in the SM background prediction, as described in Section~\ref{sec:systematics}. The data are observed to be in agreement with the SM prediction.

\section{Signal regions} \label{sec:sr}

The multilepton events that have been selected in the seven channels following the description in Sections~\ref{sec:objects} and \ref{sec:eventselec} are now categorized into two alternative SRs. 
This categorization is done either in a model-independent way, based on the characteristics of the SM backgrounds, or in a model-dependent way, based on the output of BDTs trained specifically for particular signal hypotheses.

\begin{figure*}[hbt!]
\centering
\includegraphics[width=0.49\textwidth]{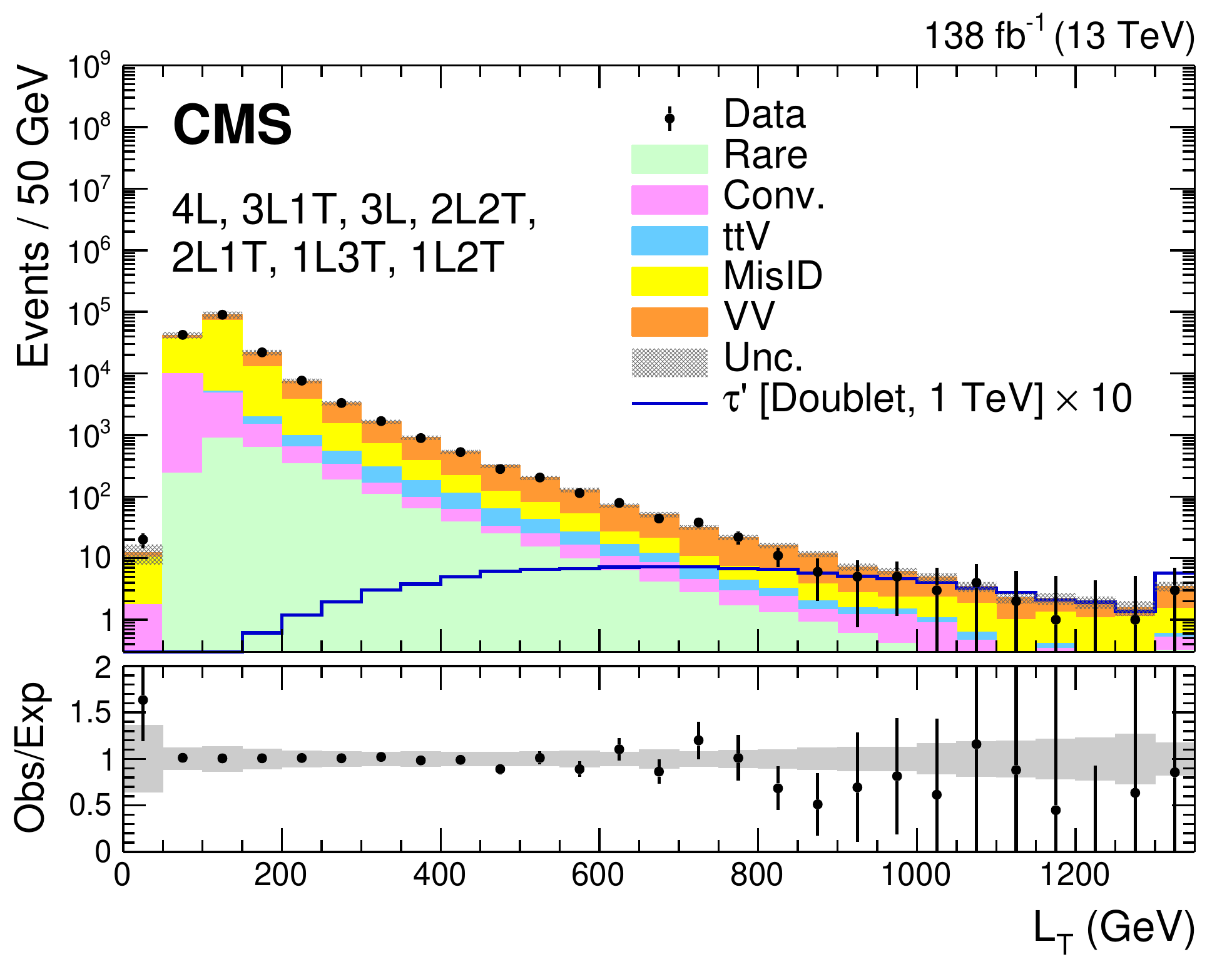}
\includegraphics[width=0.49\textwidth]{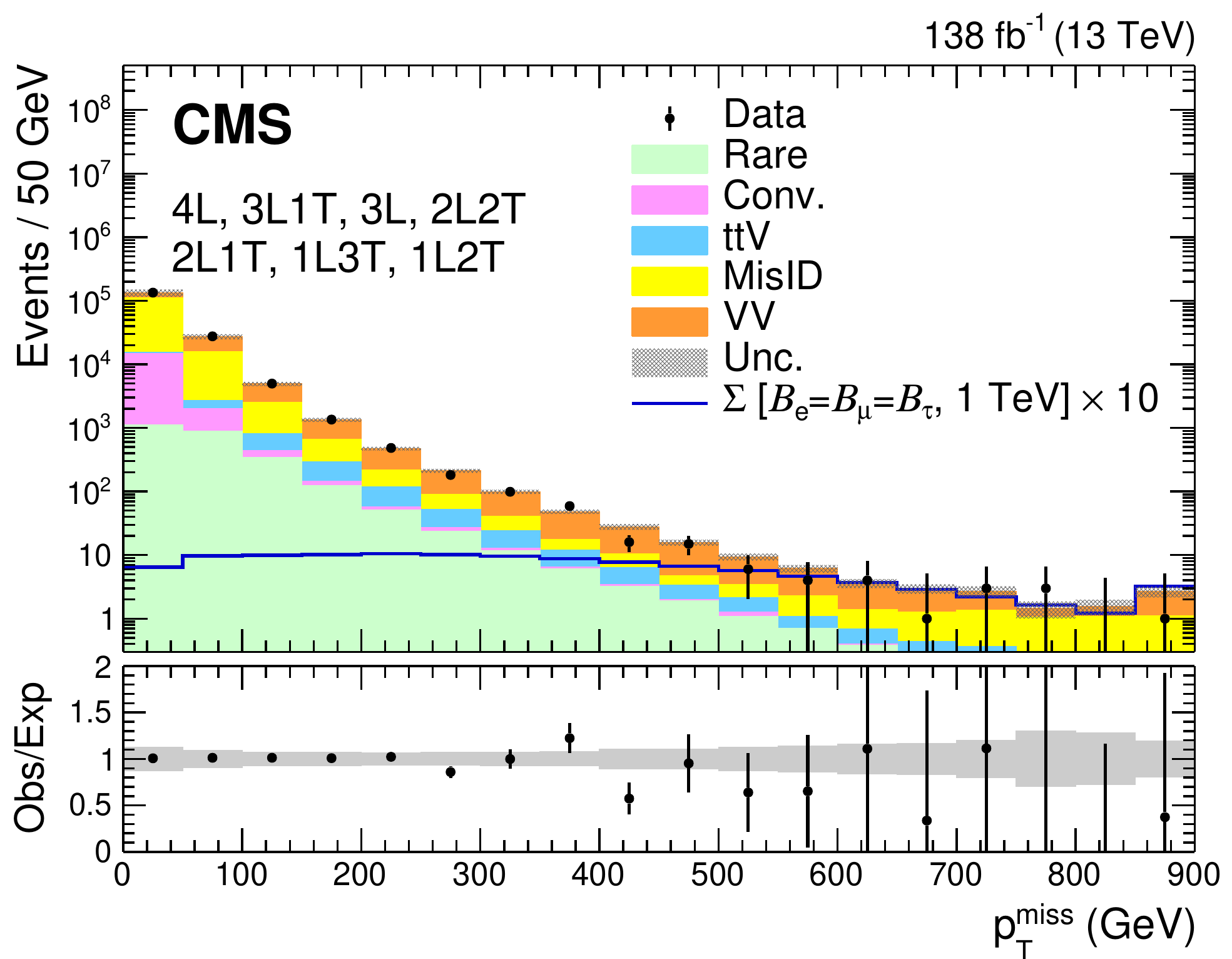}
\includegraphics[width=0.49\textwidth]{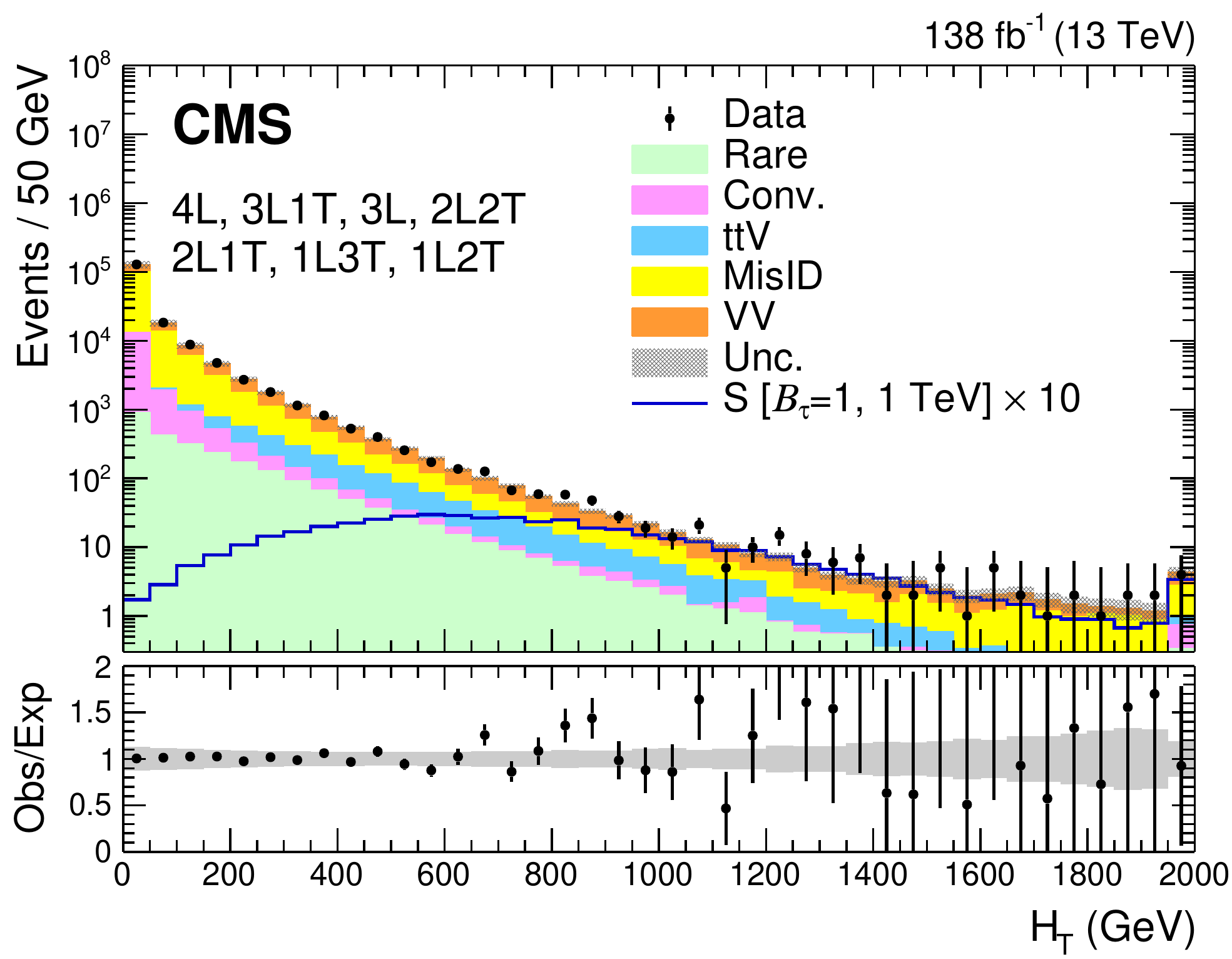}
\includegraphics[width=0.49\textwidth]{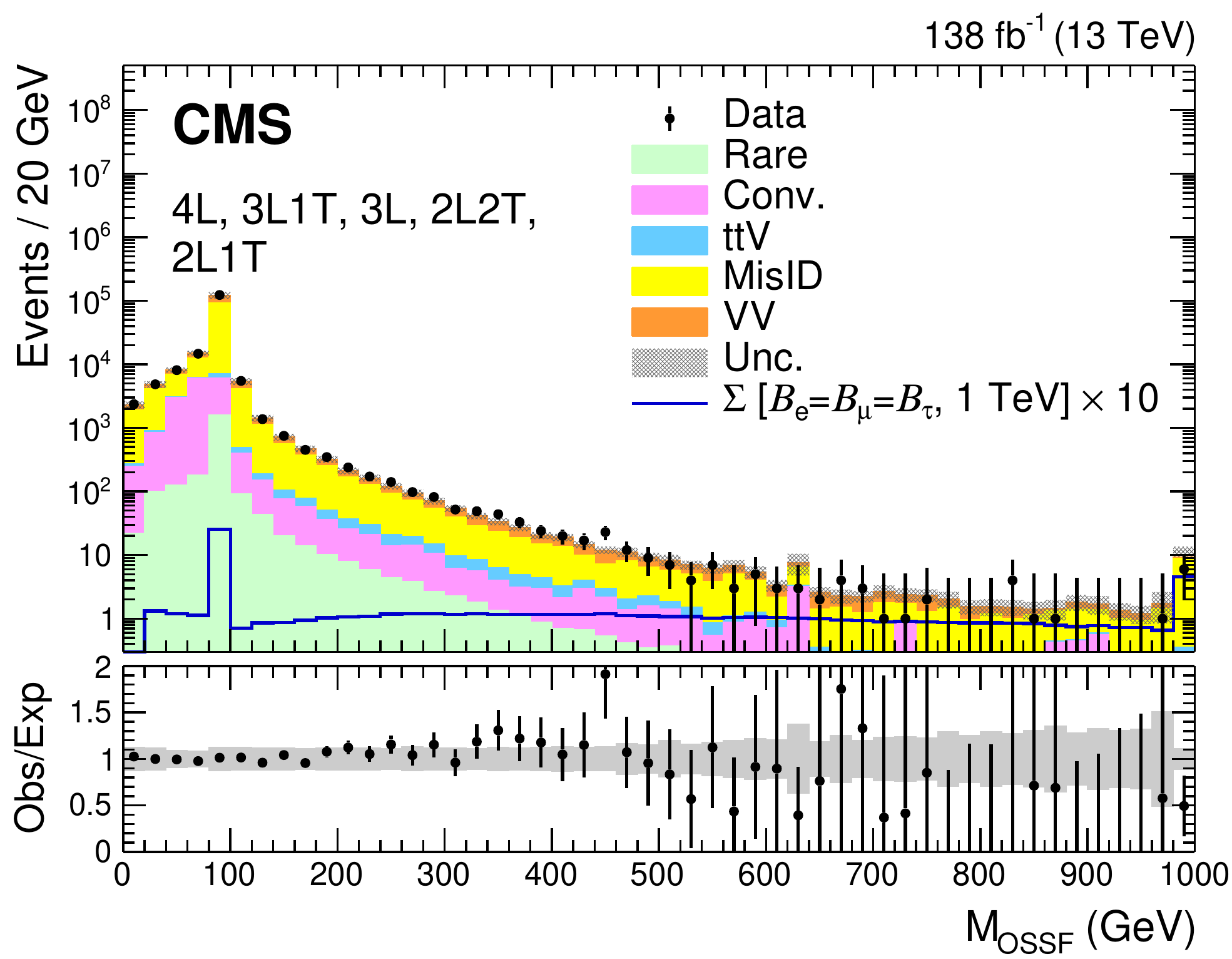}
\caption{\label{fig:SRcomb} The distributions of \LT (upper left), \ptmiss (upper right), and $\HT$ (lower left) in all seven multilepton channels, and \mossf (lower right) in channels with at least one light lepton pair (4L, 3L1T, 3L, 2L2T, and 2L1T). The rightmost bin contains the overflow events in each distribution. The lower panel shows the ratio of observed events to the total expected background prediction. The gray band on the ratio represents the sum of statistical and systematic uncertainties in the SM background prediction. As illustrative examples, a signal hypothesis for the production of the vector-like \PGt lepton in the doublet scenario of $m_{\vltau}=1\TeV$ and a scalar leptoquark coupled to a top quark and a \PGt lepton of $m_{\PS}=1\TeV$ are overlaid in the \LT and $\HT$ distributions respectively. Similarly, a signal hypothesis for the production of the type-III seesaw heavy fermions in the flavor-democratic scenario of $m_{\Sigma}=1\TeV$ is overlaid for the \ptmiss and \mossf distributions.
}
\end{figure*}

Figure~\ref{fig:SRcomb} illustrates \LT, \ptmiss, and $\HT$ distributions in the full multilepton phase space, and the \mossf distribution in channels with at least one OSSF light lepton pair. 
In each distribution, a benchmark signal hypothesis distribution is overlaid to allow a comparison of shapes between signal and background. 
The plots include the sum of statistical and systematic uncertainties in the SM background prediction, as described in Section~\ref{sec:systematics}, and the data are found to be in agreement with the SM prediction.

\subsection{Model-independent selections}
The model-independent SRs are defined by splitting the channels into various
lepton charge and flavor combinations, mass variables, and kinematic regions depending on the dominant SM background processes.
This categorization allows the complete utilization of multilepton events collected, such that any event that does not populate a CR contributes to an SR. 
Explicitly, events selected for the CRs, which are used in the estimation of major SM backgrounds as described in Section~\ref{sec:backgrounds}, are not used in any of the SRs.

The SRs are designed to separate regions where signs of BSM models could appear from regions dominant in SM background processes.
The most easily distinguishable feature is the presence of a \PZ boson candidate, determined using OSSF$n$ and \mossf. 
The $\ZZ$ and the $\WZ$ processes can be separated by additional requirements on \MT in the event. 
Similarly, the MisID background can be separated using the minimum lepton \pt. 
Further selections on \nbj give SR regions that have significant contributions from $\ttZ$.

\begin{table*}
\centering
\topcaption{Fundamental scheme of event categorization, as a function of lepton charge combinations and mass variables. 
The mass categorizations refer to masses of OSSF pairs if present, and of OSDF pairs otherwise, as explained in the text.
For categorization  purposes, all possible opposite-sign dielectron and dimuon pair masses in the event are considered,
whereas only the largest mass in the event is considered for all other opposite-sign pairs.
Only the dielectron and dimuon pairs are considered to tag events as OnZ.
The 1L3T OSSF0 and OSSF1 events are combined into a single category.
Disallowed categories are marked with ``\NA''.}
\label{tab:simplifiedtable}
\cmsTable{
\begin{scotch}{l l c   l l l c   l l l l c   l l l c }
                      &                       & $~$ & \multicolumn{3}{l}{OSSF0}      & $~$ & \multicolumn{4}{l}{OSSF1}        & $~$ & \multicolumn{3}{l}{OSSF2}            \\
                      &                       &     & BelowZ & AboveZ  & SS          &     & OnZ & BelowZ & AboveZ  & MixedZ  &     & Single-OnZ & Double-OnZ & OffZ       \\
\hline   
\multirow{2}{*}{3L}   & Low $\pt/\MT$         &     &  A1   &  A1       & A2          &     & A3  & A4     & A5      & A6      &     & \NA        &  \NA       & \NA        \\
                      & High $\pt/\MT$        &     &  A7   &   A7      & A8          &     & A9  & A10    & A11     & A12     &     & \NA        &  \NA       & \NA        \\[0.5ex]
\multirow{2}{*}{2L1T} & Low \pt             &     &  B1    & B2      & B3          &     & B4  & B5     & B6      & \NA     &     & \NA        &  \NA       & \NA        \\
                      & High \pt            &     &  B7    & B8      & B9          &     & B10 & B11    & B12     & \NA     &     & \NA        &  \NA       & \NA        \\[0.5ex]
1L2T                  &                       &     &  C1    & C2      & C3          &     & \NA & C4     & C5      & \NA     &     & \NA        &  \NA       & \NA        \\[1.5ex]

4L                    &                       &     &  D1   &  D1       & D1            &     & D2   & D3    &   D3     &   D3      &     & D4         &  D5        & D6         \\[0.5ex]
3L1T                  &                       &     &  E1   &    E1     &   E1          &     & E2   & E3    &   E3     &   E3      &     & \NA        &  \NA       & \NA        \\[0.5ex]
2L2T                  &                       &     &  F1   &    F1     &   F1          &     & F2  &  F2      &  F2      & \NA     &     & F3         &  \NA       & F4         \\[0.5ex]
1L3T                  &                       &     &  G1   &    G1     &   G1         &     & \NA  & G1    &  G1      & \NA     &     & \NA        &  \NA       & \NA        \\
\end{scotch}
}
\end{table*}

Based on the idea of the broad categorization described above, we have a fundamental scheme with 43 orthogonal selections labeled A1--G1, as summarized in Table~\ref{tab:simplifiedtable}. 
The primary classification is done based on OSSF$n$, with $n$ being 2, 1, or 0.
We also define another scheme, labeled the advanced scheme, which builds on the fundamental scheme, but adds further categories.
Each of the 43 fundamental scheme categories is first split into up to three \PQb tag multiplicity regions.
The categories with 0 \PQb tag, 1 \PQb tag, and 2 or more \PQb tag multiplicities are denoted by 0\PB, 1\PB, and 2\PB respectively, in all the subsequent tables and figures.
Furthermore, each category in a given \PQb tag multiplicity
region is split in up to four bins, using a binary low/high \ptmiss criterion and an $\HT$ requirement.
This results in a total of 204 orthogonal categories.

Events where an OSSF light lepton pair is not found, but an OSSF $\PGt\PGt$ pair is found, are categorized as BelowZ or AboveZ with respect to the \PZ pole mass ($91\GeV$) using the $\PGt\PGt$ pair mass. 
This is done since a resonance will not appear at the \PZ pole mass because of the neutrino emitted in the \PGt lepton decays.
In OSSF0 events, an OSDF pair is sought, and the event is categorized as BelowZ or AboveZ (as for $\PGt\PGt$ events) based on the OSDF pair with the largest mass.
Events with no OSSF or OSDF pairs are classified as same-sign (SS) events.

The 3L channel is further split into two categories, based on the value of either \MT or the minimum lepton \pt. 
In the 3L OnZ channel, an $\MT>125\GeV$ criterion is used for a binary low or high classification, whereas a lepton $\pt>25\GeV$ criterion is used for the rest of the 3L channel. 
The 2L1T channel is split into a similar binary classification based on the \tauh candidate $\pt>50\GeV$ criterion.

In order to be sensitive to a large class of BSM models in each of the 43 categories of the fundamental scheme, an $\LTmet$ distribution is obtained in $200\GeV$ wide bins, with
the last bin being inclusive for all higher values. 
This results in 156 $\LTmet$ bins.
The combined spectrum (across all 43 categories) gives the fundamental $\LTmet$ table. 
The width of bins in the spectrum is chosen to provide smooth and
monotonic expected background behavior, while still retaining sensitivity to nonresonant models. 
The first and last bins of the $\LTmet$ distribution are
chosen with the requirement that the per-bin expected background yield is more than 0.5 events to ensure robustness in statistical interpretations.

The second table, labeled the fundamental \ST table, is identical to the first table except that we use the \ST distribution in each category, where \ST is the scalar sum of \LT, $\HT$, and \ptmiss, also in $200\GeV$ wide bins, resulting in 257 bins. 
This table provides sensitivity to signal models with energetic jets, such as leptoquarks, whereas the $\LTmet$ table is optimized for
models without significant hadronic activity, such as vector-like lepton and type-III seesaw scenarios.

For the third and final table, we use \ST as the final discriminating variable, binned in $200\GeV$ increments, in the advanced scheme categorization resulting in 805 bins. This table, labeled the advanced \ST table, provides improved sensitivity to a wide array of
BSM signals with masses at the electroweak scale. 

The binning in all the schemes is described in Tables~\ref{tab:binningtable3L}--\ref{tab:binningtableQuad}. 
Each table is produced separately for each year of data collection, resulting in a total of 468 bins in the fundamental $\LTmet$ table scheme, 771 bins in the fundamental \ST table scheme, and 2415 bins in the advanced \ST table scheme, for the combined 2016--2018 data set.

\begin{table*}
\centering
\topcaption{The binning of $\LTmet$ and \ST distributions for the fundamental scheme in the 3L channel, and the binning of \ST distribution for the advanced scheme in the 3L channel. The categorization is described in Table~\ref{tab:simplifiedtable}. 
The ranges, as well the \ptmiss and $\HT$ requirements, are given in GeV. 
The first bins in the $\LTmet$ or \ST range contain the underflow, the last bins contain the overflow.
} \label{tab:binningtable3L}
\cmsTable{
  \begin{scotch}{llllllllllllll}
    &  \multicolumn{4}{c}{Fundamental Tables}&   \multicolumn{4}{l}{Advanced Table}&   \multicolumn{3}{l}{Advanced Table}&    \multicolumn{2}{l}{Advanced Table}\\
    &  \multicolumn{2}{l}{$\LTmet$}&  \multicolumn{2}{l}{\ST}&   \multicolumn{4}{l}{0\PB}&    \multicolumn{3}{l}{1\PB} &     \multicolumn{2}{l}{2\PB} \\
    Cat.&  Range&  Bins&  Range& Bins& \ptmiss& $\HT$& \ST Range& Bins& \ptmiss& \ST Range& Bins& \ST Range& Bins\\
    \hline
    A1&     $[0,800]$& 1--4 &  $[50,1650]$&  1--8 &    $<$125&  $<$150&  $[100,700]$&  1--3&  $<$125&  $[100,1100]$&  189--193&  $[50,1250]$&   291--296 \\
    &   & & & & $<$125 & $>$150  & $[200,1000]$ & 4--7&  & & & &    \\
    &   & & & & $>$125 & $<$150  & $[250,650]$  & 8--9& $>$125 & $[150,1350]$ & 194--199&  &           \\
    &   & & & & $>$125 & $>$150  & $[350,1150]$ & 10--13&  & & & &              \\[0.75ex]
    A2&   $[50,450]$&  5--6 &  $[150,750]$ &  9--11 &   \NA & \NA & $[150,750]$ & 14--16   & \NA  & \NA & \NA &  \NA  & \NA          \\[0.75ex]
    A3 &  $[50,1650]$& 7--14 & $[150,2750]$& 12--24 & $>$125 & $<$150 & $[50,1450]$ & 17--23 &  $>$125&  $[300,2100]$ &  200--208 &  $[350,1750]$ &  297--303 \\
    &  & & & & $>$125 & $>$150 & $[250,2650]$ & 24--35 &  & & & &            \\[0.75ex]
    A4&   $[100,900]$ & 15--18 &  $[50,1850]$  & 25--33   & $<$125 & $<$150  & $[50,650]$  & 36--38 & $<$125  & $[0,1200]$&  209--214 &  $[100,1300]$&  304--309 \\
    &  & & & &  $<$125 & $>$150   & $[150,1350]$  & 39--44&  & & & &            \\
    &  & & & &  $>$125 & $<$150   & $[100,700]$  &  45--47 & $>$125  & $[100,1500]$  & 215--221& &           \\
    &  & & & &  $>$125 & $>$150   & $[300,1500]$  & 48--53&  & & & &              \\[0.75ex]
    A5&   $[150,1150]$& 19--23 &  $[0,1800]$  &  34--42 &   $<$125 & $<$150   & $[0,1000]$  & 54--58 & $<$125 & $[100,1100]$& 222--226 &  $[100,1300]$& 310--315 \\
    &  & & & & $<$125 & $>$150   & $[150,1150]$  & 59--63   &  & & & &            \\
    &  & & & & $>$125 & $<$150   & $[100,900]$  & 64--67   & $>$125  & $[200,1200]$  & 227--231& &           \\
    &  & & & & $>$125 & $>$150   & $[300,1300]$  & 68--72   &  & & & &             \\[0.75ex]
    A6&  $[50,850]$&  24--27 &  $[0,1400]$  & 43--49   & $<$125 & $<$150 & $[50,650]$ & 73--75 & $<$125&  $[150,950]$ &  232--236 &  $[300,1100]$&  316--319 \\
    &  & & & & $<$125 & $>$150   & $[200,1000]$  & 76--79   & & & & &           \\
    &  & & & & $>$125 & $<$150   & $[200,800]$  & 80--82   & $>$125  & $[350,1150]$  & 237--239& &           \\
    &  & & & & $>$125 & $>$150   & $[500,1100]$  & 83--85   & & & & &              \\[0.75ex]
    A7&  $[0,1000]$ & 28--32 &  $[150,1750]$ & 50--57   &  $<$125 & $<$150 & $[50,650]$ & 86--88 & $<$125 &  $[150,1150]$& 240--244 &  $[150,1350]$ & 320--325 \\
    &  & & & & $<$125 & $>$150   & $[150,950]$  & 89--92   &  & & & &            \\
    &  & & & & $>$125 & $<$150   & $[150,750]$  & 93--95   & $>$125 & $[350,1350]$ & 245--249& &          \\
    &  & & & & $>$125 & $>$150   & $[350,1350]$ & 96--100   &  & & & &               \\[0.75ex]
    A8&  $[100,500]$&  33--34 & $[50,650]$  &  58--60   &  \NA & \NA   & $[50,650]$  & 101--103   &  \NA  & \NA & \NA &  \NA  & \NA          \\[0.75ex]
    A9&  $[150,1350]$& 35--40 & $[150,2150]$ & 61--70   &  $<$125 & $<$150 & $[150,2150]$  & 104--108 & $<$125&  $[300,1300]$ &  250--254 &  $[450,1250]$& 326--329 \\
    &  & & & & $<$125 & $>$150   & $[400,1800]$  & 109--115   &  & & & &            \\
    &  & & & & $>$125 & $<$150   & $[200,1000]$  & 116--119   & $>$125  & $[350,1350]$  & 255--259   & &          \\
    &  & & & & $>$125 & $>$150   & $[500,1700]$  & 120--125   &  & & & &               \\[0.75ex]
    A10& $[100,1100]$ & 41--45 & $[0,1800]$ & 71--79   & $<$125 & $<$150  & $[0,800]$  & 126--129 & $<$125&  $[150,1150]$&  260--264 &  $[250,1250]$&  330--334 \\
    &  & & & & $<$125 & $>$150   & $[200,1400]$  & 130--135   &  & & & &            \\
    &  & & & & $>$125 & $<$150   & $[150,950]$  & 136--139  & $>$125  & $[300,1300]$  & 265--269  & &           \\
    &  & & & & $>$125 & $>$150   & $[300,1500]$  & 140--145   &  & & & &               \\[0.75ex]
    A11 & $[0,1400]$ & 46--52 &  $[50,2050]$  & 80--89   &  $<$125 & $<$150   & $[50,1250]$  & 146--151   & $<$125 & $[200,1400]$ &  270--275 &  $[200,1600]$&  335--341 \\
    &  & & & & $<$125 & $>$150   & $[200,1600]$  & 152--158   &  & & & &            \\
    &  & & & & $>$125 & $<$150   & $[200,1200]$  & 159--163   & $>$125  & $[300,1500]$  & 276--281  & &          \\
    &  & & & & $>$125 & $>$150   & $[400,1800]$  & 164--170   &  & & & &               \\[0.75ex]
    A12& $[100,1100]$ & 53--57 &  $[150,1750]$  & 90--97   &  $<$125 & $<$150   & $[100,900]$  & 171--174 & $<$125&  $[100,1100]$&  282--286 &  $[350,1150]$&  342--345 \\
    &  & & & & $<$125 & $>$150   & $[250,1450]$  & 175--180   &  & & & &            \\
    &  & & & & $>$125 & $<$150   & $[300,900]$  & 181--183   & $>$125  & $[500,1300]$  & 287--290  & &           \\
    &  & & & & $>$125 & $>$150   & $[450,1450]$  & 184--188   &  & & & &               \\
\end{scotch}
}
\end{table*}

\begin{table*}
\centering
\topcaption{The binning of $\LTmet$ and \ST distributions for the fundamental scheme in the 2L1T channel, and the binning of \ST distribution for the advanced scheme in the 2L1T channel. The categorization is described in Table~\ref{tab:simplifiedtable}.
The ranges, as well as the \ptmiss and $\HT$ requirements, are given in GeV. 
The first bins in the $\LTmet$ or \ST range contain the underflow, and the last bins contain the overflow.
} \label{tab:binningtable2L1T}
\cmsTable{
  \begin{scotch}{llllllllllllll}

    & \multicolumn{4}{c}{Fundamental Tables}& \multicolumn{4}{l}{Advanced Table}&  \multicolumn{3}{l}{Advanced Table}& \multicolumn{2}{l}{Advanced Table}\\
    & \multicolumn{2}{l}{$\LTmet$}&  \multicolumn{2}{l}{\ST}&    \multicolumn{4}{l}{0\PB}&     \multicolumn{3}{l}{1\PB}&   \multicolumn{2}{l}{2\PB} \\
    Cat.&  Range&  Bins&  Range& Bins& \ptmiss& $\HT$& \ST Range& Bins& \ptmiss& \ST Range& Bins& \ST Range& Bins\\
    \hline
    B1& $[100,700]$&  58--60&   $[100,1300]$&   98--103&    $<$100&  $<$150&  $[150,550]$&  346--347&   $<$100&  $[0,800]$&   489--492&   $[150,950]$&   571--574 \\
    &       &           &            &            &         $<$100& $>$150&   $[250,850]$&  348--350&  & & & & \\
    &       &           &            &            &         $>$100& $<$150&   $[0,600]$  &  351--353&   $>$100&  $[100,1100]$& 493--497&   &          \\
    &       &           &            &            &         $>$100&  $>$150&   $[350,1150]$&   354--357&   & & & &  \\[0.75ex]
    B2& $[50,850]$&   61--64&   $[0,1600]$&    104--111&    $<$100& $<$150&   $[50,650]$&   358--360&   $<$100&  $[50,1050]$& 498--502&   $[150,1150]$&  575--579  \\
    &  & & & & $<$100 & $>$150   & $[300,1100]$  & 361--364    & & & & &           \\
    &  & & & &  $>$100 & $<$150   & $[200,800]$  & 365--367   & $>$100  & $[250,1250]$  & 503--507  & &          \\
    &  & & & &  $>$100 & $>$150   & $[300,1300]$  & 368--372     & & & & &              \\[0.75ex]
    B3& $[100,500]$&  65--66&   $[150,750]$&  112--114&     \NA&    \NA&  $[150,750]$&    373--375&   \NA& \NA& \NA& \NA& \NA \\[0.75ex]
    B4&   $[150,950]$& 67--70& $[0,1800]$ & 115--123&  $>$100 & $<$150& $[50,750]$&   376--379&   $>$100  &$[200,1200]$&  508--512&  $[250,1050]$&   580--583   \\
    &  & & & & $>$100 & $>$150   & $[400,1600]$  & 380--385     & & & & &          \\[0.75ex]
    B5&   $[100,700]$& 71--73 & $[50,1250]$ & 124--129&   $<$100&  $<$150&  $[150,550]$&  386--387&   $<$100&  $[50,850]$&  513--516 & $[50,1050]$&  584--588    \\
    &  & & & &  $<$100 & $>$150   & $[100,1100]$  & 388--392&    & & & &           \\
    &  & & & &   $>$100& $<$150   & $[0,600]$  & 393--395  & $>$100  & $[100,900]$  & 517--520 &   &    \\
    &  & & & &   $>$100& $>$150   & $[300,1100]$  & 396--399&    & & & &            \\[0.75ex]
    B6&   $[0,1000]$& 74--78&   $[150,1550]$&   130--136&   $<$100& $<$150& $[50,850]$&  400--403&    $<$100&  $[150,1150]$&  521--525&  $[250,1050]$&   589--592  \\
    &  & & & &  $<$100 & $>$150   & $[200,1200]$  & 404--408   & & & & &            \\
    &  & & & &   $>$100 & $<$150   & $[200,800]$  & 409--411  & $>$100  & $[250,1250]$  & 526--530  & &          \\
    &  & & & &   $>$100 & $>$150   & $[300,1300]$  & 412--416   & & & & &              \\[0.75ex]

    B7&   $[100,700]$& 79--81& $[50,1250]$&   137--142&   $<$100&  $<$150&   $[150,550]$&   417--418&  $<$100&  $[100,700]$&  531--533&  $[400,800]$&  593--594  \\
    &  & & & &  $<$100 & $>$150   & $[150,750]$  & 419--421   & & & & &           \\
    &  & & & &   $>$100 & $<$150   & $[150,550]$  & 422--423   & $>$100  & $[250,1050]$  & 534--537  & &          \\
    &  & & & &   $>$100 & $>$150   & $[350,1150]$  & 424--427     & & & & &              \\[0.75ex]
    B8&   $[0,1000]$&  82--86& $[150,1750]$&  143--150&   $<$100&  $<$150&   $[150,750]$&   428--430&  $<$100&  $[100,1100]$& 538--542&  $[350,1350]$& 595--599   \\
    &  & & & & $<$100 & $>$150   & $[250,1050]$  & 431--434   & & & & &           \\
    &  & & & &  $>$100 & $<$150   & $[150,950]$  & 435--438   & $>$100  & $[250,1450]$  & 543--548  & &          \\
    &  & & & &  $>$100 & $>$150   & $[250,1450]$  & 439--444     & & & & &              \\[0.75ex]
    B9&   $[100,500]$&  87--88&  $[100,700]$&  151--153 &  \NA & \NA &       $[100,700]$&   445--447&   \NA  & \NA & \NA & \NA  & \NA          \\[0.75ex]
    B10&  $[250,1250]$& 89--93&  $[200,2000]$& 154--162 &  $>$100 & $<$150 & $[100,1100]$&  448--452&   $>$100&  $[250,1250]$&  549--553 & Incl.&  600    \\
    &  & & & & $>$100 & $>$150   & $[400,2000]$  & 453--460   & & & &  &           \\[0.75ex]
    B11&   $[100,900]$& 94--97& $[50,1450]$&   163--169&  $<$100 & $<$150   & $[100,700]$  & 461--463   &$<$100  &$[100,900]$  &554--557 & $[250,950]$  &  601--603  \\
    &  & & & & $<$100 & $>$150   & $[250,1050]$  & 464--467   & & & & &            \\
    &  & & & &  $>$100 & $<$150   & $[150,750]$  & 468--470   & $>$100  & $[250,1050]$  & 558--561  & &           \\
    &  & & & &  $>$100 & $>$150   & $[400,1200]$  & 471--474   & & & & &               \\[0.75ex]
    B12&  $[50,1050]$& 98--102& $[150,1750]$&  170--177&   $<$100&  $<$150&  $[100,900]$&   475--478&  $<$100&   $[200,1000]$&  562--565 &  $[400,1000]$  &  604--606  \\
    &  & & & &  $<$100 & $>$150   & $[350,1150]$  & 479--482   &  & & & &           \\
    &  & & & &   $>$100 & $<$150   & $[300,900]$  & 483--485   & $>$100  & $[350,1350]$  & 566--570  & &          \\
    &  & & & &   $>$100 & $>$150   & $[600,1200]$  & 486--488     & & & & &              \\
\end{scotch}
}
\end{table*}

\begin{table*}
\centering
\topcaption{The binning of $\LTmet$ and \ST distributions for the fundamental scheme in the 1L2T channel, and the binning of \ST distribution for the advanced scheme in the 1L2T channel. The categorization is described in Table~\ref{tab:simplifiedtable}.
The ranges, as well as the \ptmiss and $\HT$ requirements, are given in GeV. 
The first bins in the $\LTmet$ or \ST range contain the underflow, and the last bins contain the overflow.
} \label{tab:binningtable1L2T}
\cmsTable{
\begin{scotch}{llllllllllllll}
    &  \multicolumn{4}{c}{Fundamental Tables}&   \multicolumn{4}{l}{Advanced Table}&   \multicolumn{3}{l}{Advanced Table}&    \multicolumn{2}{l}{Advanced Table}\\
    &  \multicolumn{2}{l}{$\LTmet$}&  \multicolumn{2}{l}{\ST}&   \multicolumn{4}{l}{0\PB}&    \multicolumn{3}{l}{1\PB} &     \multicolumn{2}{l}{2\PB} \\
    Cat.&  Range&  Bins&  Range& Bins& \ptmiss& $\HT$& \ST Range& Bins& \ptmiss& \ST Range& Bins& \ST Range& Bins\\
    \hline
    C1& $[100,500]$& 103--104 & $[100,900]$ & 178--181 &   $<$75 & $<$75 & $[0,400]$ & 607--608 & $<$75&  $[100,500]$ & 657--658 & $[200,600]$& 684--685 \\
    &  & & & & $<$75 & $>$75   & $[200,600]$  & 609--610   &  & & & &            \\
    &  & & & & $>$75 & $<$75   & Incl.  & 611 & $>$75  & $[100,700]$  & 659--661  & &   \\
    &  & & & & $>$75 & $>$75   & $[150,750]$  & 612--614   &  & & & &               \\[0.75ex]
    C2& $[150,750]$& 105--107 & $[100,1100]$ & 182--186 &  $<$75 & $<$75 & $[150,750]$& 615--616 & $<$75&  $[50,650]$ & 662--664& $[100,900]$& 686--689 \\
    &  & & & & $<$75 & $>$75   & $[100,700]$  & 617--619   &  & & & &            \\
    &  & & & & $>$75 & $<$75   & $[150,750]$  & 620--622   & $>$75  & $[200,800]$  & 665--667  & &          \\
    &  & & & & $>$75 & $>$75   & $[300,900]$  & 623--625   &  & & & &               \\[0.75ex]
    C3& $[50,450]$ &  108--109 & $[150,550]$  &  187--188  & \NA & \NA & $[150,550]$  & 626--627   & \NA  & \NA & \NA & \NA  & \NA \\[0.75ex]
    C4& $[50,850]$ &  110--113 & $[100,1700]$ &  189--196  & $<$75 & $<$75   & $[0,600]$  & 628--630   & $<$75  &$[0,800]$  & 668--671 & $[200,1000]$&  690--693 \\
    &  & & & & $<$75 & $>$75   & $[150,950]$  & 631--634   &  & & & &         \\
    &  & & & & $>$75 & $<$75   & $[150,750]$  & 635--637   & $>$75  & $[150,1150]$  & 672--676  & &           \\
    &  & & & & $>$75 & $>$75   & $[150,1350]$  & 638--643   &  & & & &           \\[0.75ex]
    C5& $[50,850]$ & 114--117 &  $[150,1350]$  & 197--202  &   $<$75 & $<$75 & $[0,600]$  & 644--646 & $<$75&  $[150,750]$  & 677--679 & $[200,800]$&  694--696 \\
    &  & & & & $<$75 & $>$75   & $[250,850]$  & 647--649   &  & & & &           \\
    &  & & & & $>$75 & $<$75   & $[50,650]$  & 650--652   & $>$75  & $[200,1000]$  & 680--683  & &         \\
    &  & & & & $>$75 & $>$75   & $[300,1100]$  & 653--656   &  & & & &              \\
\end{scotch}
}
\end{table*}

\begin{table*}
\centering
\topcaption{The binning of $\LTmet$ and \ST distributions for the fundamental scheme in the 4L, 3L1T, 2L2T, and 1L3T channels, and the binning of \ST distribution for the advanced scheme in the 4L, 3L1T, 2L2T, and 1L3T channels. The categorization is described in Table~\ref{tab:simplifiedtable}.
The ranges, as well as the \ptmiss and $\HT$ requirements, are given in GeV. 
The first bins in the $\LTmet$ or \ST range contain the underflow, and the last bins contain the overflow. 
For the 3L1T and 2L2T channels, multiple categories in the 1\PB or 2\PB selections are combined. 
These bins are marked with a single- or a double-dagger. 
For the 1L3T channel, all the \PQb tag categories are combined and the corresponding bins are marked with an asterisk.
} \label{tab:binningtableQuad}
\cmsTable{
\begin{scotch}{llllllllllllll}
    &  \multicolumn{4}{c}{Fundamental Tables}&   \multicolumn{4}{l}{Advanced Table}&   \multicolumn{3}{l}{Advanced Table}&    \multicolumn{2}{l}{Advanced Table}\\
    &  \multicolumn{2}{l}{$\LTmet$}&  \multicolumn{2}{l}{\ST}&   \multicolumn{4}{l}{0\PB}&    \multicolumn{3}{l}{1\PB} &     \multicolumn{2}{l}{2\PB} \\
    Cat.&  Range&  Bins&  Range& Bins& \ptmiss& $\HT$& \ST Range& Bins& \ptmiss& \ST Range& Bins& \ST Range& Bins\\
    \hline
    D1&  Incl.&  118 &  Incl.  &  203 &   \NA & \NA  & Incl.  & 697   & \NA  & \NA & \NA & \NA  & \NA  \\[0.75ex]
    D2&  $[150,950]$ & 119--122 & $[0,1400]$ &  204--210 &   $<$75 & $<$50 &  $[150,550]$&  698--699&   $<$75 &  $[200,800]$ & 739--741 & $[400,1000]$&  765--767 \\
    &  & & & & $<$75 & $>$50  & $[200,1000]$ & 700--703   &  & & & &            \\
    &  & & & & $>$75 & $<$50  & $[100,700]$  & 704--706   & $>$75  & $[300,1100]$  & 742--745  & &           \\
    &  & & & & $>$75 & $>$50  & $[250,1050]$ & 707--710   &  & & & &               \\[0.75ex]
    D3&  $[150,750]$ & 123--125 & $[150,950]$ & 211--214 &   $<$75 & $<$50 & $[0,400]$&  711--712 &  \NA &  $[250,850]$ & 746--748 & Incl.&  768 \\
    &  & & & & $<$75 & $>$50   & Incl.  & 713   &  & & & &            \\
    &  & & & & $>$75 & \NA   & Incl.  & 714     &  & & & &           \\[0.75ex]
    D4& $[50,1250]$  & 126--131 & $[100,1500]$ & 215--221 &   $<$75 & $<$50 & $[0,1000]$ & 715--719 & $<$75 &  $[100,900]$& 749--752 & $[250,1050]$ & 769--772 \\
    &  & & & & $<$75 & $>$50   & $[150,1150]$  & 720-724   &  & & & &            \\
    &  & & & & $>$75 & $<$50   & $[150,750]$  & 725--727   & $>$75  & $[250,1050]$  & 753--756  & &           \\
    &  & & & & $>$75 & $>$50   & $[400,1200]$  & 728--731   &  & & & &               \\[0.75ex]
    D5&  $[100,700]$ & 132--134& $[50,1050]$ & 222--226   & \NA & \NA   & \NA  & \NA   & $<$75 &  $[100,900]$ & 757--760 & Incl. &  773 \\
    &  & & & & \NA & \NA   & \NA  & \NA   & $>$75  & Incl.  & 761   & &          \\[0.75ex]
    D6&  $[0,800]$  & 135--138 & $[100,1100]$  & 227--231  &   $<$75 & $<$50 & $[0,600]$  & 732--734 &  \NA & $[150,750]$  & 762--764 & Incl.&  774 \\
    &  & & & & $<$75 & $>$50   & $[150,750]$  & 735--737  &  & & & &            \\
    &  & & & & $>$75 & \NA   & Incl.  & 738   & &  &  & &           \\ [1.5ex]
    E1& $[100,500]$ & 139--140 & $[250,650]$  & 232--233 & \NA & \NA & Incl.  & 775   & \NA  &  $[150,950]$ &795--798~$\dagger$ & $[250,850]$& 802--804~$\dagger$ \\[0.75ex]
    E2& $[100,900]$ & 141--144 & $[50,1250]$  & 234--238 & \NA & \NA & $[100,1100]$  & 776--780 & \NA & $[150,950]$ & 795--798~$\dagger$ & $[250,850]$& 802--804~$\dagger$ \\[0.75ex]
    E3& $[150,750]$ & 145--147 & $[0,1000]$  & 239--244 & \NA & \NA & $[0,800]$  & 781--784   & \NA  & $[150,950]$ &795--798~$\dagger$ & $[250,850]$&   802--804~$\dagger$ \\[1.5ex]
    F1&   Incl.  & 148& $[50,450]$ & 245--246   & \NA & \NA   & Incl.  & 785&  \NA  & $[200,800]$ & 799--801~$\ddagger$ & Incl.  & 805~$\ddagger$  \\[0.75ex]
    F2&   $[150,550]$  & 149--150 & $[100,700]$  & 247--249   & \NA & \NA   & $[150,550]$  & 786--787   & \NA  & $[200,800]$  &799--801~$\ddagger$ & Incl. & 805~$\ddagger$ \\[0.75ex]
    F3&   $[100,700]$  & 151--153 & $[150,950]$  & 250--253   & \NA & \NA   & $[100,900]$  & 788--791   & \NA  & $[200,800]$  &799--801~$\ddagger$ & Incl. & 805~$\ddagger$ \\[0.75ex]
    F4&   $[150,550]$  & 154--155 & $[200,800]$  & 254--256   & \NA & \NA   & $[150,550]$  & 792--793   & \NA  & $[200,800]$   &799--801~$\ddagger$ & Incl. & 805~$\ddagger$ \\[0.75ex]
    G1&   Incl.  &  156 &  Incl.  &  257   &  \NA & \NA   & Incl.  & 794$^*$   & \NA  & Incl. & 794$^*$ & Incl.  & 794$^*$     \\
\end{scotch}
}
\end{table*}

\clearpage

\subsection{Model-dependent selections}

The model-dependent SRs are defined by employing BDTs that are trained to discriminate a specific signal from the SM backgrounds. 
We have used the BDT implementation from the \textsc{tmva} package~\cite{Helge:TMVA}. 
Individual BDTs for specific model scenarios and for each year of data collection are trained to discriminate the signal process from the major SM backgrounds ($\WZ$, $\ZZ$, DY, $\ttbar$, and $\ZG$). 

\subsubsection{Discriminant training}

The BDT training process consider all multilepton events that pass the event selection, and are performed separately for each year of data collection.
Events passing the CR selections are removed from the training process, but are used to validate the modeling of BDT input variables and the outputs of the trained BDTs.

For each of the three data-taking periods, BDTs are trained using statistically independent simulated event samples of signal and background from the other two periods.
The misidentified lepton background contributions used for training the BDT are taken from the DY and $\ttbar$ MC samples; hence the training does not employ the sideband events in data used to predict the misidentified lepton backgrounds.

The properties of the targeted BSM models vary considerably across the probed 0.1 to 2.0\TeV mass range, and may depend explicitly on lepton flavor. 
To address this, we define small windows in signal mass, combining a few neighboring signal mass hypotheses in a single training, yielding three mass-range-specific BDTs for each signal. 

For the vector-like lepton model, a single BDT is trained using both the doublet and singlet scenarios.
For the type-III seesaw model, separate BDTs are trained for the flavor-democratic ($\mathcal{B}_{\Pe}=\mathcal{B}_{\PGm}=\mathcal{B}_{\PGt}$) scenario and for the $\mathcal{B}_{\PGt}=1$ scenario. 
Similarly, for the leptoquark model, two separate BDTs are trained for the models with couplings to \PGt leptons ($\mathcal{B}_{\PGt}=1$) and light leptons ($\mathcal{B}_{\Pe}+\mathcal{B}_{\PGm}=1$).

A combination of up to 48 object- and event-level quantities are used as input variables to the model-specific trainings.
These include \pt, invariant masses, angular variables, lepton charge and flavor, and \PQb-tagged jet multiplicities, as described in Section~\ref{sec:eventselec}.
A full list of the quantities used in the BDT training process is provided in Table~\ref{tab:allbdtvar}. 

\begin{table*}[h]
\centering
\topcaption{Input variables used for the BDTs trained for the various BSM models.
Note that the indices $i,j$ run over the leptons of all flavors ($i,j=1,2,3,4$) in a given event.
If a given variable is not defined in a given channel, the variable is set to a nonphysical default value for signal and background processes, and plays no role in training.
}\label{tab:allbdtvar}
\cmsTable{
\begin{scotch}{ l  l  l  l  }
Variable type & \multicolumn{3}{c} {Used for}  \\
 & All signals&  Vector-like lepton &  Seesaw and leptoquarks\\\hline
Event & $\HT$, \ptmiss, \nbj, \ml & \ql & \LT, $\pt^{\mathrm{i}}/\LT$, $\LT/\ST$, $\HT/\ST$, $\ptmiss/\ST$ \\ [0.75ex]
Lepton & $\pt^{\mathrm{i}}$, $\pt^{\text{OSSF}}$ & & \\[0.75ex]
Angular& \dRmin& {\small Max, Min:} $\Delta\phi^{\mathrm{i}}$, ~{\small Max, Min:} $\Delta\phi^{\mathrm{ij}}$ & {\small Max:} $\Delta\eta^{\mathrm{ij}}$ \\[0.75ex]
Mass & $\MT^{\mathrm{i}}$ & $M^{\mathrm{ij}}$, $\MT^{\mathrm{12}}$, $\MT^{\mathrm{13}}$, $\MT^{\mathrm{23}}$ & \\
\end{scotch}
}
\end{table*}

All BDTs used for each BSM model have 800 trees with a maximum depth of 10, and utilize a minimum node size of 1.5\% with 10 steps during node cut optimization.
The \textit{GradientBoost} algorithm is chosen for boosting the trees.
The BDT hyperparameters, as well as the choices of training strategy described here have been optimized to give the largest background
rejection for a given signal efficiency in the training samples. 
This optimization is done while ensuring that the performance of the BDTs in orthogonal testing data sets matches the training performance, and that the performance in testing data sets does not change significantly for small changes in the BDT hyperparameters.

To summarize, for the vector-like lepton model, three mass ranges and thus three BDTs are trained per year of data taking. 
For the type-III seesaw and leptoquark models, three mass ranges with two flavor scenarios in each range are considered,
giving six BDTs each per year.

\subsubsection{Discriminant application}

The BDT of a given signal model training provides a score in the range of $(-1,1)$, with the interval close to 1 associated
with the highest sensitivity to the signal.
Therefore, we define a number of variable-width regions across the BDT score, with narrower regions defined on the high score side.
To further increase signal sensitivity, the BDT scores in the three-lepton (3L, 2L1T, 1L2T) channels are combined into a single distribution; similarly the BDT scores in the four-lepton (4L, 3L1T, 2L2T, 1L3T) channels are combined into one distribution as well. 
This procedure is applied to each year separately in order to achieve optimal signal-to-noise ratio, with bin widths chosen in each year to obtain uniformly increasing expected background yields. This defines the BDT regions in which we perform counting experiments. 

We denote signal-specific BDTs by \textit{SS}, \textit{VLL}, and \textit{LQ} for the type-III seesaw, vector-like lepton, and leptoquark models, respectively. 
The various mass ranges are denoted by \textit{L} (low), \textit{M} (medium), and \textit{H} (high). 
For each signal mass hypothesis, the performance from every mass-range BDT of that signal model is considered, and the BDT that gives the best expected exclusion is chosen for that mass in the evaluation. 
Occasionally this leads to mismatched training and application mass ranges. 
For example, a vector-like \PGt lepton of $m_{\vltau}=500\GeV$ is used in the \textit{VLL-M} training, but the best expected performance at $m_{\vltau}=500\GeV$ is achieved from the \textit{VLL-H} training, and therefore is used for its application. 
This behavior can be attributed to the larger acceptance and population of events from higher mass samples in the tails of sensitive variables such as \LT, \ptmiss etc., which benefits the signal versus background separation. 
A summary of all these mass ranges for the individual BDTs used in the training and for the application for each model can be found in Table~\ref{tab:BDTmassranges}.

\begin{table*}[h]
  \centering
    \topcaption{ Signal mass points as used in the training of BDTs and the masses for which the specific trained BDT is applied in the SRs according to the best sensitivity. The labels \textit{L}, \textit{M}, and \textit{H} denote low, medium, and high mass ranges, respectively.}
    \begin{scotch}{l l l}
      BDT & Trained using masses (\GeVns{})&  Applied for masses (\GeVns{})\\ \hline
      Type-III seesaw \\
      \textit{SS-L} & 200, 300& 200, 300\\
      \textit{SS-M} & 400, 550, 700, 850&  400, 550\\
      \textit{SS-H} & 1000, 1250& 700 and higher \\ [1.5ex]
      Vector-like lepton \\
      \textit{VLL-L} & 100, 150, 200& 100, 150, 200\\
      \textit{VLL-M} & 300, 500&  250, 300, 350, 400\\
      \textit{VLL-H} & 650, 700, 800&  450 and higher\\ [1.5ex]
      Leptoquarks \\
      \textit{LQ-L} & 300, 400& 300, 400\\
      \textit{LQ-M} & 500, 600, 700&  500, 600, 700\\
      \textit{LQ-H} & 1200, 1300, 1400& 800 and higher\\
    \end{scotch}
  \label{tab:BDTmassranges}
\end{table*}

\begin{figure}[hbt!]
\centering
\includegraphics[width=0.49\textwidth]{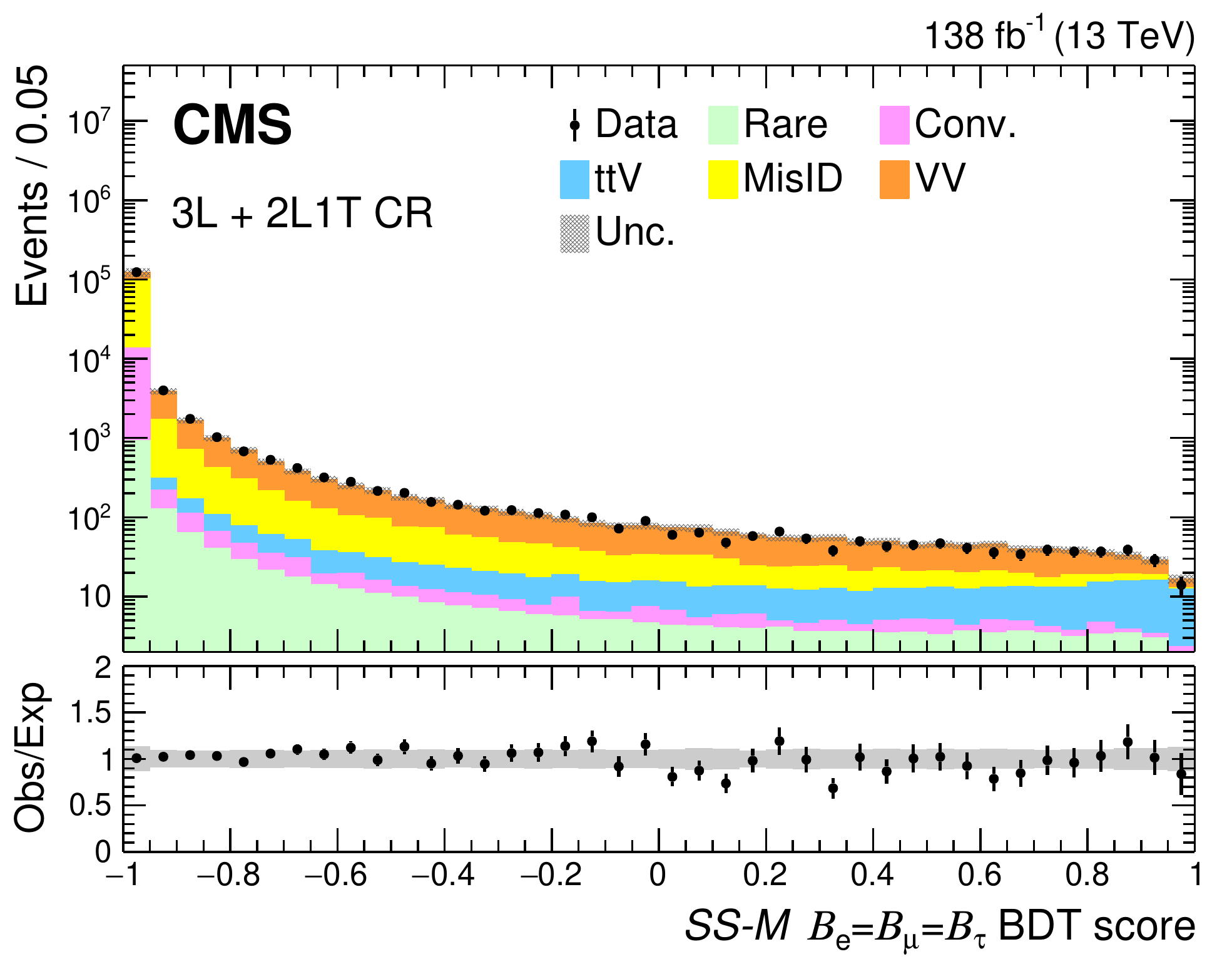}
\includegraphics[width=0.49\textwidth]{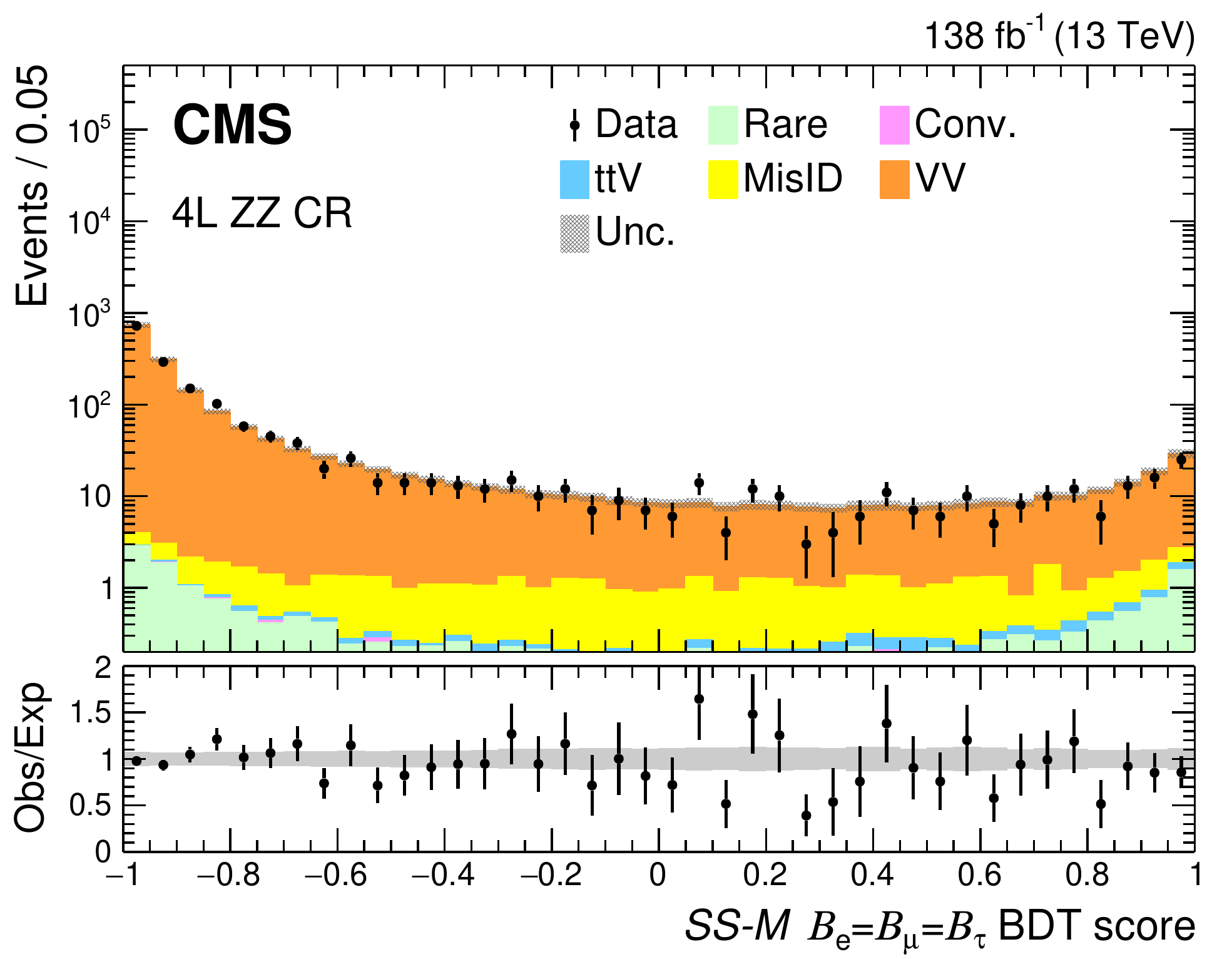}
\caption{\label{fig:ControlBDT} Distributions of BDT score from the \textit{SS-M} $\mathcal{B}_{\Pe}=\mathcal{B}_{\PGm}=\mathcal{B}_{\PGt}$ BDT are shown for the 3L+2L1T CR (\cmsLeft), and the 4L $\ZZ$ CR (\cmsRight). The 3L+2L1T CR consists of the 3L OnZ, 3L \ZG, and 2L1T MisID CRs. The lower panel shows the ratio of observed events to the total expected background prediction. The gray band on the ratio represents the sum of statistical and systematic uncertainties in the SM background prediction.
}
\end{figure}

\begin{figure}[hbt!]
\centering
\includegraphics[width=0.49\textwidth]{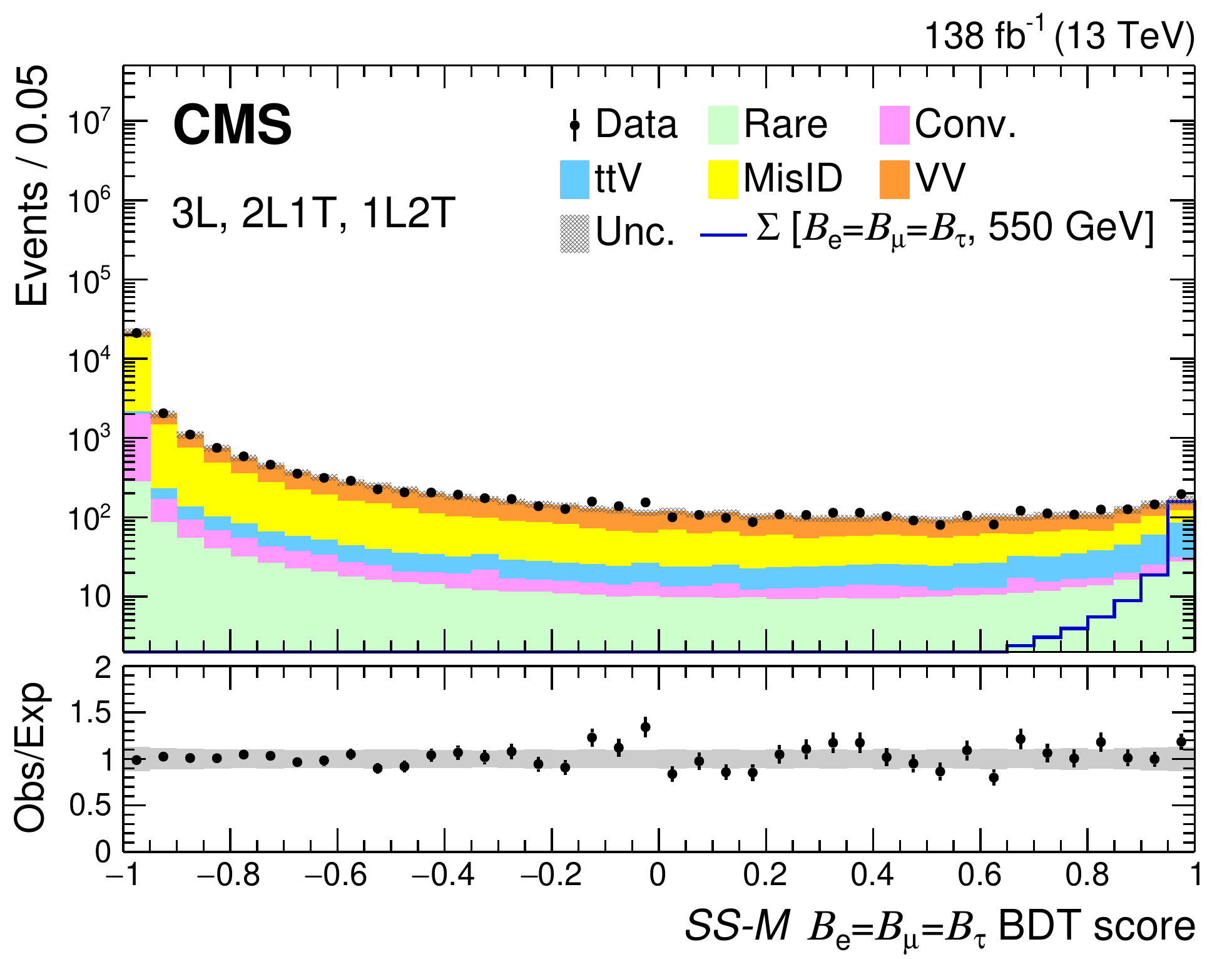}
\includegraphics[width=0.49\textwidth]{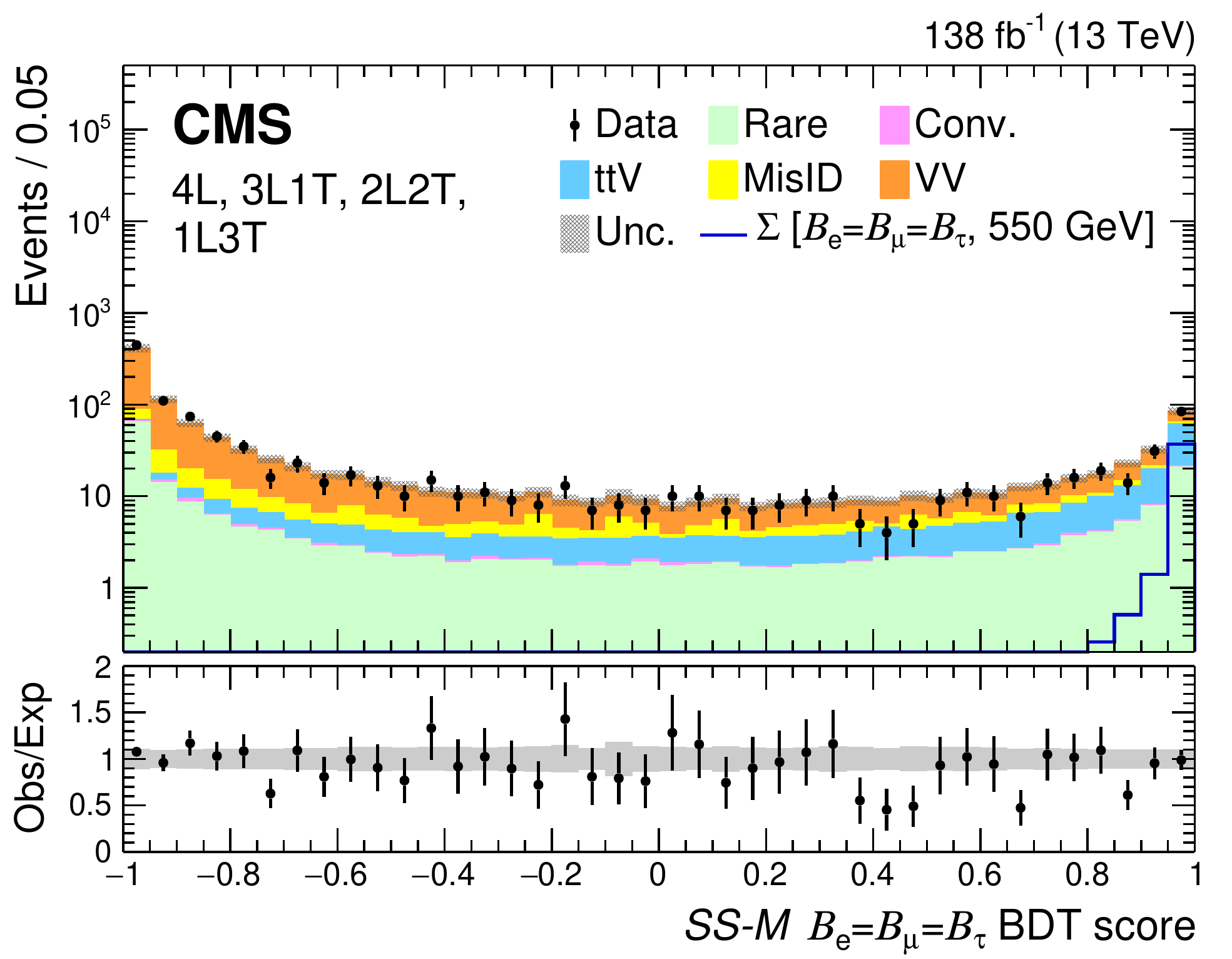}
\caption{\label{fig:SRinclBDT} Distributions of BDT score from the \textit{SS-M} $\mathcal{B}_{\Pe}=\mathcal{B}_{\PGm}=\mathcal{B}_{\PGt}$ BDT are shown for events lying outside the control regions in the three-lepton (3L, 2L1T, 1L2T) channels (\cmsLeft), and the four-lepton (4L, 3L1T, 2L2T, 1L3T) channels (\cmsRight). The lower panel shows the ratio of observed events to the total expected background prediction. The gray band on the ratio represents the sum of statistical and systematic uncertainties in the SM background prediction. For illustration, an example signal hypothesis for the production of the type-III seesaw heavy fermions in the flavor-democratic scenario for $m_{\Sigma}=550\GeV$ is also overlaid.}
\end{figure}

Figure~\ref{fig:ControlBDT} shows the output from the \textit{SS-M} BDT in the flavor-democratic scenario, with statistical and systematic uncertainties in the SM background prediction, as described in Section~\ref{sec:systematics}. The BDT output is shown in the 4L $\ZZ$ CR, and in the combined 3L OnZ, 3L \ZG\ and 2L1T MisID CRs, and the data are observed to be in good agreement with the expected SM  background prediction. The output distributions of the same BDT in the three-lepton (3L, 2L1T, 1L2T) and four-lepton (4L, 3L1T, 2L2T, 1L3T) channels, for events that do not fall in any of the control regions, are shown in Fig.~\ref{fig:SRinclBDT}. Such distributions are then transformed to the variable-width bins as explained above, to form the BDT regions for performing the final search. These are shown in Section~\ref{sec:results}.

\section{Systematic uncertainties} \label{sec:systematics}

All background and signal estimates have uncertainties due to the finite number of events in simulated samples or data sidebands.
These statistical uncertainties are typically small, as we utilize large MC samples or data sidebands, but are nonetheless propagated to the analysis.

The fractional inclusive normalization uncertainties for $\WZ$, $\ZG$, $\ttZ$, and $\ZZ$ backgrounds are 3--5, 10, 15--25, and 4--5\% respectively, in all three years of data collection.
For all other background processes, a $50\%$ flat systematic uncertainty is assigned to the theoretical cross section estimates at LO or NLO to cover any higher-order effects, as well as the renormalization and factorization scale uncertainties.

Systematic uncertainties arise from the corrections applied to the background and signal simulations for lepton reconstruction, isolation, and trigger efficiencies; {\cPqb} tagging efficiency; 
pileup modeling; electron and jet energy resolution; and electron, muon, \PGt leptons, jet, and unclustered energy scale measurements. 
The uncertainties in such corrections typically correspond to 1--10\% variation in the simulation-based irreducible background and signal yields across all SRs. 
Similarly, uncertainties due to choices of factorization and renormalization scales~\cite{Cacciari:2003fi} and PDFs~\cite{Ball:2014uwa,Ball:2017nwa} are also evaluated for signal and dominant irreducible background processes, yielding variations smaller than 10\% in the SRs.
The uncertainties in the diboson \pt and jet multiplicity modeling in the $\WZ$ and $\ZZ$ MC samples typically yield variations in the range of 5--15 and 5--30\%, respectively. 
The electron charge misidentification rate is also corrected in a dedicated DY-enriched dielectron selection of data events, and a $30\%$ relative uncertainty is assigned to it.
The integrated luminosities of the 2016, 2017, and 2018 data-taking periods are individually known with uncertainties in the 1.2--2.5\% range~\cite{CMS:2021xjt,CMS:2018elu,CMS:2019jhq}, while the total 2016--2018 integrated luminosity has an uncertainty of 1.6\%. 

The uncertainty in the misidentified lepton background, which is estimated from data via the matrix method, is dominated by the uncertainties in the lepton misidentification rates.
The relative statistical uncertainties in the measurement of the misidentification rates are typically in the 10--30\% range. 
As we use an extrapolation of electron and muon misidentification rate measurements for $\pt>50\GeV$ and \tauh misidentification
rate measurements for $\pt>80\GeV$, we double these uncertainties and assign a flat $60\%$ relative uncertainty for all high-\pt leptons.
In summary, lepton misidentification rates have typical relative uncertainties of 10, 30, and 60\% in the low, medium, and high lepton \pt regions, respectively, where low is defined as ($10<\pt<20\GeV$ for light leptons, $10<\pt<30\GeV$ for \tauh), medium is ($20<\pt<50\GeV$ for light leptons, $30<\pt<80\GeV$ for \tauh), and high is ($\pt>50\GeV$ for light leptons, $\pt>80\GeV$ for \tauh).
These result in variations in the range of 20--50\% of the misidentified lepton background contribution estimates, and these nuisances are also kept uncorrelated in each of the three data-taking periods.
In addition, we consider process-dependent uncertainties in the lepton misidentification rates. 
These are estimated by comparing the misidentification rates observed in the DY- and $\ttbar$-enriched measurements, and are typically in the range of 5--25\% and correlated across the data-taking periods.

In order to account for different compositions of misidentified lepton origins in multilepton events with and without \PQb-tagged jets, the systematic uncertainties in the misidentified lepton backgrounds in different table categories and \PQb-tagged jet multiplicities, as well as BDT regions, are taken to be uncorrelated. The uncertainties in the diboson \pt and jet multiplicity modeling in the $\WZ$ and $\ZZ$ MC samples are taken to be uncorrelated between the different channels. All other systematic uncertainties are taken to be correlated across all table categories, BDT regions, and channels. 

The uncertainty sources, the affected processes, the resulting uncertainty in the yield of those processes, and the correlations across the data-taking periods are summarized in Table~\ref{tab:systematics}.

\begin{table*}
\centering
\topcaption{Sources, magnitudes, effective variations, and correlation properties of systematic uncertainties in the SRs.
Uncertainty sources marked as ``Yes'' in the Correlation column have their nuisance parameters correlated across the 3 years of data collection. 
} \label{tab:systematics}
\cmsTable{
\begin{scotch}{l l l l l l}
Uncertainty source                    & Magnitude      & Type         & Processes                 & Variation   & Correlation \\ \hline
Statistical                                 & 1--100\%      & per event    & All MC samples        & 1--100\%   & No  \\ 
Integrated luminosity                                  & 1.2--2.5\%    & per event    & Conv./Rare/Signal     & 1.2--2.5\% & Yes  \\ 
Electron/Muon reco., ID, and iso. efficiency & 1--5\%        & per lepton   & All MC samples        & 2--5\%     & No  \\
\tauh reco., ID, and iso. efficiency           & 5--15\%       & per lepton   & All MC samples        & 5--25\%    & No  \\
Lepton displacement efficiency              & 1--2\%        & per lepton   & All MC samples        & 3--5\%     & No  \\
Trigger efficiency                          & 1--4\%        & per lepton   & All MC samples        & $<$3\%    & No  \\
$\PQb$ tagging efficiency                            & 1--10\%       & per jet      & All MC samples        & 2--5\%     & No  \\ 
Pileup                                      & 5\%          & per event    & All MC samples        & $<$3\%    & Yes \\
PDF, fact./renorm. scale                    & $<$20\%      & per event    & All MC samples        & $<$10\%   & Yes \\
Jet energy scale                            & 1--10\%       & per jet      & All MC samples        & $<$5\%    & No  \\
Unclustered energy scale                    & 1--25\%       & per event    & All MC samples        & $<$2\%    & No  \\
Electron energy scale and resolution        & $<$2\%       & per lepton & All MC samples        & $<$5\%    & Yes \\ 
Muon energy scale and resolution            & 2\%          & per lepton     & All MC samples        & $<$5\%    & No  \\
\tauh energy scale                            & $<$10\%      & per lepton & All MC samples        & $<$5\%    & No  \\ 
Electron charge misidentification           & 30\%         & per lepton & All MC samples        & $<$25\%   & No  \\ [1ex]
$\WZ$ normalization                            & 3--5\%        & per event    & $\WZ$                 & 3--5\%     & No  \\
$\ZZ$ normalization                            & 4--5\%        & per event    & $\ZZ$                 & 4--5\%     & No  \\
$\ttZ$ normalization                       & 15--25\%      & per event    & $\ttZ$                & 15--25\%   & No  \\
Conversion normalization                    & 10--50\%      & per event    & $\ZG$/Conv.           & 10--50\%   & No  \\
Rare normalization                          & 50\%         & per event    & Rare                  & 50\%      & No  \\
Prompt and misidentification rates          & 20--60\%      & per lepton   & MisID                 & 20--50\%   & No  \\
DY-$\ttbar$ process dependence             & 5--25\%      & per lepton   & MisID                 & 5--25\%   & Yes  \\
Diboson jet multiplicity modeling           & $<$30\%      & per event    & $\WZ/\ZZ$             & 5--30\%    & No  \\
Diboson \pt modeling                        & $<$30\%      & per event    & $\WZ/\ZZ$             & 5--15\%    & No  \\
\end{scotch}}
\end{table*}

\section{Results} \label{sec:results}

\begin{figure*}[hbt!]
\centering
\includegraphics[width=0.9\textwidth]{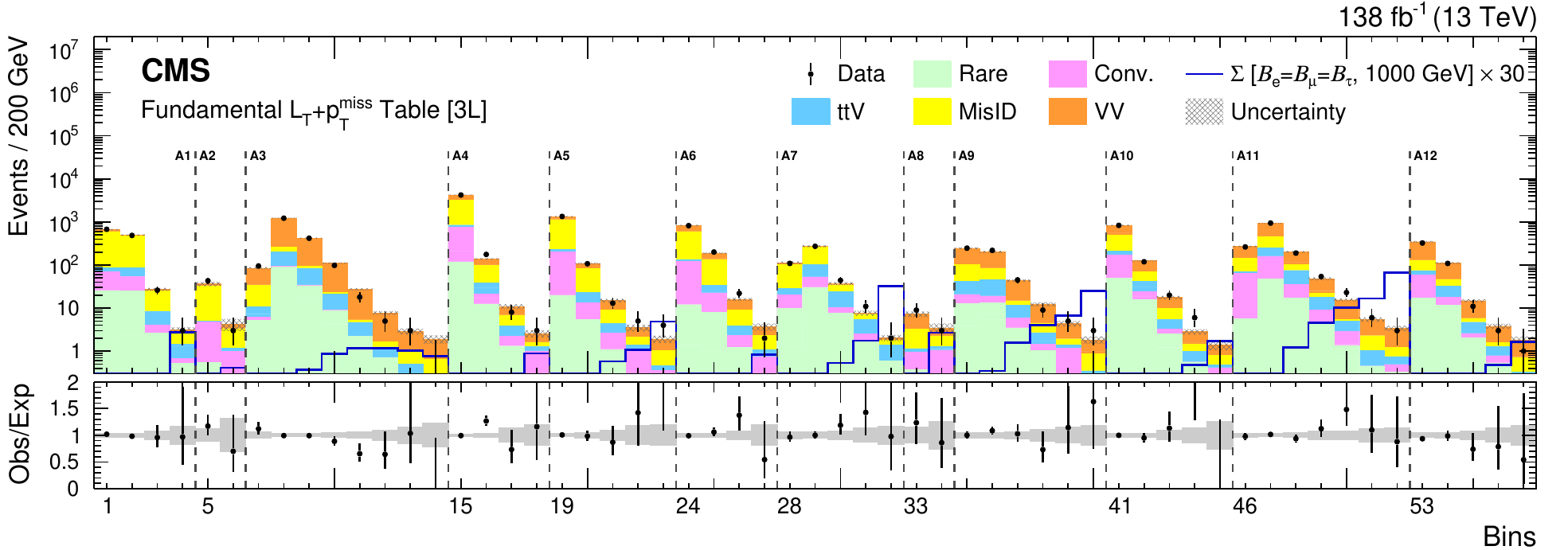}
\includegraphics[width=0.9\textwidth]{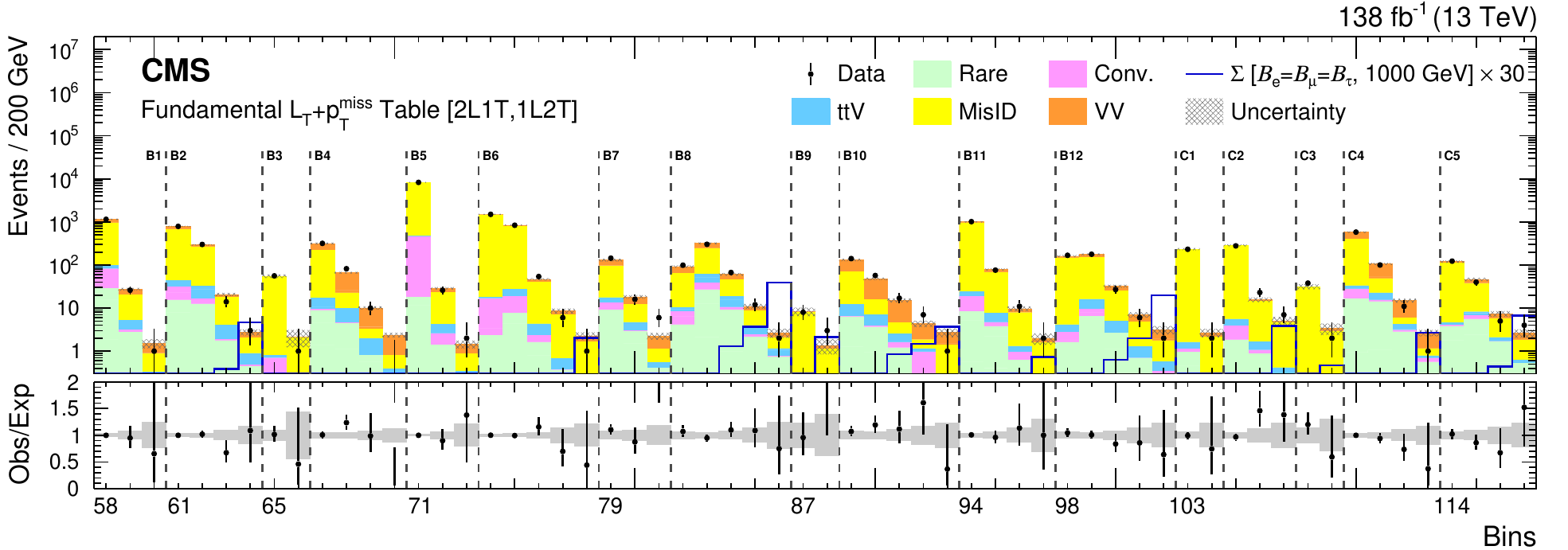}
\includegraphics[width=0.9\textwidth]{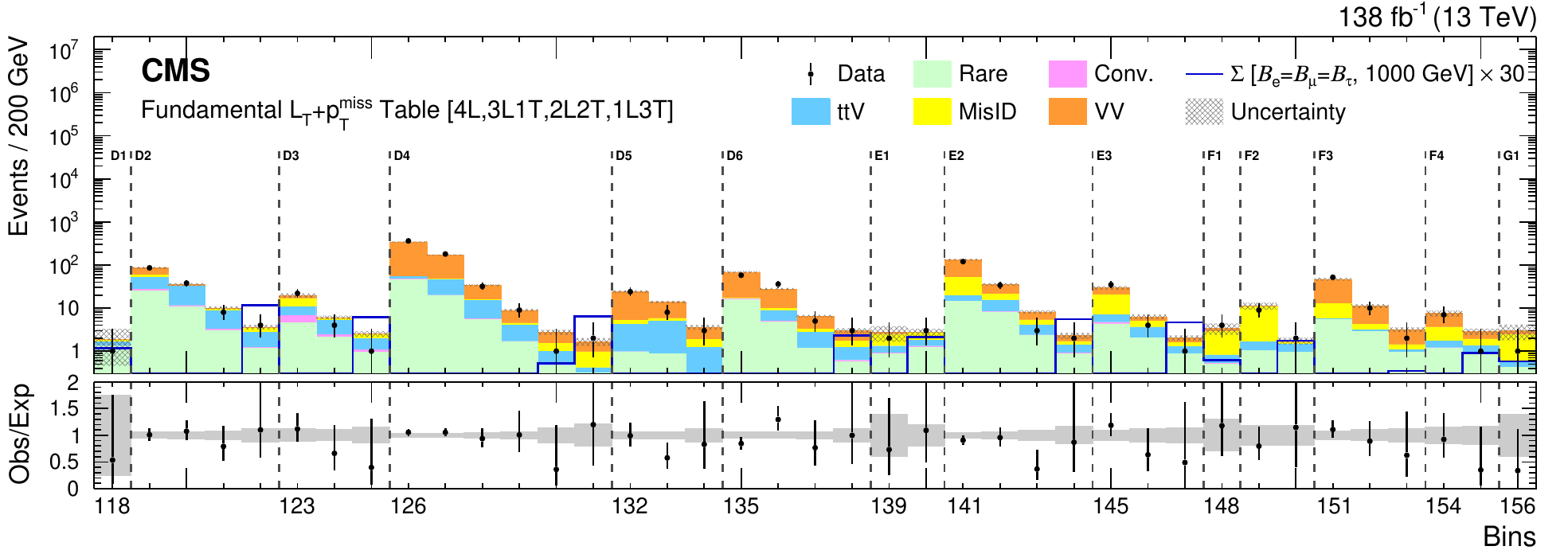}
\caption{\label{fig:resultLTMETTable} The SR distributions of the fundamental $\LTmet$ table for the combined 2016--2018 data set. The detailed description of the bin numbers can be found in Tables~\ref{tab:binningtable3L}--\ref{tab:binningtableQuad}. The lower panel shows the ratio of observed events to the total expected background prediction. The gray band on the ratio represents the sum of statistical and systematic uncertainties in the SM background prediction. The expected SM background distributions and the uncertainties are shown after fitting the data under the background-only hypothesis.
For illustration, an example signal hypothesis for the production of the type-III seesaw heavy fermions in the flavor-democratic scenario for $m_{\Sigma}=1\TeV$, before the fit, is also overlaid. 
}
\end{figure*}

\begin{figure*}[hbt!]
\centering
\includegraphics[width=0.9\textwidth]{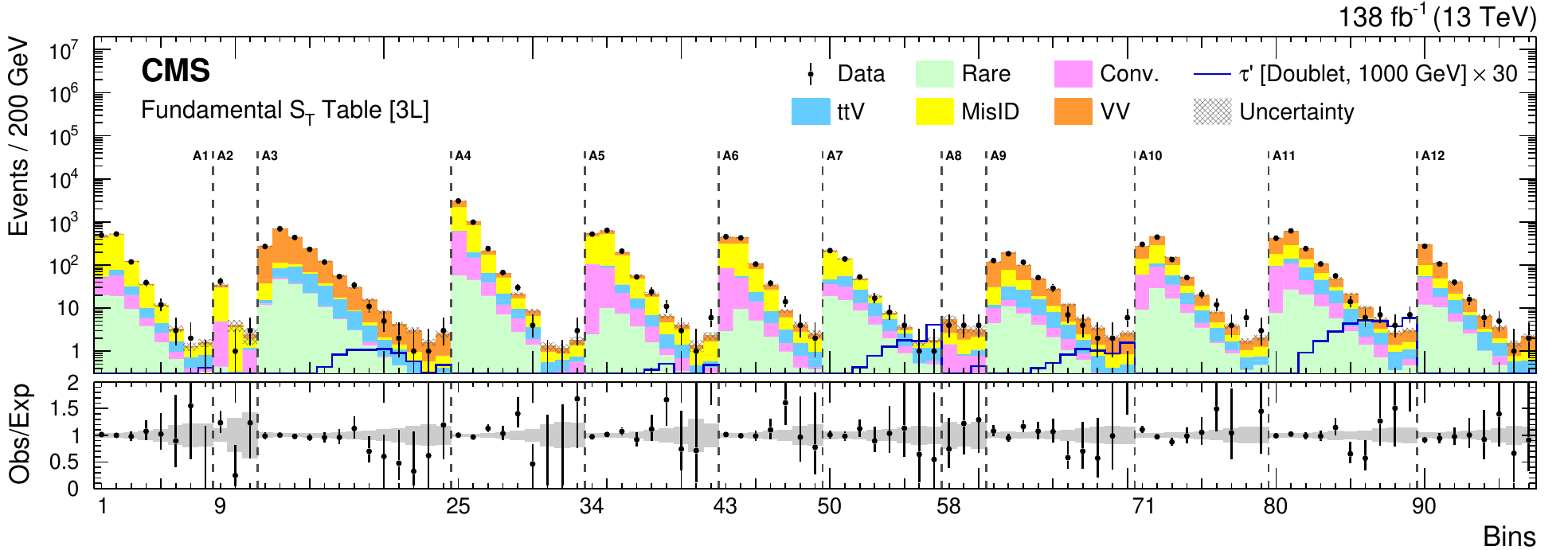}
\includegraphics[width=0.9\textwidth]{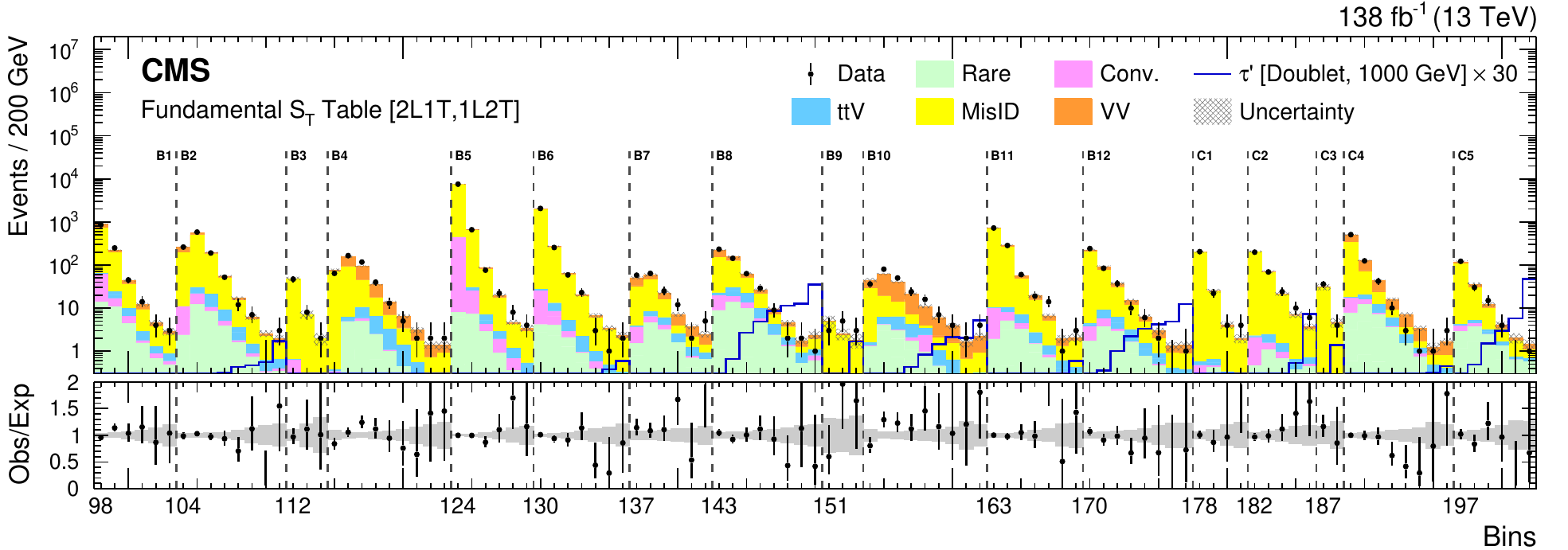}
\includegraphics[width=0.9\textwidth]{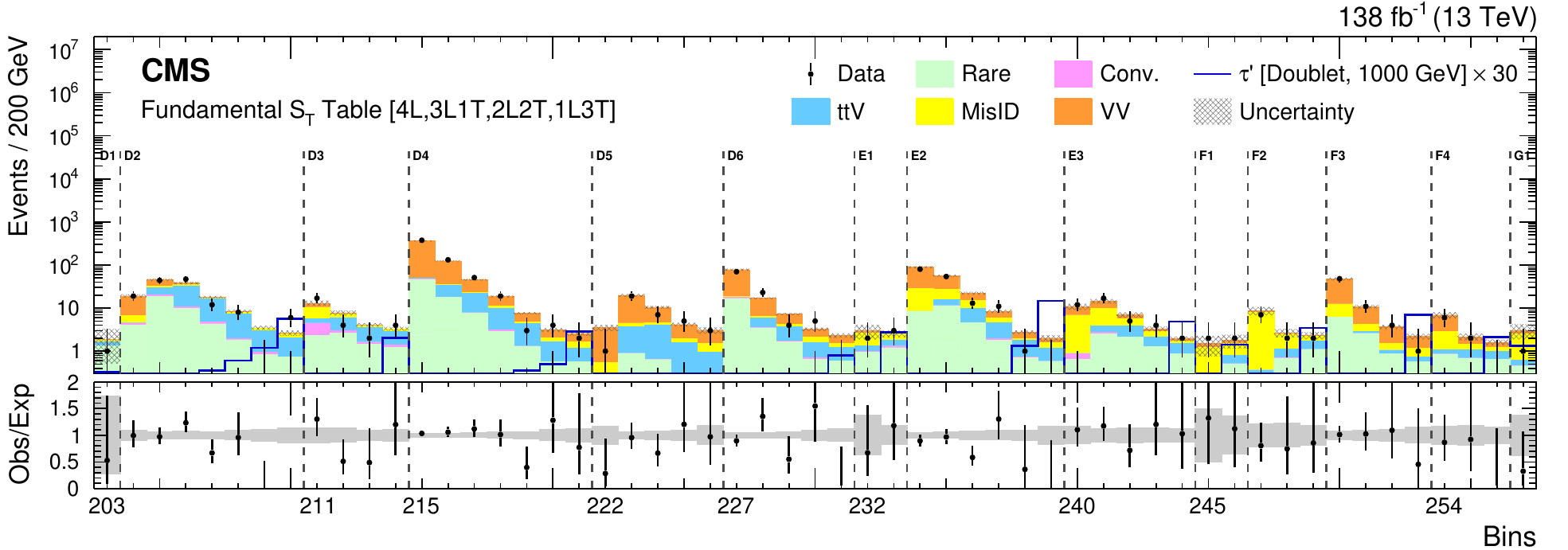}
\caption{\label{fig:resultSTTable} The SR distributions of the fundamental \ST table for the combined 2016--2018 data set. 
The detailed description of the bin numbers can be found in Tables~\ref{tab:binningtable3L}--\ref{tab:binningtableQuad}. The lower panel shows the ratio of observed events to the total expected background prediction. The gray band on the ratio represents the sum of statistical and systematic uncertainties in the SM background prediction. The expected SM background distributions and the uncertainties are shown after fitting the data under the background-only hypothesis.
For illustration, an example signal hypothesis for the production of the vector-like \PGt lepton in the doublet scenario for $m_{\vltau}=1\TeV$, before the fit, is also overlaid.
}
\end{figure*}

\begin{figure*}[hbt!]
\centering
\includegraphics[width=0.9\textwidth]{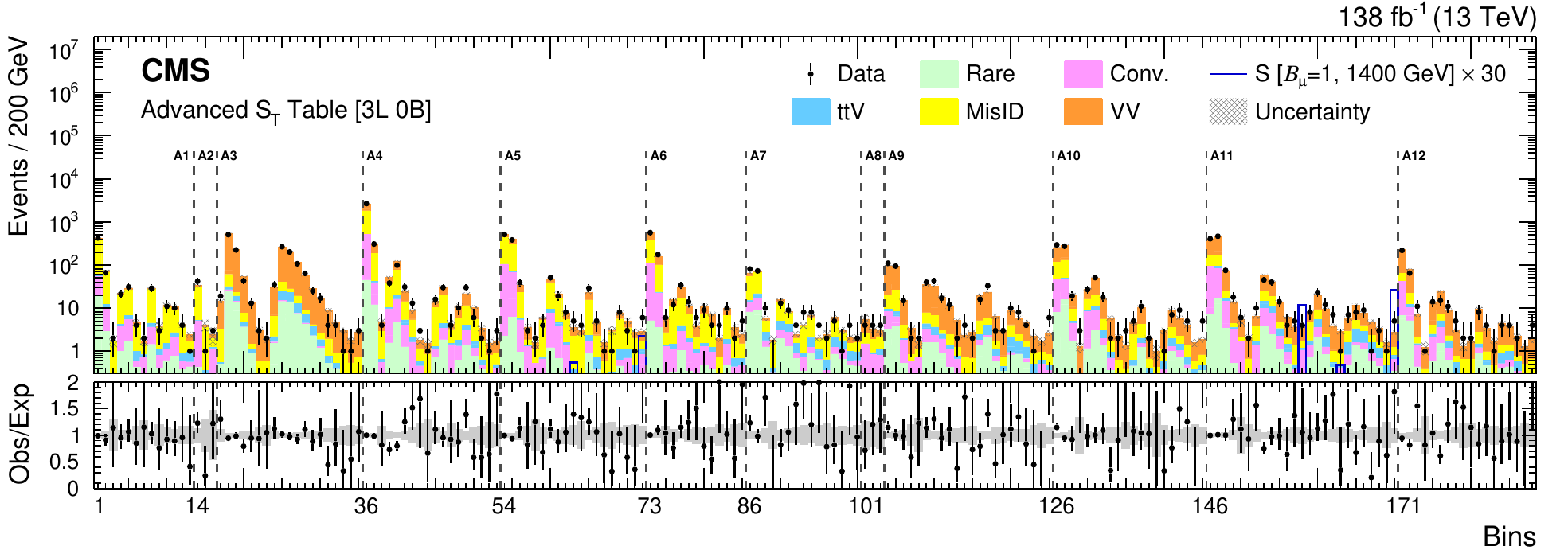}
\includegraphics[width=0.9\textwidth]{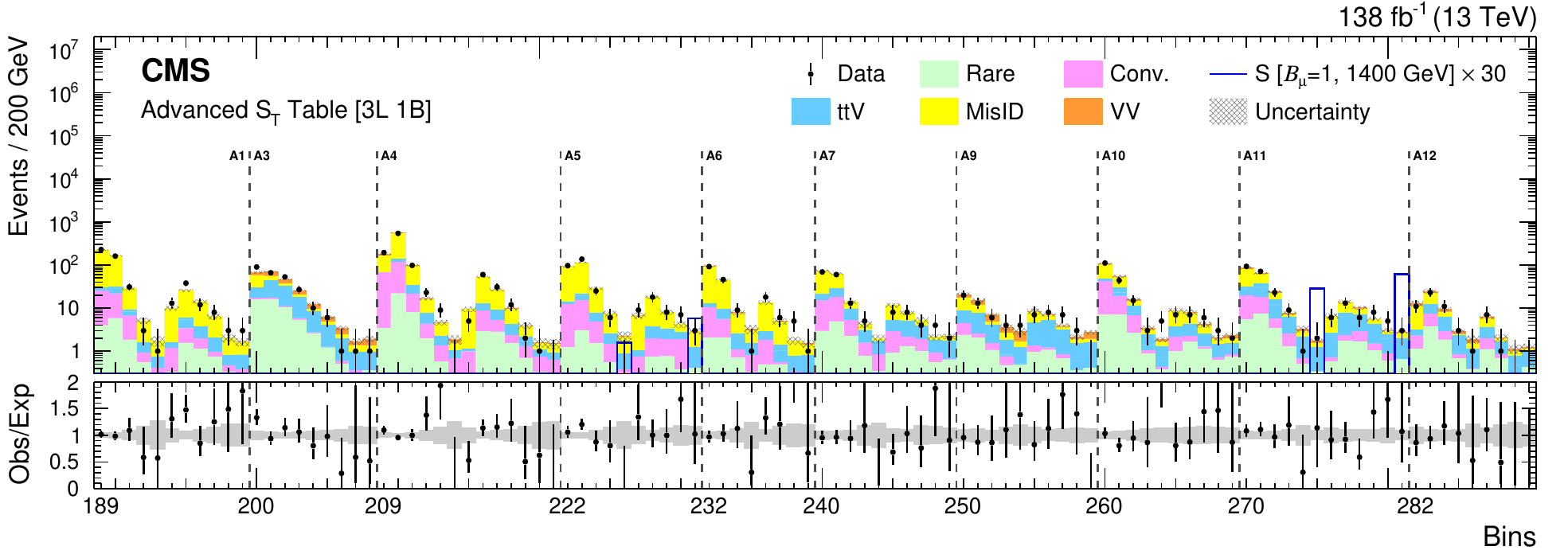}
\includegraphics[width=0.9\textwidth]{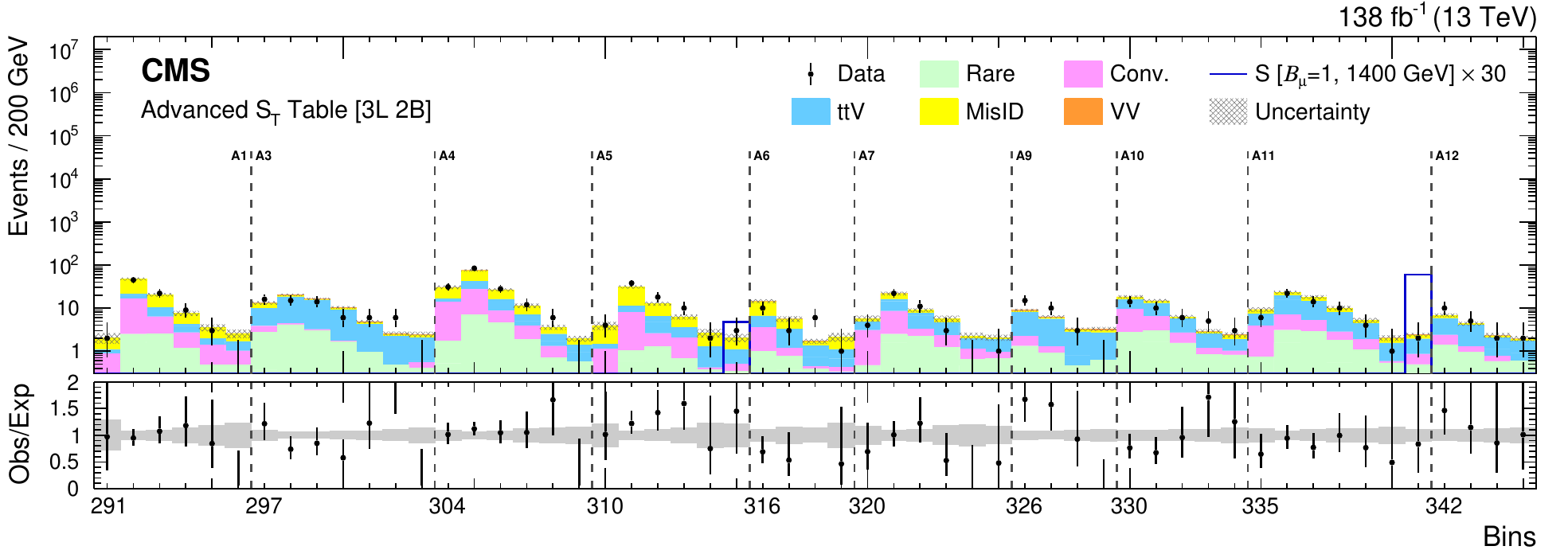}
\caption{\label{fig:resultAdvTable3L} The 3L SR distributions of the advanced \ST table for the combined 2016--2018 data set. The detailed description of the bin numbers can be found in Table~\ref{tab:binningtable3L}. The lower panel shows the ratio of observed events to the total expected background prediction. The gray band on the ratio represents the sum of statistical and systematic uncertainties in the SM background prediction. The expected SM background distributions and the uncertainties are shown after fitting the data under the background-only hypothesis.
For illustration, an example signal hypothesis for the production of the scalar leptoquark coupled to a top quark and a muon for $m_{\PS}=1.4\TeV$, before the fit, is also overlaid.
}
\end{figure*}

\begin{figure*}[hbt!]
\centering
\includegraphics[width=0.9\textwidth]{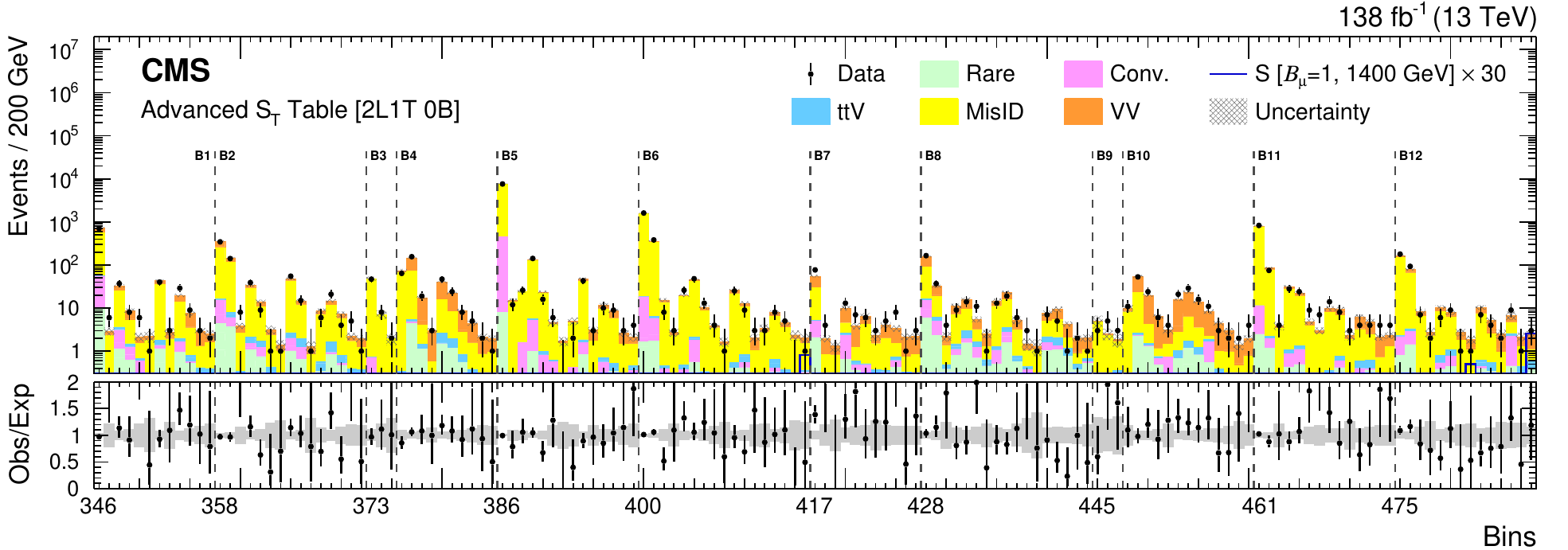}
\includegraphics[width=0.9\textwidth]{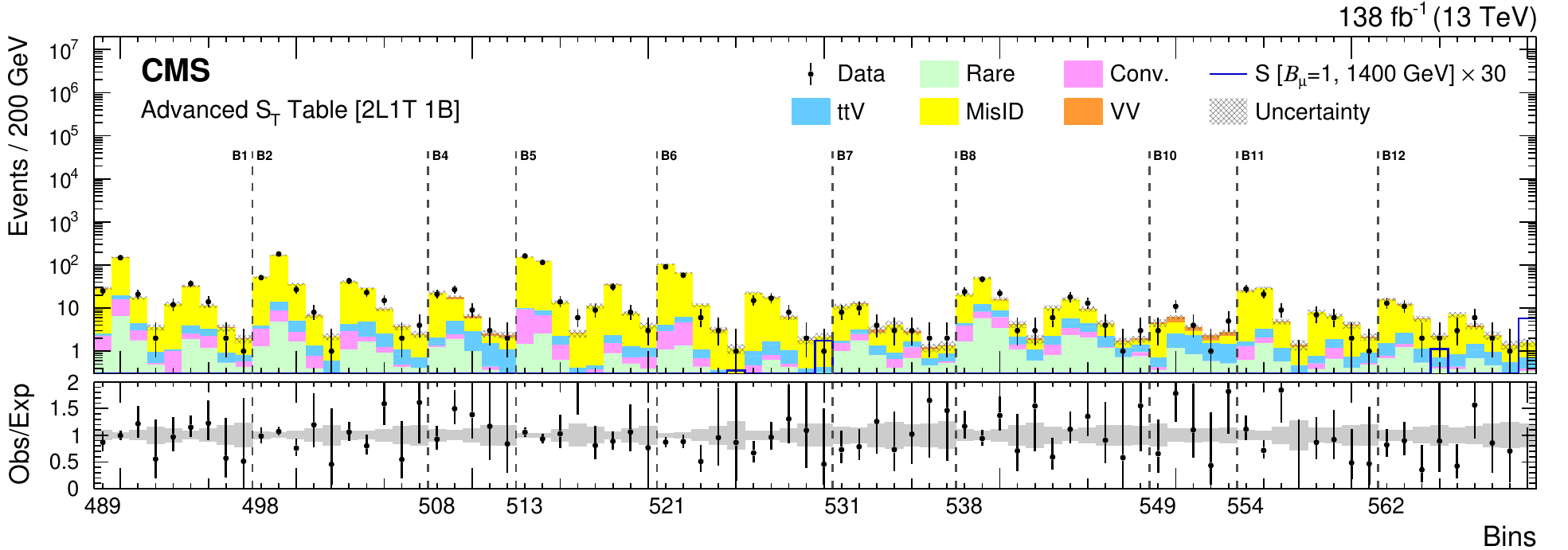}
\includegraphics[width=0.9\textwidth]{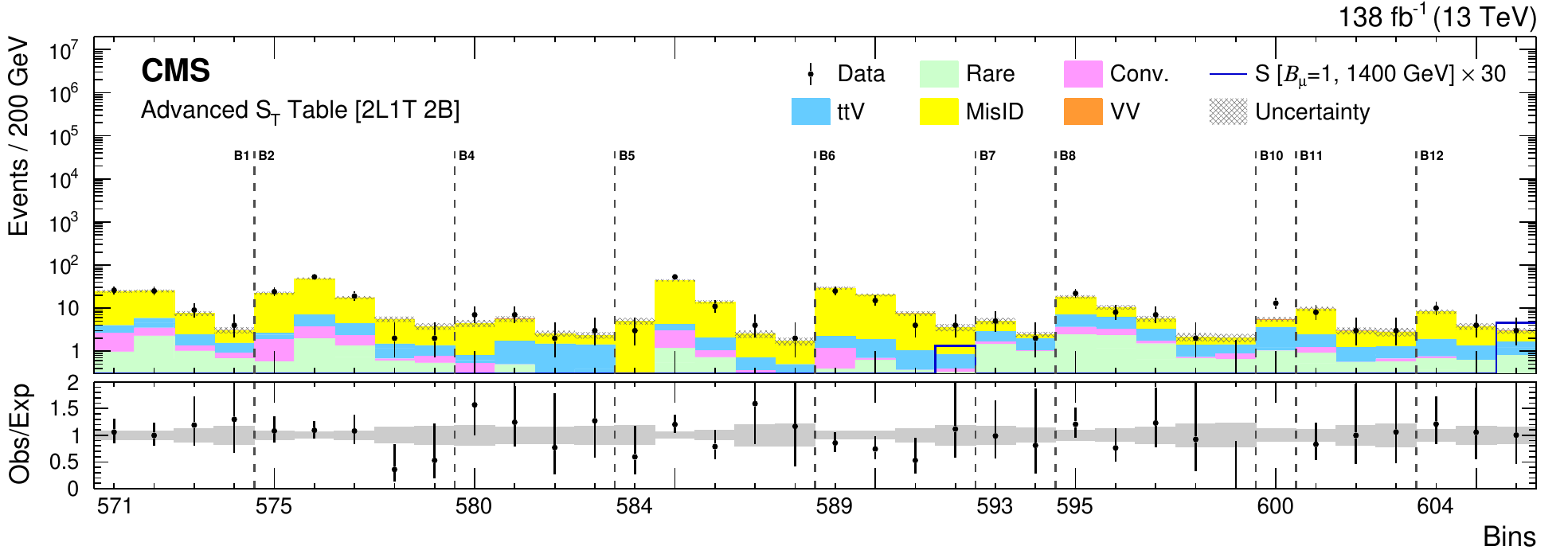}
\caption{\label{fig:resultAdvTable2L1T} The 2L1T SR distributions of the advanced \ST table for the combined 2016--2018 data set. The detailed description of the bin numbers can be found in Table~\ref{tab:binningtable2L1T}. The lower panel shows the ratio of observed events to the total expected background prediction. The gray band on the ratio represents the sum of statistical and systematic uncertainties in the SM background prediction. The expected SM background distributions and the uncertainties are shown after fitting the data under the background-only hypothesis.
For illustration, an example signal hypothesis for the production of the scalar leptoquark coupled to a top quark and a muon for $m_{\PS}=1.4\TeV$, before the fit, is also overlaid.
}
\end{figure*}

\begin{figure*}[hbt!]
\centering
\includegraphics[width=0.9\textwidth]{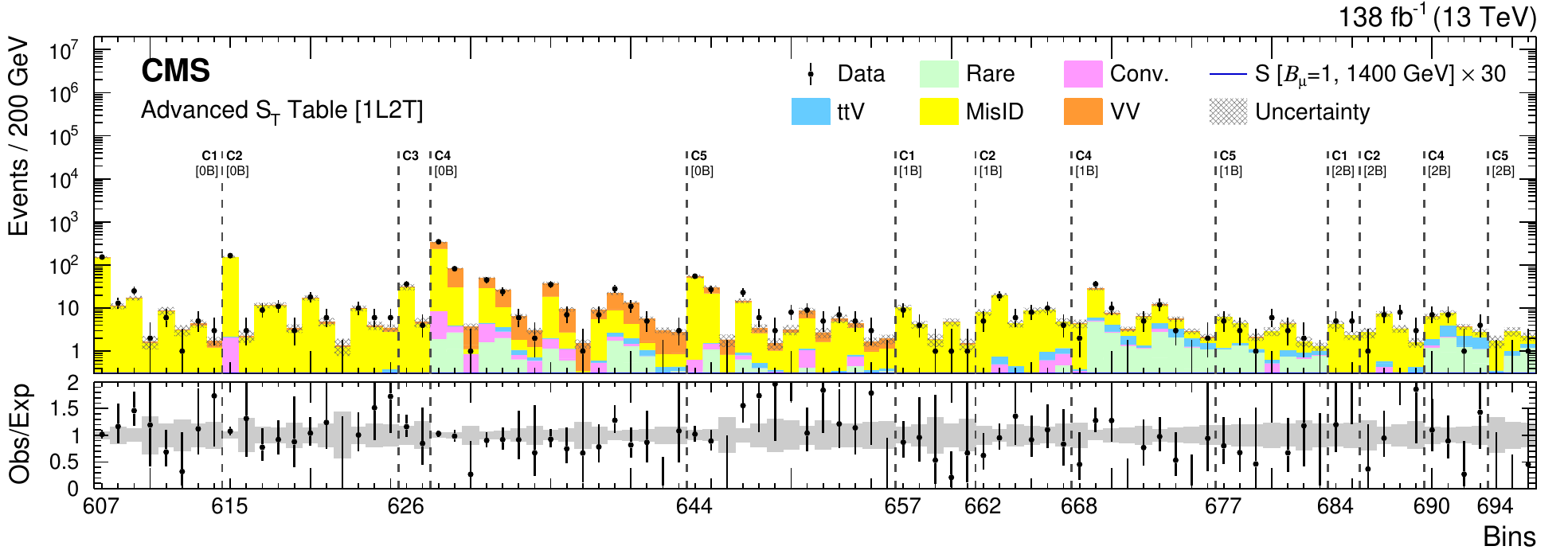}
\caption{\label{fig:resultAdvTable1L2T} The 1L2T SR distribution of the advanced \ST table for the combined 2016--2018 data set. The detailed description of the bin numbers can be found in Table~\ref{tab:binningtable1L2T}. The lower panel shows the ratio of observed events to the total expected background prediction. The gray band on the ratio represents the sum of statistical and systematic uncertainties in the SM background prediction. The expected SM background distributions and the uncertainties are shown after fitting the data under the background-only hypothesis.
An example signal hypothesis for the production of the scalar leptoquark coupled to a top quark and a muon for $m_{\PS}=1.4\TeV$, before the fit, is also overlaid. For this category, the signal yield is negligible and is not visible in the figure.
}
\end{figure*}

\begin{figure*}[hbt!]
\centering
\includegraphics[width=0.9\textwidth]{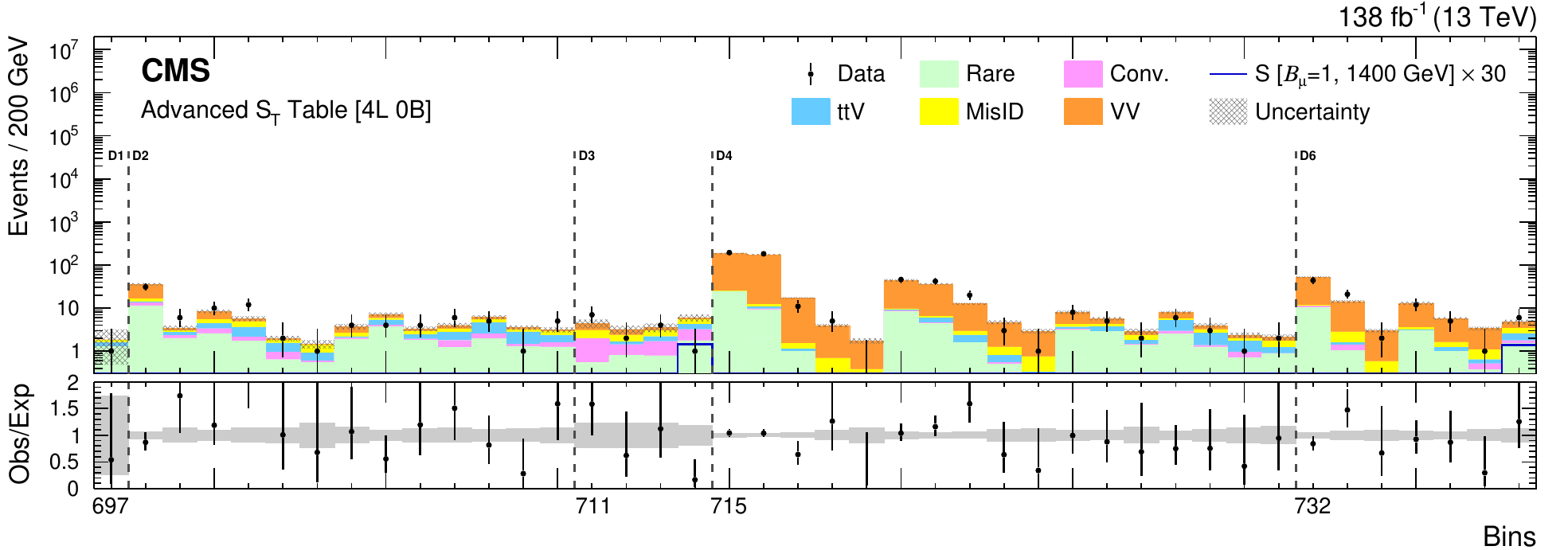}
\includegraphics[width=0.9\textwidth]{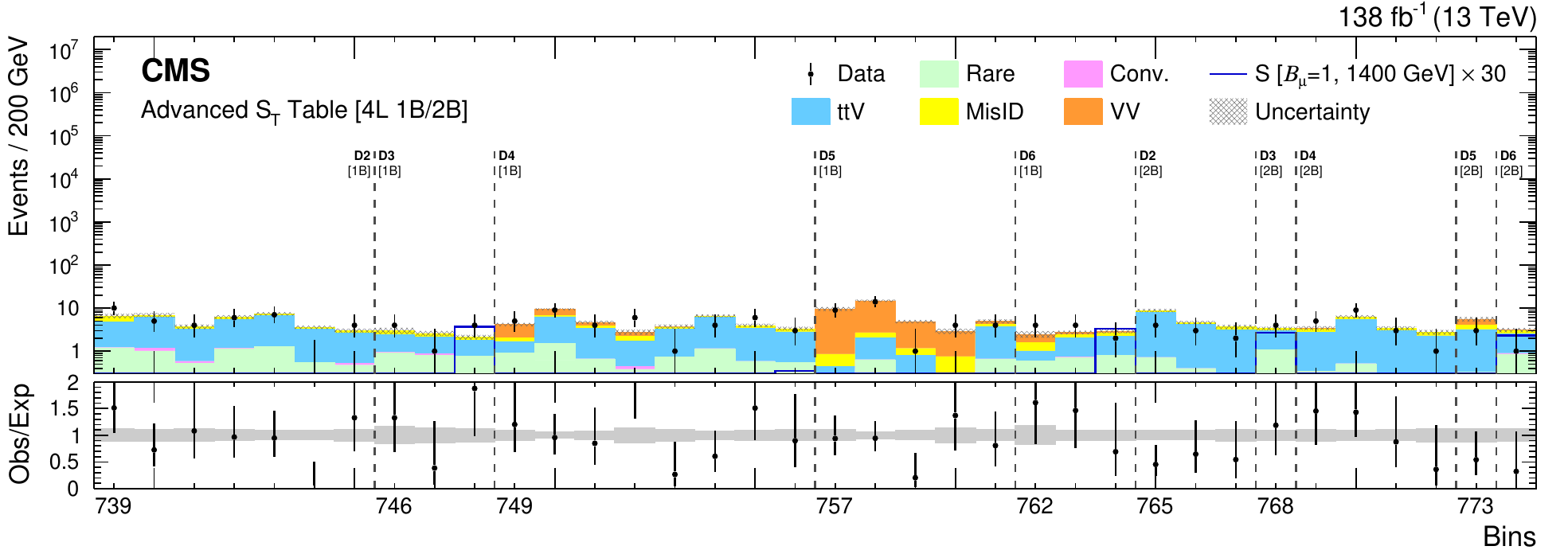}
\includegraphics[width=0.9\textwidth]{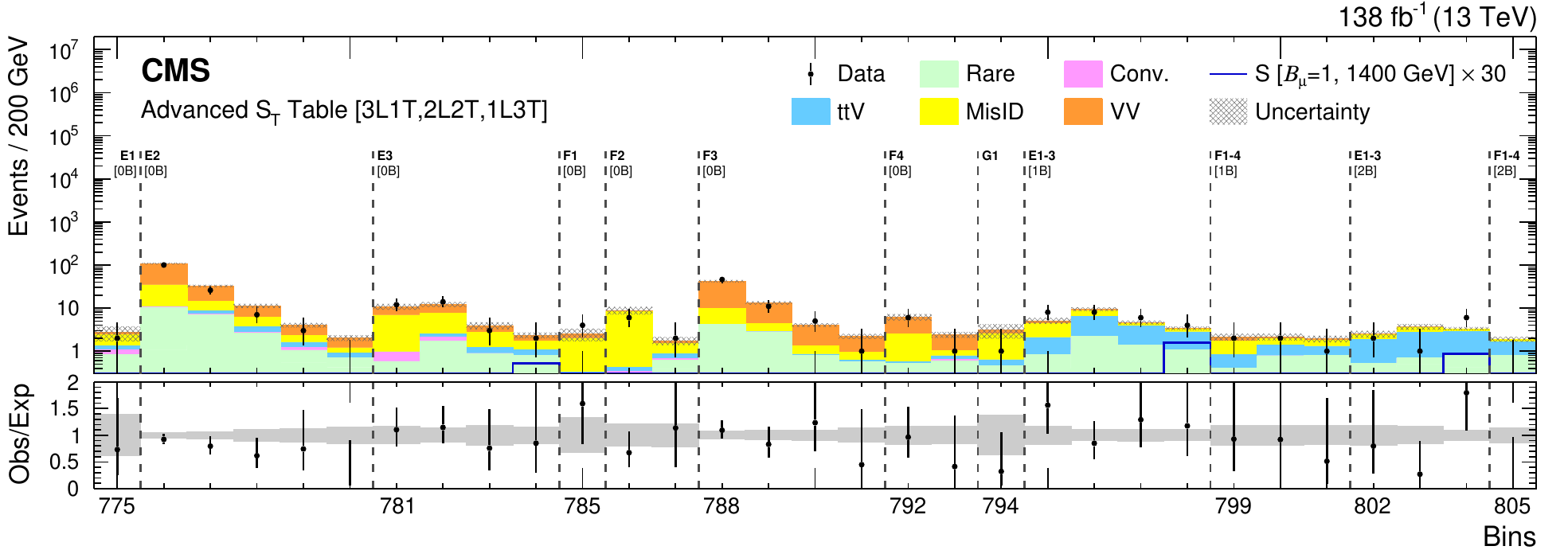}
\caption{\label{fig:resultAdvTableQuad} The 4L, 3L1T, 2L2T and 1L3T SR distributions of the advanced \ST table for the combined 2016--2018 data set. The detailed description of the bin numbers can be found in Table~\ref{tab:binningtableQuad}. The lower panel shows the ratio of observed events to the total expected background prediction. The gray band on the ratio represents the sum of statistical and systematic uncertainties in the SM background prediction. The expected SM background distributions and the uncertainties are shown after fitting the data under the background-only hypothesis.
For illustration, an example signal hypothesis for the production of the scalar leptoquark coupled to a top quark and a muon for $m_{\PS}=1.4\TeV$, before the fit, is also overlaid.
}
\end{figure*}

The $\LTmet$ distributions of the fundamental table scheme for all channels with the combined 2016--2018 data set are shown in Fig.~\ref{fig:resultLTMETTable},
where each histogram bin corresponds to an orthogonal $\LTmet$ bin in regions defined by the fundamental table scheme.
Similarly, the \ST distributions of the fundamental table scheme and the advanced table scheme are shown in Fig.~\ref{fig:resultSTTable},
and Figs.~\ref{fig:resultAdvTable3L}--\ref{fig:resultAdvTableQuad}, respectively.

\begin{figure*}[hbt!]
\centering
\includegraphics[width=0.9\textwidth]{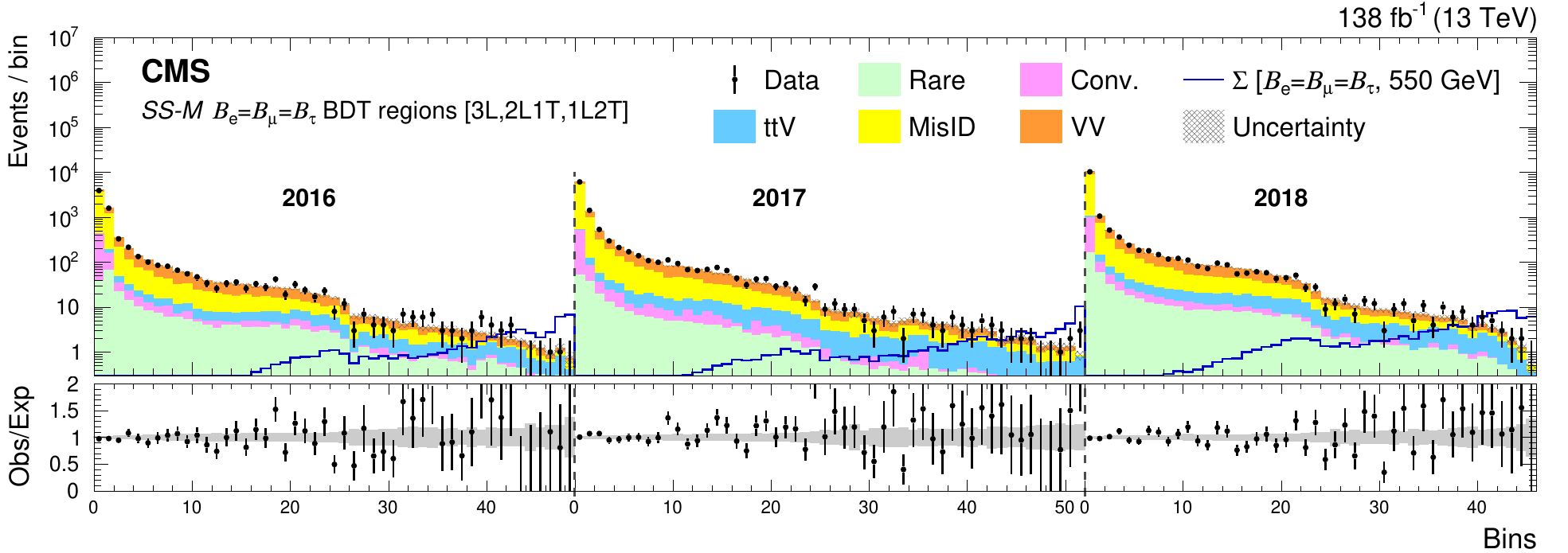}
\includegraphics[width=0.9\textwidth]{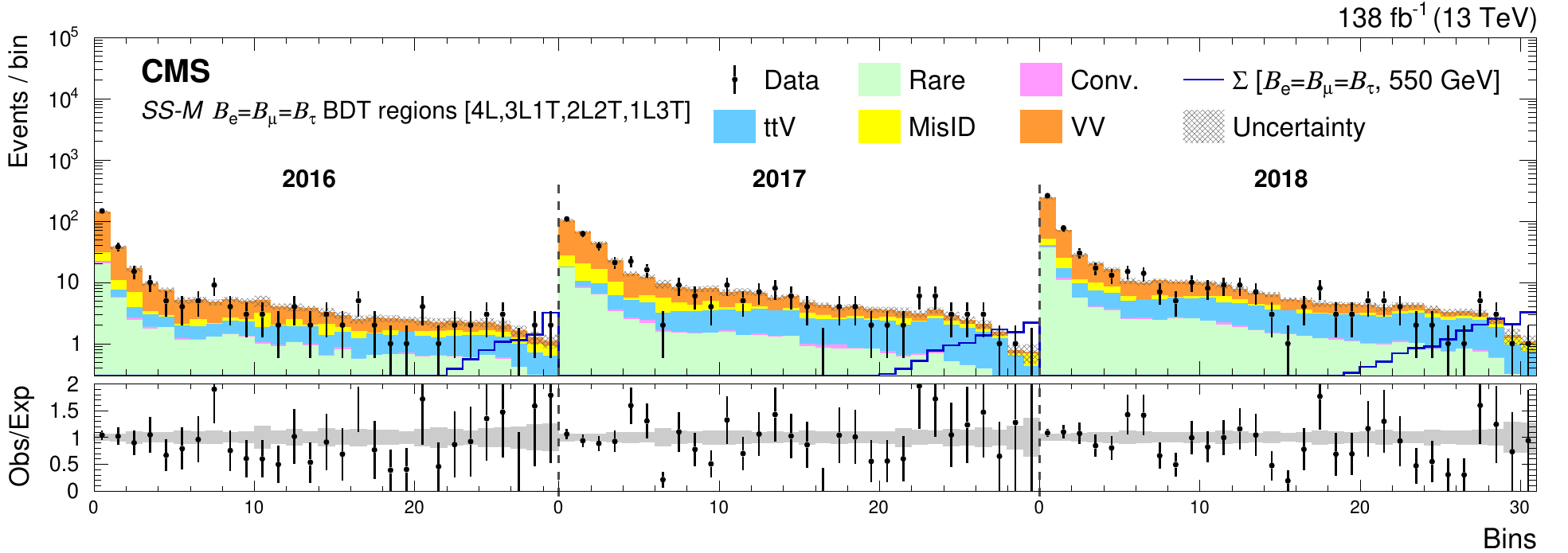}
\caption{\label{fig:resultSeesawfdMedMVAbins} The \textit{SS-M} $\mathcal{B}_{\Pe}=\mathcal{B}_{\PGm}=\mathcal{B}_{\PGt}$ BDT regions for the 3-lepton (upper) and 4-lepton (lower) channels for the 2016--2018 data sets. The lower panel shows the ratio of observed events to the total expected background prediction. The gray band on the ratio represents the sum of statistical and systematic uncertainties in the SM background prediction. The expected SM background distributions and the uncertainties are shown after fitting the data under the background-only hypothesis.
For illustration, an example signal hypothesis for the production of the type-III seesaw heavy fermions in the flavor-democratic scenario for $m_{\Sigma}=550\GeV$, before the fit, is also overlaid.
}
\end{figure*}

\begin{figure*}[hbt!]
\centering
\includegraphics[width=0.9\textwidth]{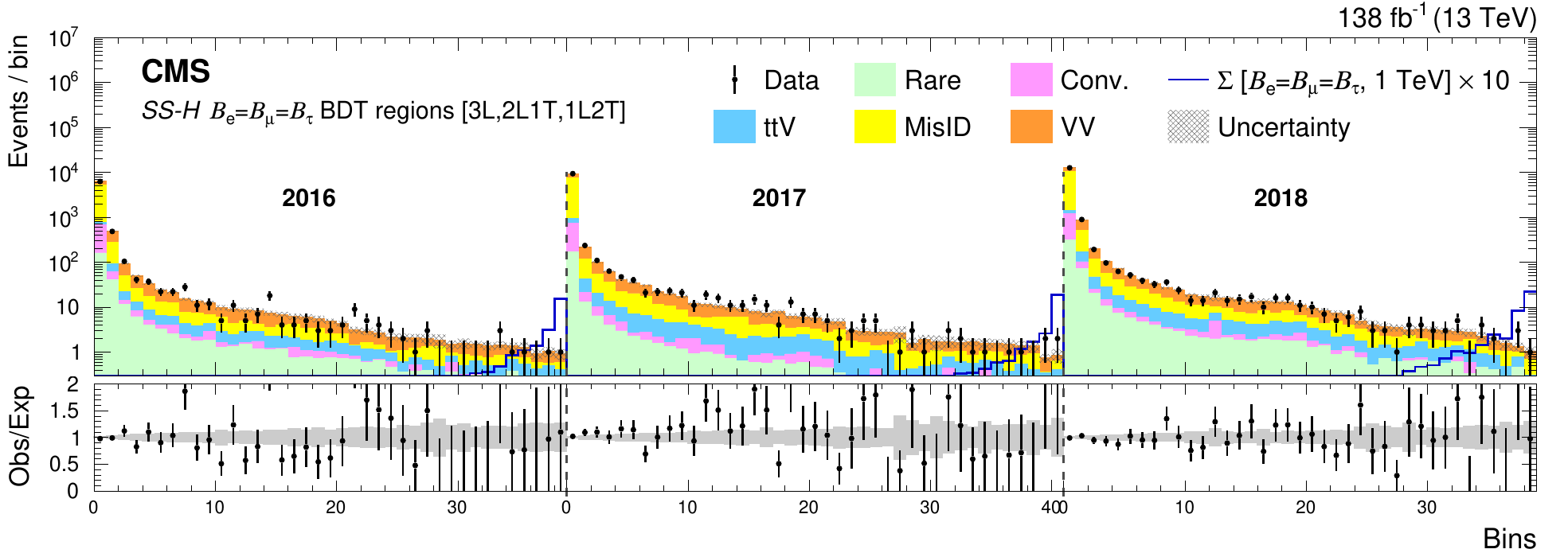}
\includegraphics[width=0.9\textwidth]{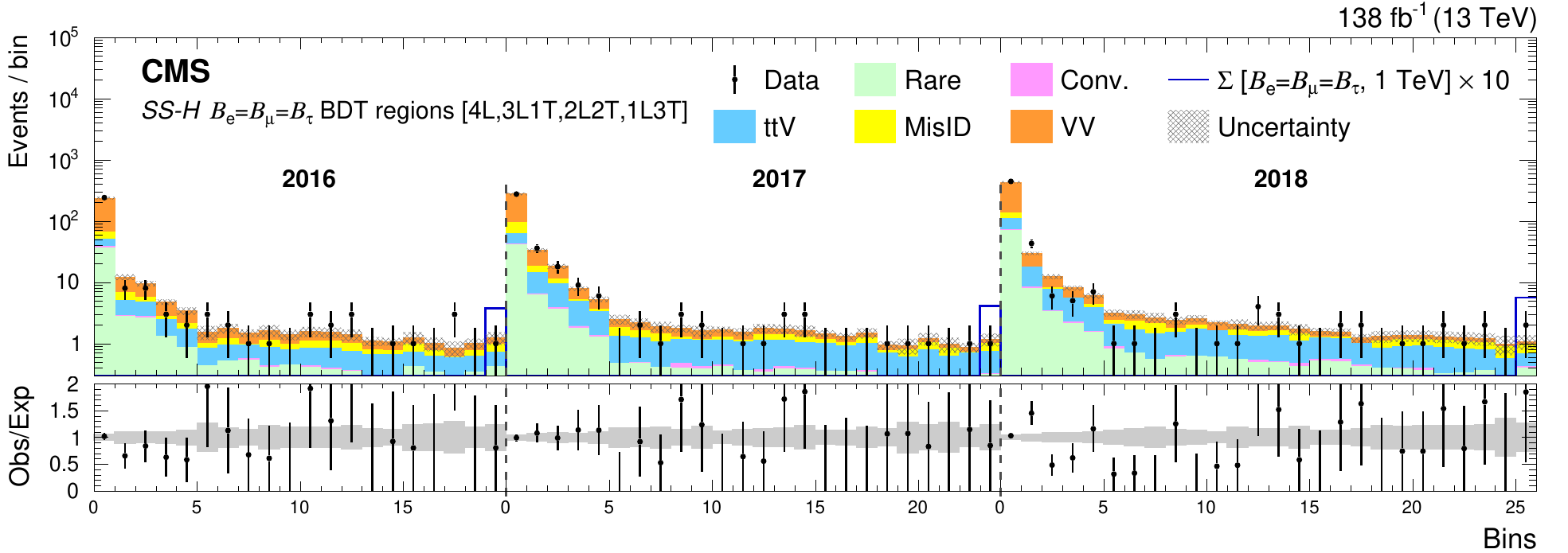}
\caption{\label{fig:resultSeesawfdHighMVAbins} The \textit{SS-H} $\mathcal{B}_{\Pe}=\mathcal{B}_{\PGm}=\mathcal{B}_{\PGt}$ BDT regions for the 3-lepton (upper) and 4-lepton (lower) channels for the 2016--2018 data sets. The lower panel shows the ratio of observed events to the total expected background prediction. The gray band on the ratio represents the sum of statistical and systematic uncertainties in the SM background prediction. The expected SM background distributions and the uncertainties are shown after fitting the data under the background-only hypothesis.
For illustration, an example signal hypothesis for the production of the type-III seesaw heavy fermions in the flavor-democratic scenario for $m_{\Sigma}=1\TeV$, before the fit, is also overlaid.
}
\end{figure*}

\begin{figure*}[hbt!]
\centering
\includegraphics[width=0.9\textwidth]{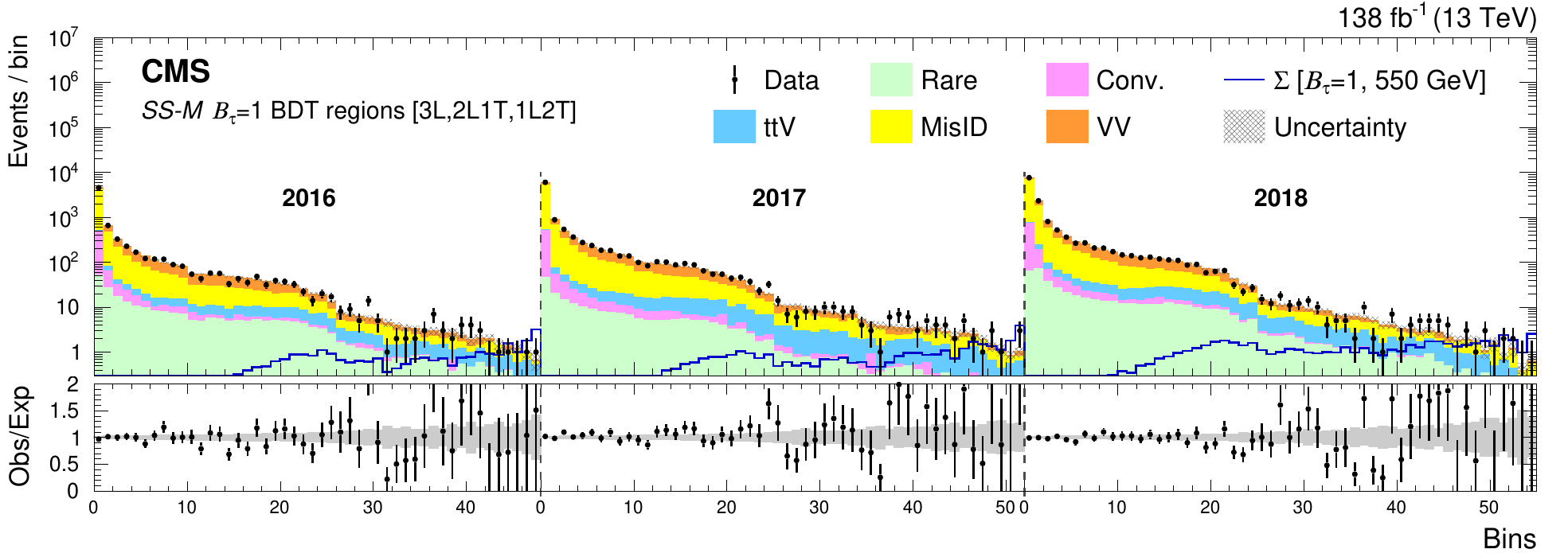}
\includegraphics[width=0.9\textwidth]{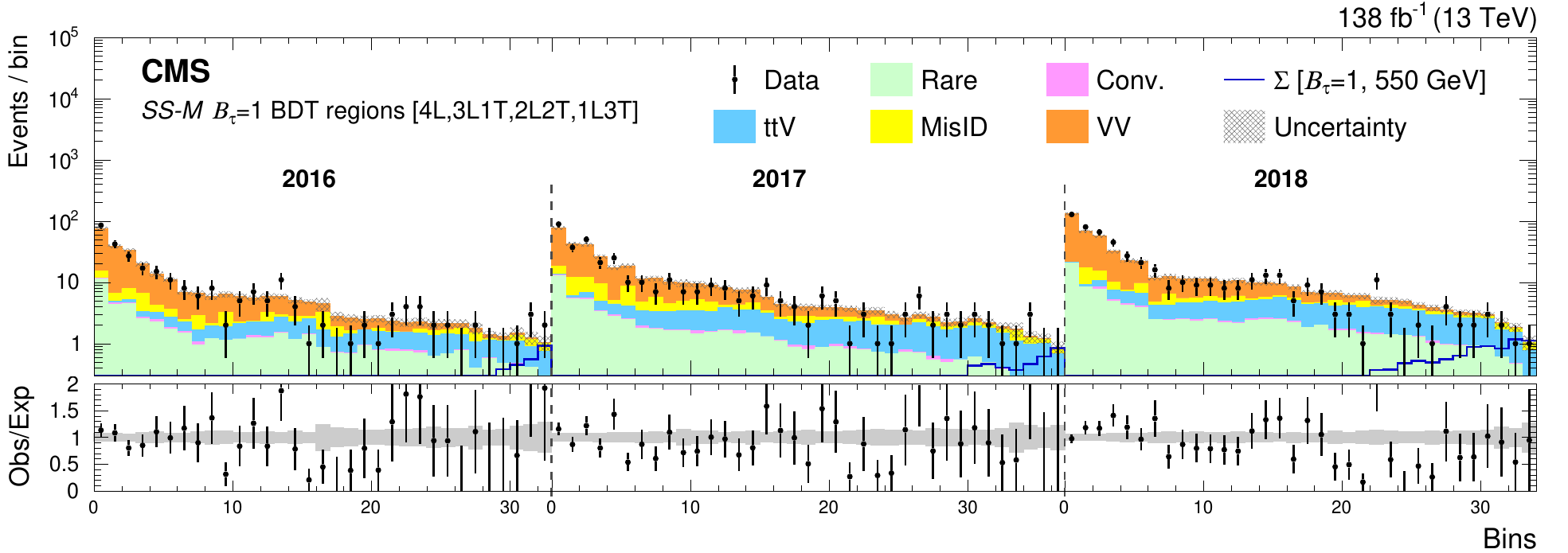}
\caption{\label{fig:resultSeesawtauMedMVAbins} The \textit{SS-M} $\mathcal{B}_{\PGt}=1$ BDT regions for the 3-lepton (upper) and 4-lepton (lower) channels for the 2016--2018 data sets. The lower panel shows the ratio of observed events to the total expected background prediction. The gray band on the ratio represents the sum of statistical and systematic uncertainties in the SM background prediction. The expected SM background distributions and the uncertainties are shown after fitting the data under the background-only hypothesis.
For illustration, an example signal hypothesis for the production of the type-III seesaw heavy fermions in the scenario with mixing exclusively to \PGt lepton for $m_{\Sigma}=550\GeV$, before the fit, is also overlaid.
}
\end{figure*}

\begin{figure*}[hbt!]
\centering
\includegraphics[width=0.9\textwidth]{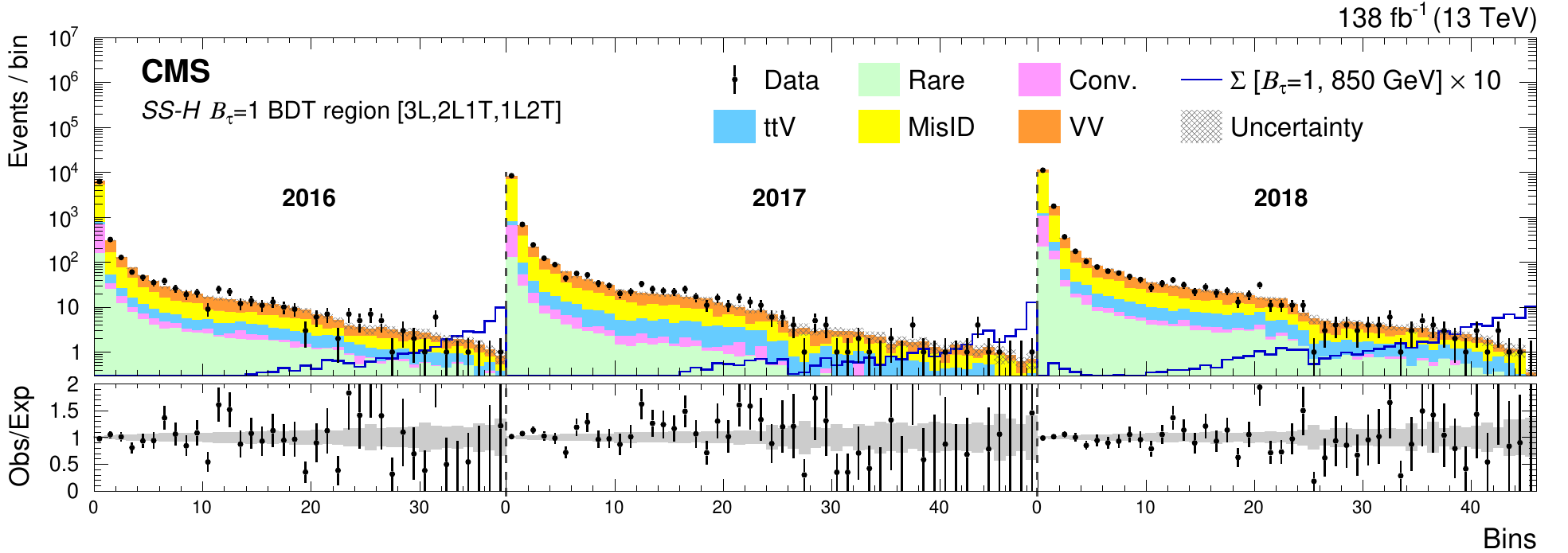}
\includegraphics[width=0.9\textwidth]{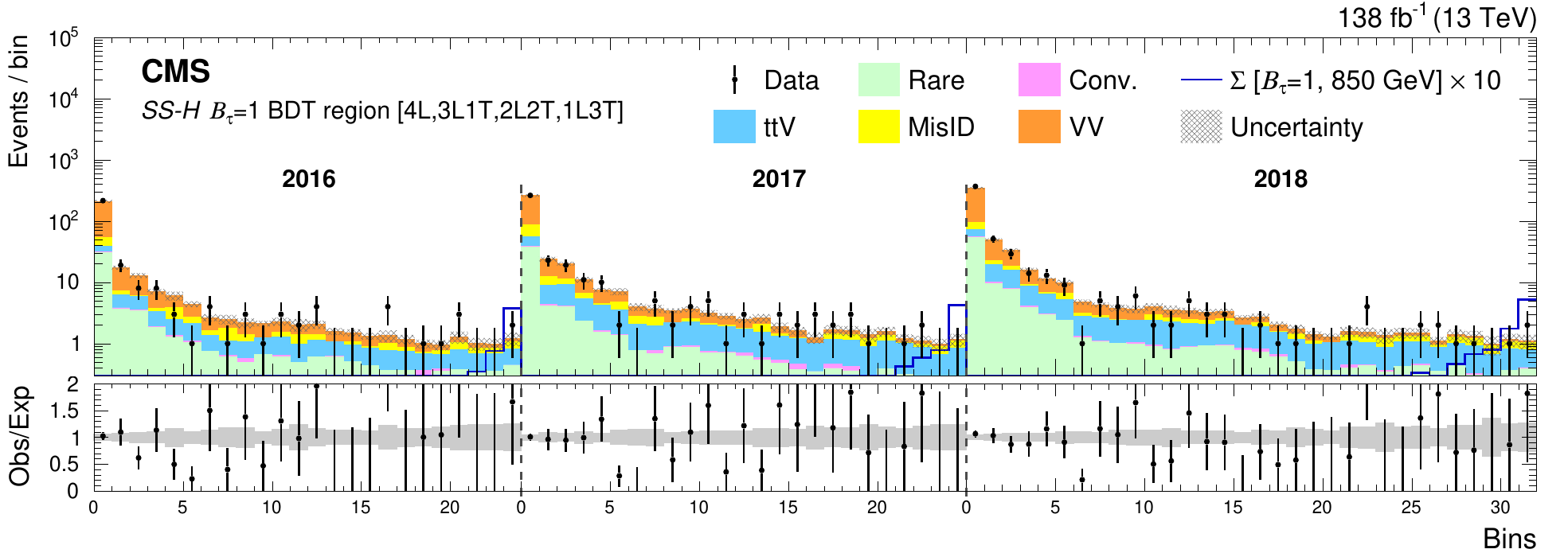}
\caption{\label{fig:resultSeesawtauHighMVAbins} The \textit{SS-H} $\mathcal{B}_{\PGt}=1$ BDT regions for the 3-lepton (upper) and 4-lepton (lower) channels for the 2016--2018 data sets. The lower panel shows the ratio of observed events to the total expected background prediction. The gray band on the ratio represents the sum of statistical and systematic uncertainties in the SM background prediction. The expected SM background distributions and the uncertainties are shown after fitting the data under the background-only hypothesis.
For illustration, an example signal hypothesis for the production of the type-III seesaw heavy fermions in the scenario with mixing exclusively to \PGt lepton for $m_{\Sigma}=850\GeV$, before the fit, is also overlaid.
}
\end{figure*}

\begin{figure*}[hbt!]
\centering
\includegraphics[width=0.9\textwidth]{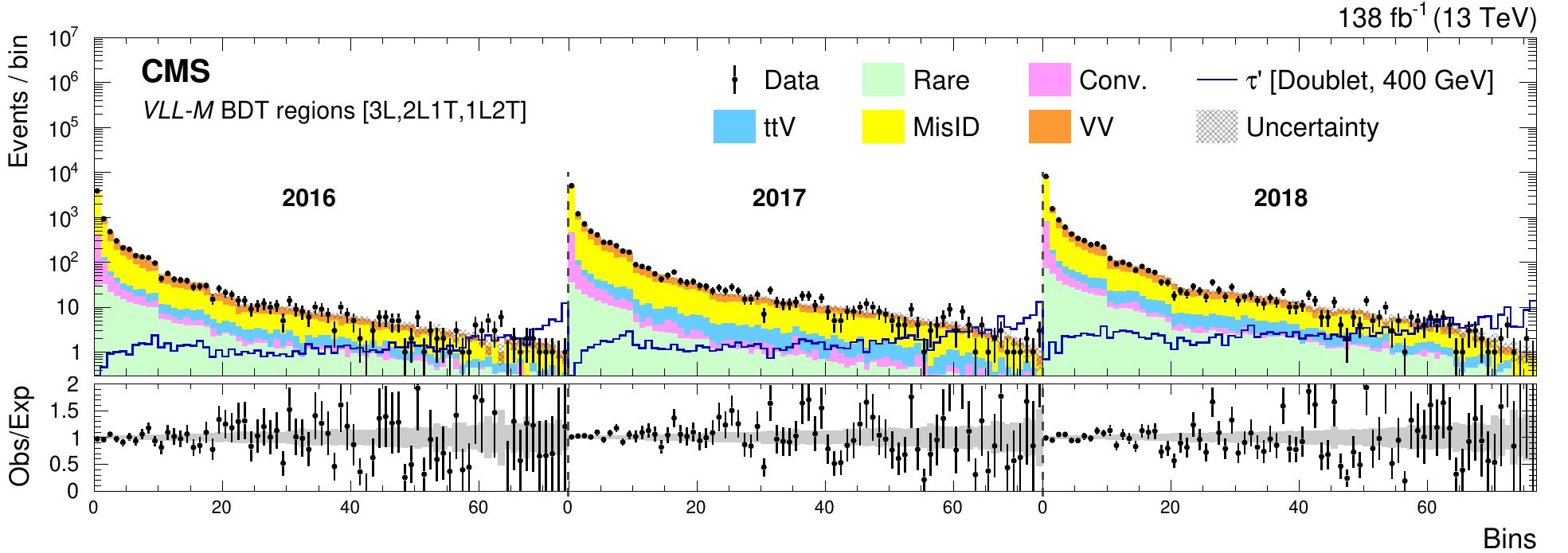}
\includegraphics[width=0.9\textwidth]{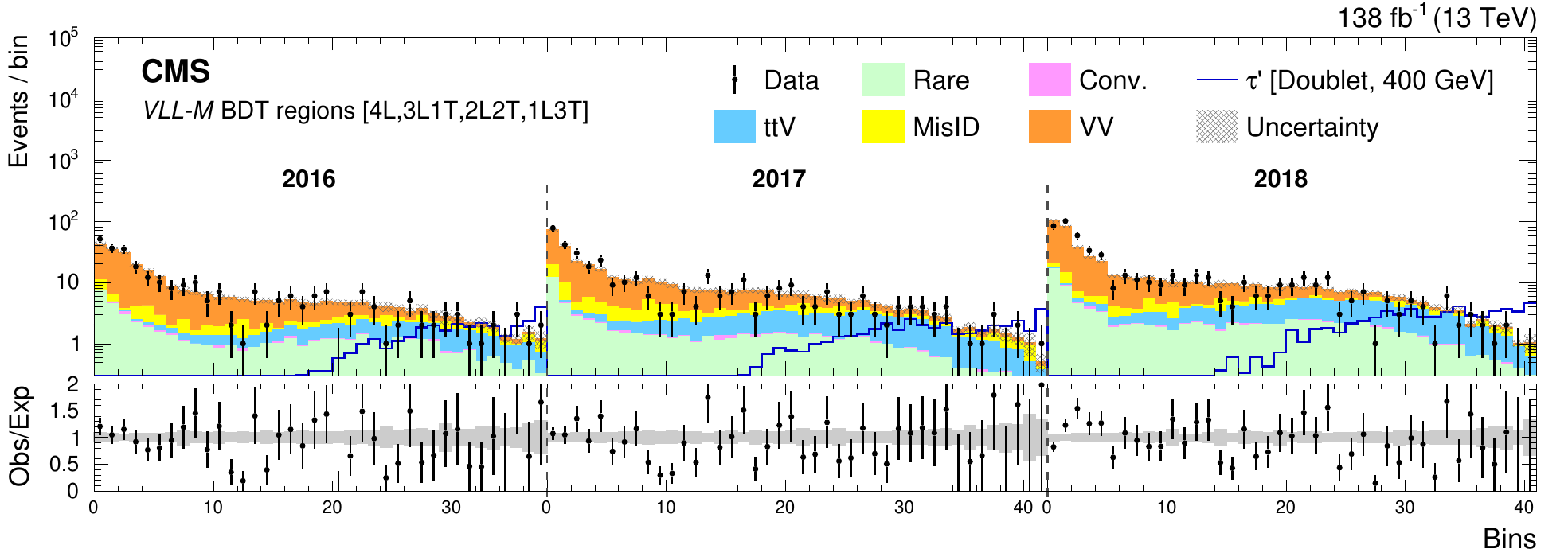}
\caption{\label{fig:resultVLLMedMVAbins} The \textit{VLL-M} BDT regions for the 3-lepton (upper) and 4-lepton (lower) channels for the 2016--2018 data sets. The lower panel shows the ratio of observed events to the total expected background prediction. The gray band on the ratio represents the sum of statistical and systematic uncertainties in the SM background prediction. The expected SM background distributions and the uncertainties are shown after fitting the data under the background-only hypothesis.
For illustration, an example signal hypothesis for the production of the vector-like \PGt lepton in the doublet scenario for $m_{\vltau}=400\GeV$, before the fit, is also overlaid.
}
\end{figure*}

\begin{figure*}[hbt!]
\centering
\includegraphics[width=0.9\textwidth]{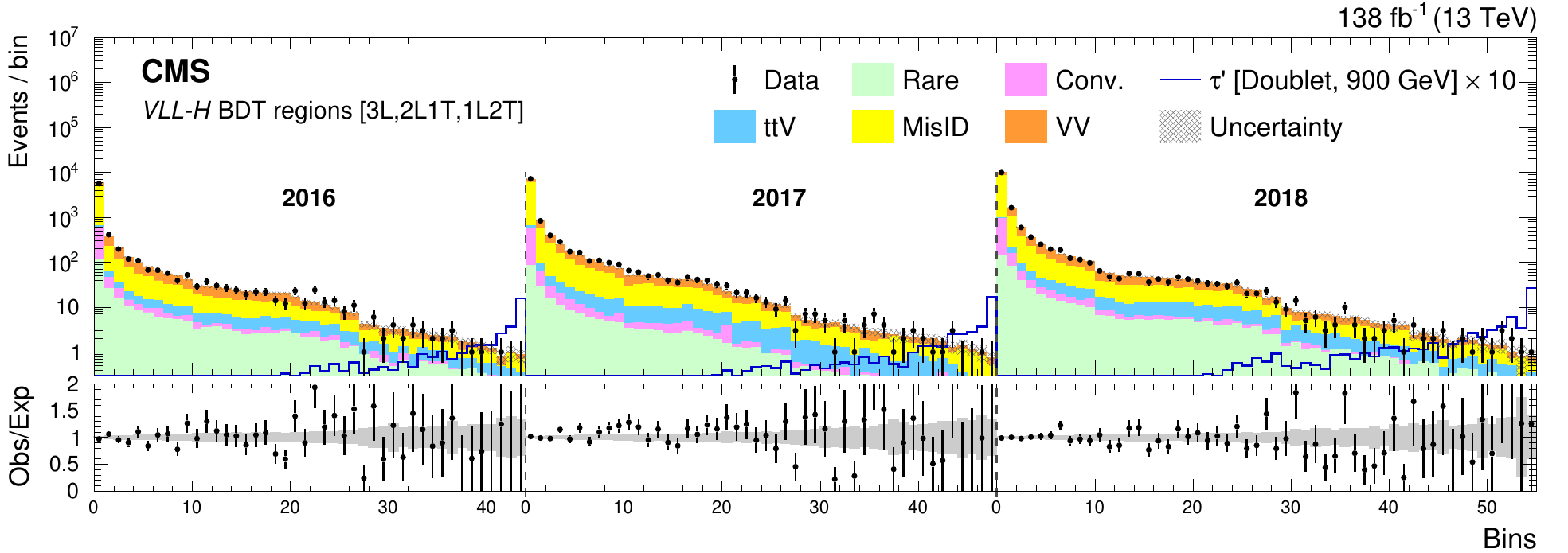}
\includegraphics[width=0.9\textwidth]{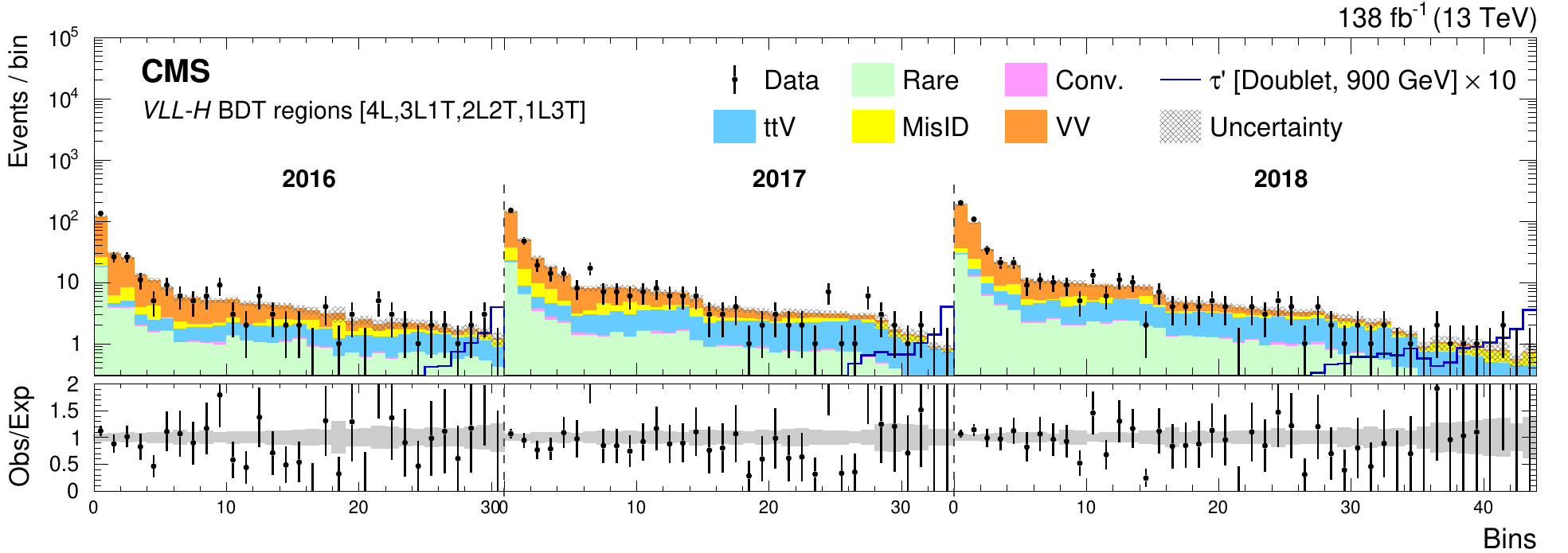}
\caption{\label{fig:resultVLLHighMVAbins} The \textit{VLL-H} BDT regions for the 3-lepton (upper) and 4-lepton (lower) channels for the 2016--2018 data sets. The lower panel shows the ratio of observed events to the total expected background prediction. The gray band on the ratio represents the sum of statistical and systematic uncertainties in the SM background prediction. The expected SM background distributions and the uncertainties are shown after fitting the data under the background-only hypothesis.
For illustration, an example signal hypothesis for the production of the vector-like \PGt lepton in the doublet scenario for $m_{\vltau}=900\GeV$, before the fit, is also overlaid.
}
\end{figure*}

\begin{figure*}[hbt!]
\centering
\includegraphics[width=0.9\textwidth]{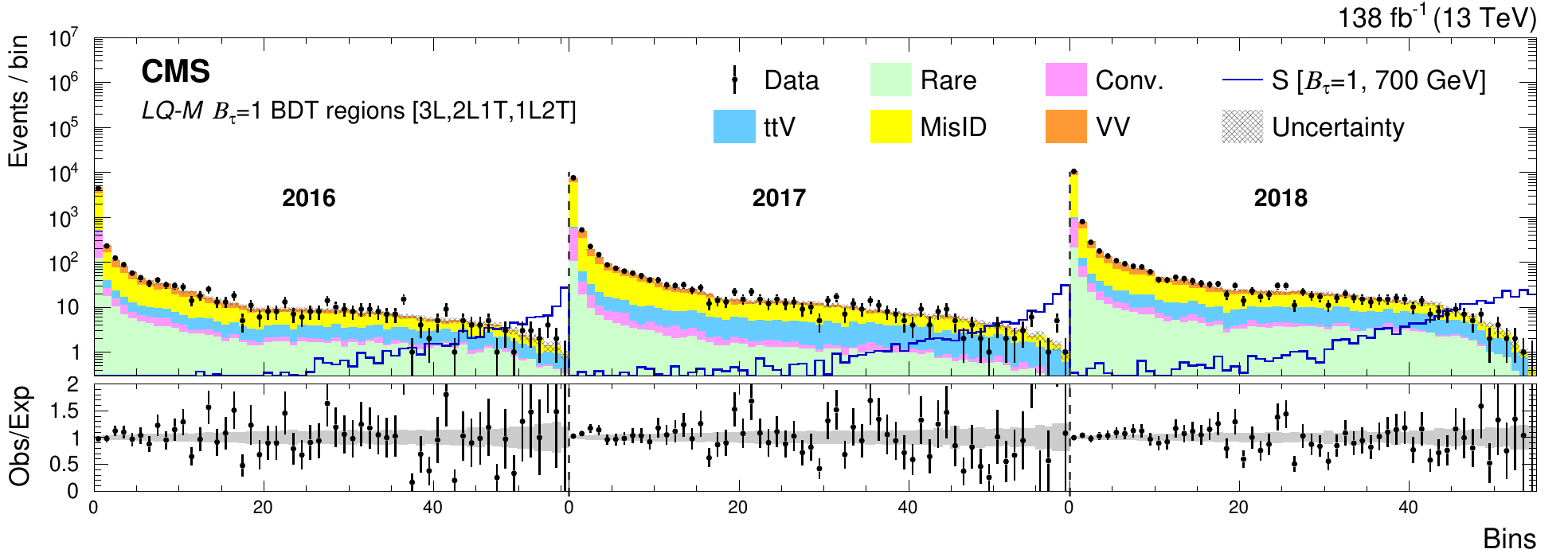}
\includegraphics[width=0.9\textwidth]{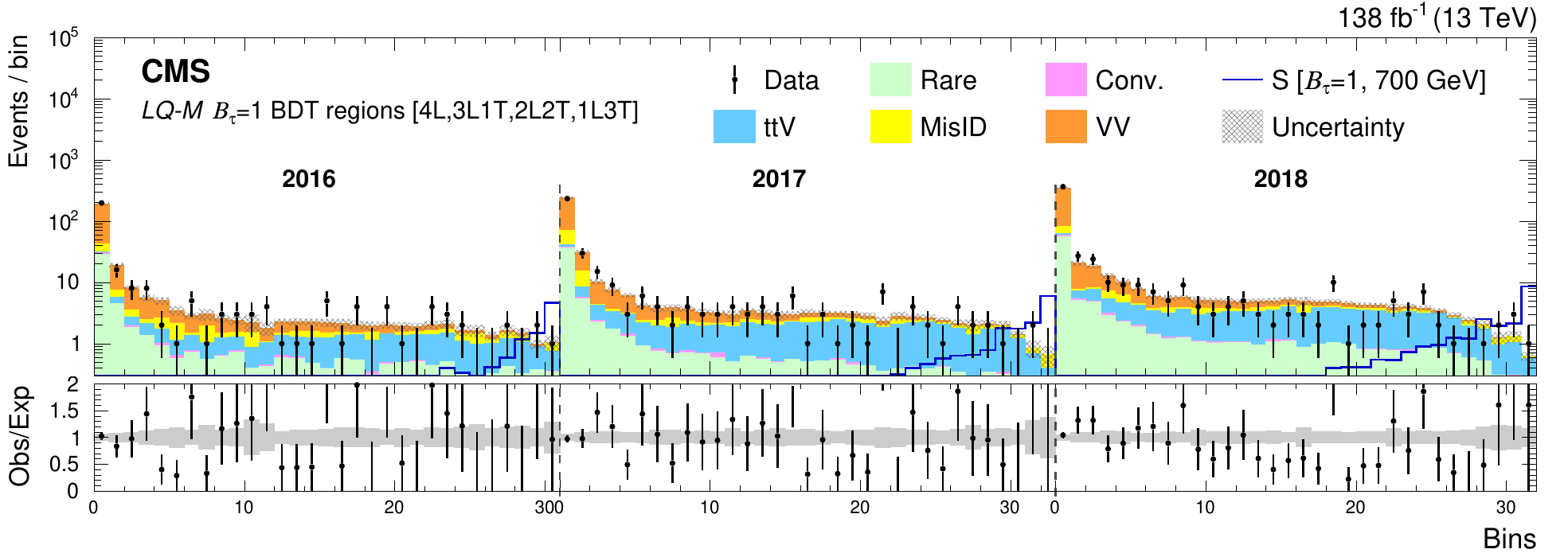}
\caption{\label{fig:resultLQtauMedMVAbins} The \textit{LQ-M} $\mathcal{B}_{\PGt}=1$ BDT regions for the 3-lepton (upper) and 4-lepton (lower) channels for the 2016--2018 data sets. The lower panel shows the ratio of observed events to the total expected background prediction. The gray band on the ratio represents the sum of statistical and systematic uncertainties in the SM background prediction. The expected SM background distributions and the uncertainties are shown after fitting the data under the background-only hypothesis.
For illustration, an example signal hypothesis for the production of the scalar leptoquark coupled to a top quark and a \PGt lepton for $m_{\PS}=700\GeV$, before the fit, is also overlaid.
}
\end{figure*}

\begin{figure*}[hbt!]
\centering
\includegraphics[width=0.9\textwidth]{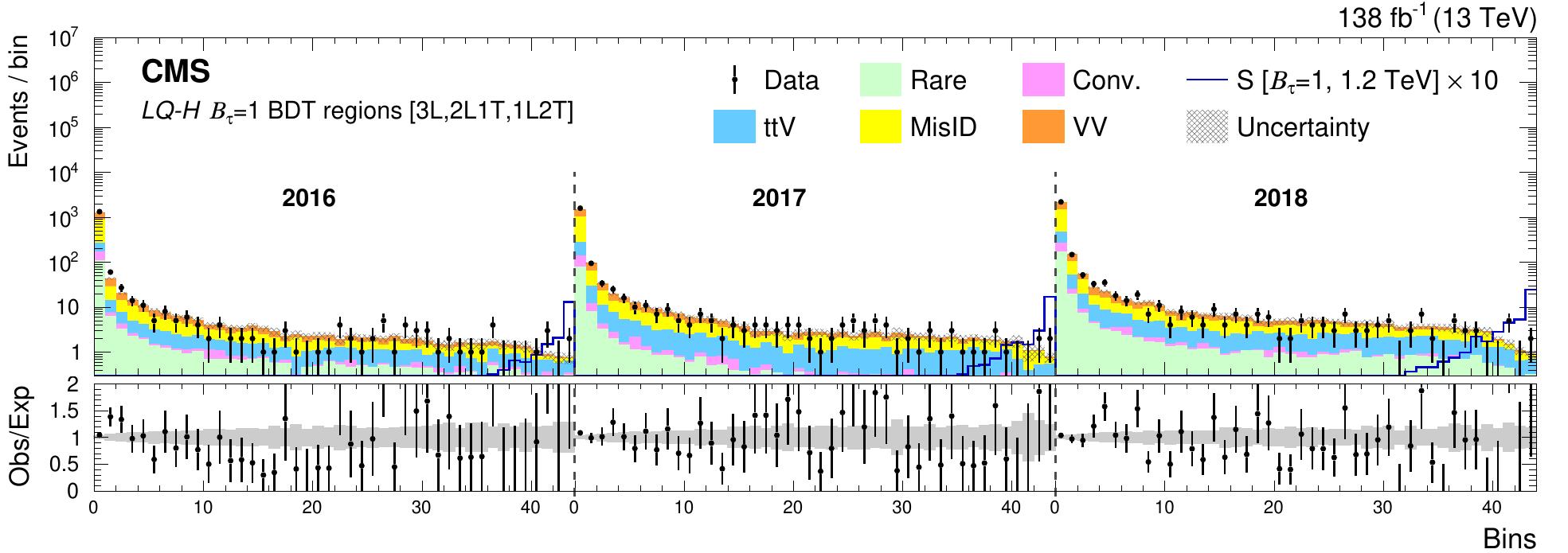}
\includegraphics[width=0.9\textwidth]{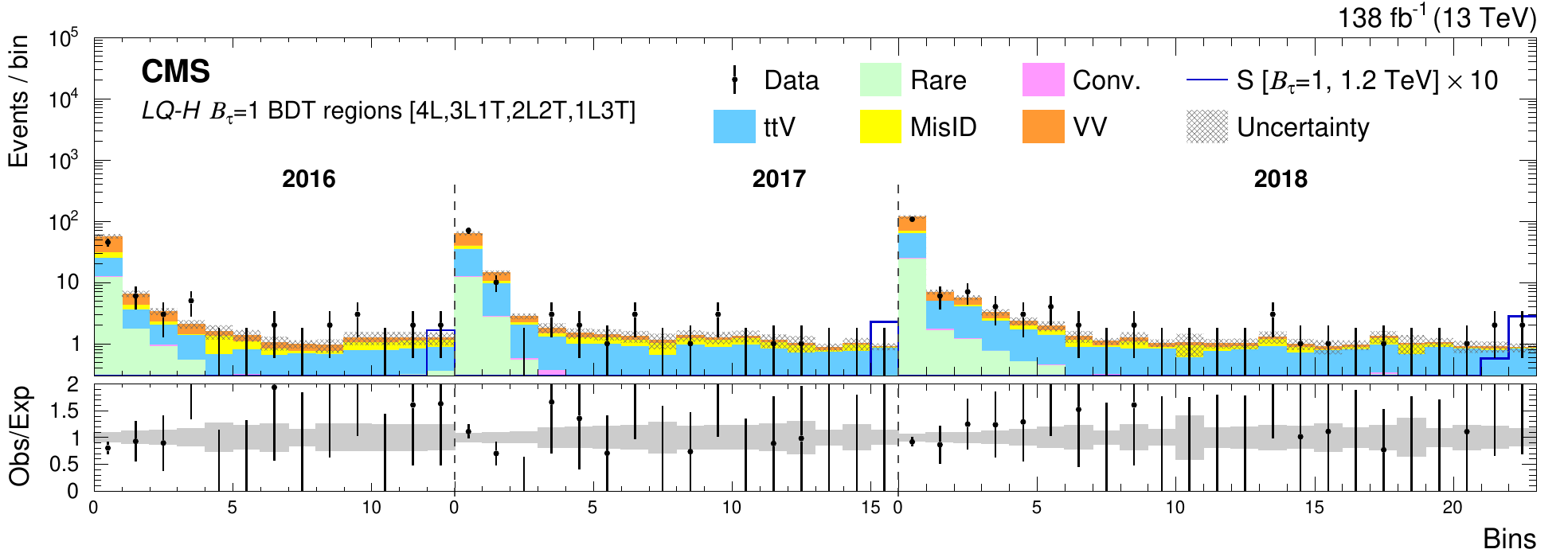}
\caption{\label{fig:resultLQtauHighMVAbins} The \textit{LQ-H} $\mathcal{B}_{\PGt}=1$ BDT regions for the 3-lepton (upper) and 4-lepton (lower) channels for the 2016--2018 data sets. The lower panel shows the ratio of observed events to the total expected background prediction. The gray band on the ratio represents the sum of statistical and systematic uncertainties in the SM background prediction. The expected SM background distributions and the uncertainties are shown after fitting the data under the background-only hypothesis.
For illustration, an example signal hypothesis for the production of the scalar leptoquark coupled to a top quark and a \PGt lepton for $m_{\PS}=1.2\TeV$, before the fit, is also overlaid.
}
\end{figure*}

\begin{figure*}[hbt!]
\centering
\includegraphics[width=0.9\textwidth]{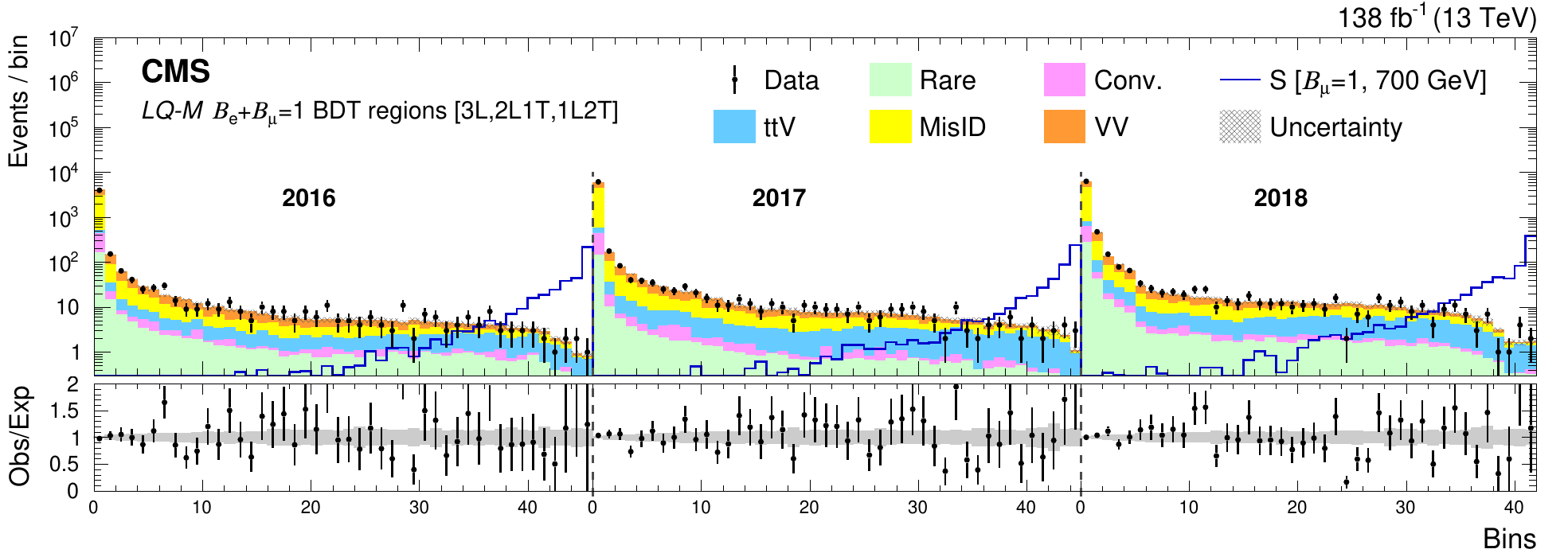}
\includegraphics[width=0.9\textwidth]{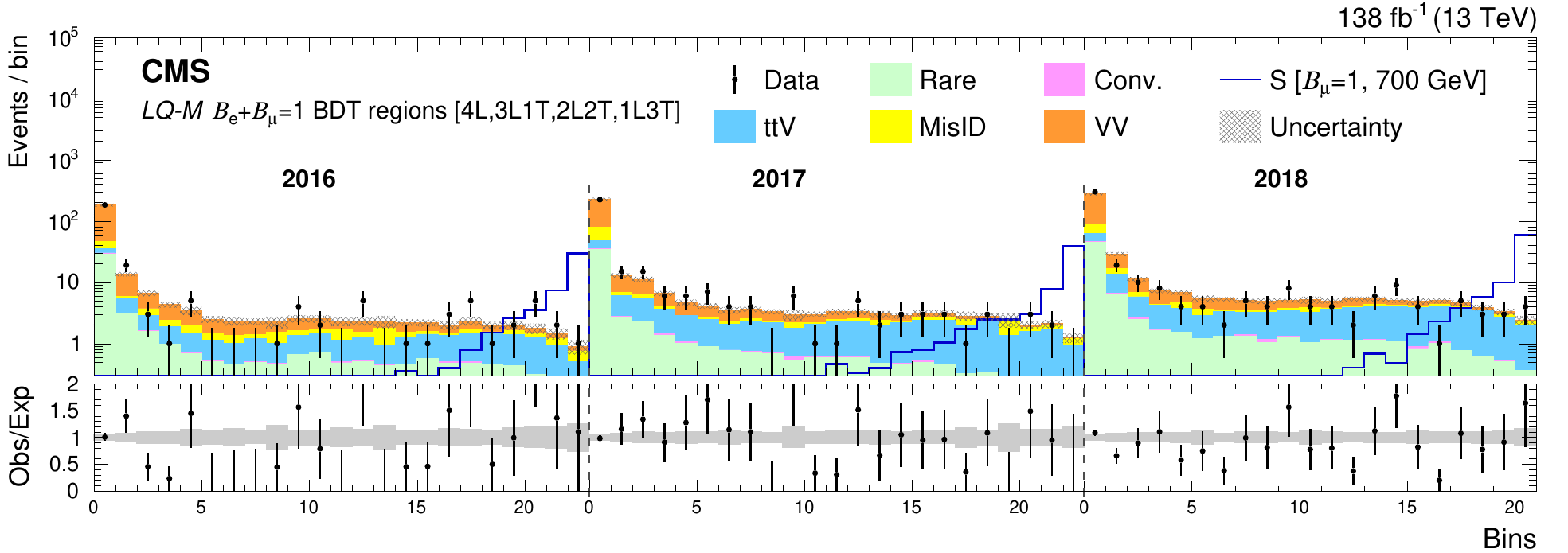}
\caption{\label{fig:resultLQllMedMVAbins} The \textit{LQ-M} $\mathcal{B}_{\Pe}+\mathcal{B}_{\PGm}=1$ BDT regions for the 3-lepton (upper) and 4-lepton (lower) channels for the 2016--2018 data sets. The lower panel shows the ratio of observed events to the total expected background prediction. The gray band on the ratio represents the sum of statistical and systematic uncertainties in the SM background prediction. The expected SM background distributions and the uncertainties are shown after fitting the data under the background-only hypothesis.
For illustration, an example signal hypothesis for the production of the scalar leptoquark coupled to a top quark and a muon for $m_{\PS}=700\GeV$, before the fit, is also overlaid.
}
\end{figure*}

\begin{figure*}[hbt!]
\centering
\includegraphics[width=0.9\textwidth]{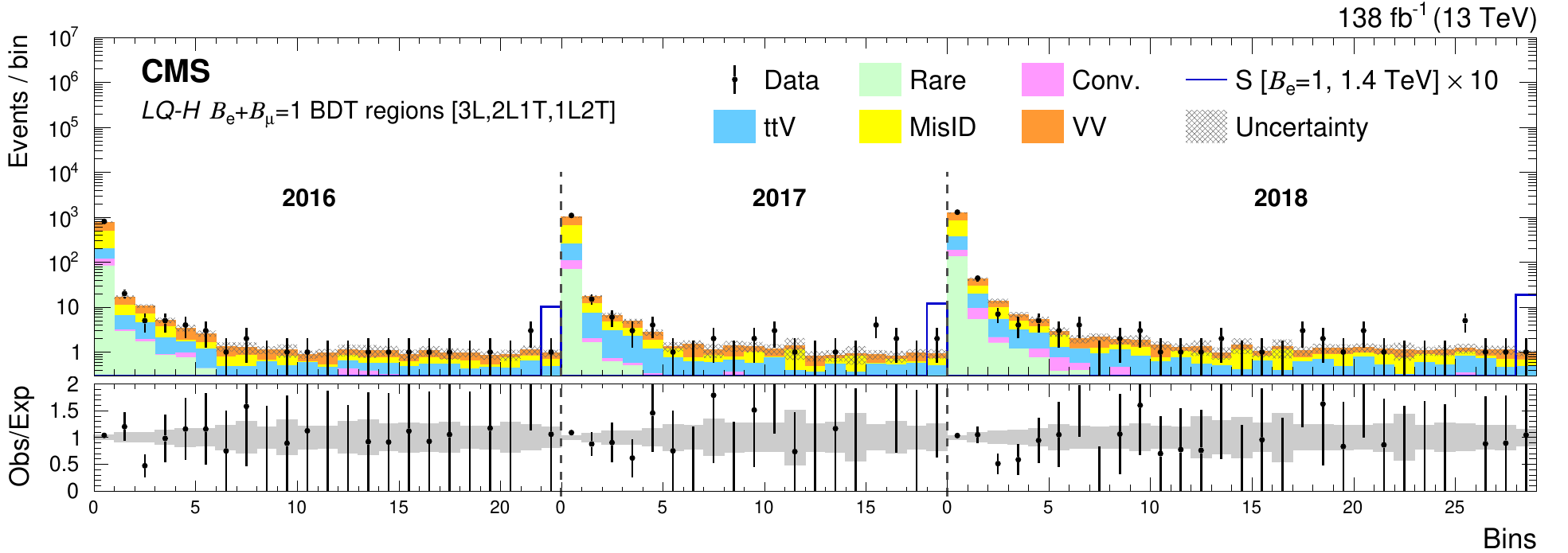}
\includegraphics[width=0.9\textwidth]{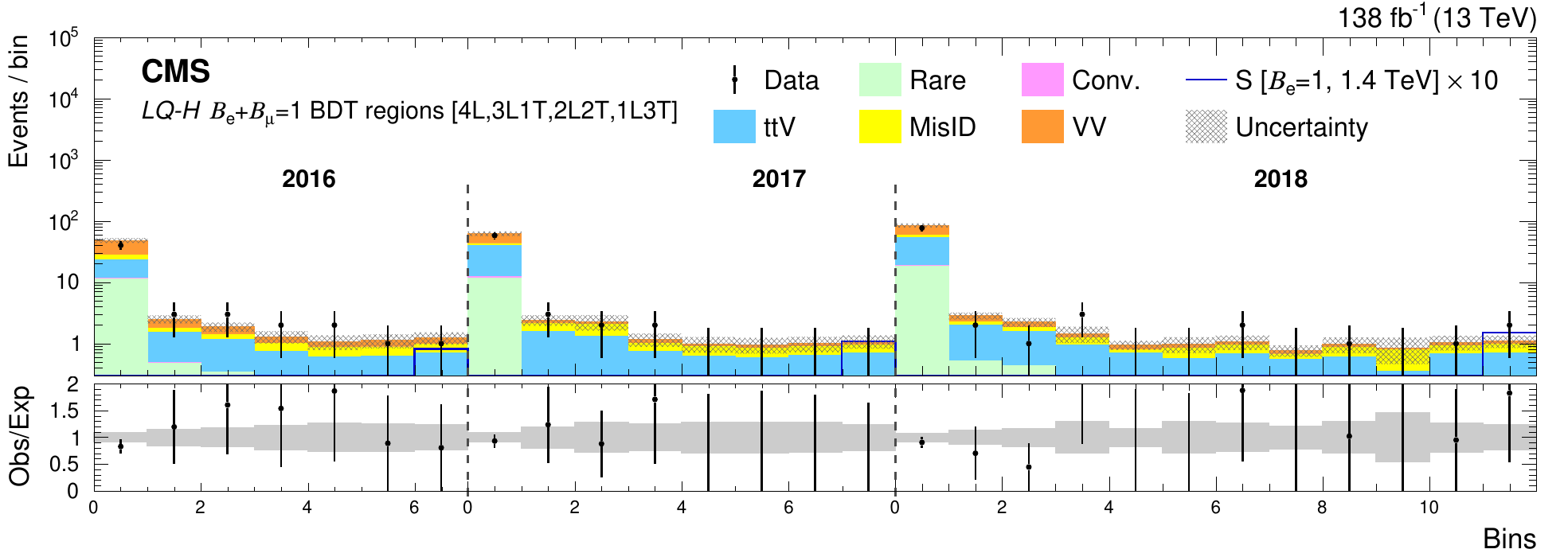}
\caption{\label{fig:resultLQllHighMVAbins} The \textit{LQ-H} $\mathcal{B}_{\Pe}+\mathcal{B}_{\PGm}=1$ BDT regions for the 3-lepton (upper) and 4-lepton (lower) channels for the 2016--2018 data sets. The lower panel shows the ratio of observed events to the total expected background prediction. The gray band on the ratio represents the sum of statistical and systematic uncertainties in the SM background prediction. The expected SM background distributions and the uncertainties are shown after fitting the data under the background-only hypothesis.
For illustration, an example signal hypothesis for the production of the scalar leptoquark coupled to a top quark and an electron for $m_{\PS}=1.4\TeV$, before the fit, is also overlaid.
}
\end{figure*}

The BDT region distributions for the two highest mass BDTs for the 3-lepton channels (3L, 2L1T, 1L2T) and the 4-lepton channels (4L, 3L1T, 2L2T, 1L3T) with the combined 2016--2018 data set for the type-III seesaw model with $\mathcal{B}_{\Pe}=\mathcal{B}_{\PGm}=\mathcal{B}_{\PGt}$ and $\mathcal{B}_{\PGt}=1$ couplings are shown in Figs.~\ref{fig:resultSeesawfdMedMVAbins}--\ref{fig:resultSeesawfdHighMVAbins} and~\ref{fig:resultSeesawtauMedMVAbins}--\ref{fig:resultSeesawtauHighMVAbins}, respectively. 
Similarly, distributions for the doublet vector-like lepton, leptoquarks with $\mathcal{B}_{\PGt}=1$ couplings, and leptoquarks with $\mathcal{B}_{\Pe}+\mathcal{B}_{\PGm}=1$ couplings are shown in Figs.~\ref{fig:resultVLLMedMVAbins}--\ref{fig:resultVLLHighMVAbins},~\ref{fig:resultLQtauMedMVAbins}--\ref{fig:resultLQtauHighMVAbins}, and~\ref{fig:resultLQllMedMVAbins}--\ref{fig:resultLQllHighMVAbins}, respectively.

Across all the channels considered, we find agreement between the data and the predictions of the SM background.
No significant excess of data consistent with the models we probe is observed. 
We perform goodness-of-fit tests in each of the model-independent SR table schemes based on the saturated model method~\cite{Baker:1983tu} to quantify the deviations between the background-only hypothesis and the observed data.
The fundamental $\LTmet$, \ST, and the advanced \ST schemes have global p-values~\cite{pvalue} of 0.67, 0.53, and 0.11, respectively, consistent with the background-only hypotheses.

Apart from the global agreement, we report the most significant local deviations as seen in the three model-independent schemes in the combined 2016--2018 data set, without considering the look-elsewhere effect~\cite{Gross:2010qma}. 
The largest local excess in the fundamental $\LTmet$ table is found in the bin 16 (3L, A4 category, OSSF1 BelowZ, low minimum lepton \pt, $300<\LTmet<500\GeV$), resulting in a data excess of 2.7 standard deviations.  We also observe a local deficit of approximately 1.7 standard deviations in the bin 143 (3L1T, E3 category, OSSF1 OnZ, $500<\LTmet<700\GeV$). 
For the fundamental \ST table, we observe the largest local excess of around 2.3 standard deviations in the bin 78 (3L, A10 category, OSSF1 BelowZ, high minimum lepton \pt, $1400<\ST<1600\GeV$) in the combined 2016--2018 data set, and similarly we also have a local deficit of 1.9 standard deviations in the bin 236 (3L1T, E2 category, OSSF1 OnZ, $650<\ST<850\GeV$). 
Finally, the largest local excess in the advanced \ST table is found in the bin 600 (2L1T, B10 category, OSSF1 OnZ, high \tauh \pt, 2\PB, Inclusive) resulting in an excess of 2.5 standard deviations, and the largest local deficits are observed in bins 40 (3L, A4 category, OSSF1 BelowZ, low minimum lepton \pt, 0\PB, low \ptmiss, high $\HT$, $350<\ST<550\GeV$) and 714 (4L, D3 category, OSSF1 OffZ, 0\PB, high \ptmiss, Inclusive) with 2.0 standard deviations each. 
These extreme local deviations are generally driven by excesses or deficits in a single year of data-taking, and are not found to be consistent across all three data-taking periods. 
All other local deviations are measured to be less significant.

The three model-independent SR table schemes are used separately to calculate the upper limits on the production cross section for the three BSM models considered here. 
For each of the three separate table schemes, the corresponding bins from Tables~\ref{tab:binningtable3L}--\ref{tab:binningtableQuad} are treated as counting experiments in each data-taking period, and are fitted simultaneously with bins for each of the three years of data collection in the statistical analysis. 
Similarly, the BDT regions corresponding to the 3- and 4-lepton channels for a specific BDT variable (depending on the model, mass-range, and flavor scenario) are treated as counting experiments in each data-taking period, and are fitted simultaneously for each of the three years of data collection.

To calculate the upper limits, we use a modified frequentist approach with the \CLs~\cite{Junk:1999kv,Read:2002hq,Cowan:2010js,ATLAS:2011tau} criterion, with a test statistic based on the binned profile likelihood, in the asymptotic approximation. 
The upper limits are calculated at 95\% \CL. 
The systematic uncertainties and their correlations as
described in Section~\ref{sec:systematics} are incorporated in the likelihood as nuisance parameters with log-normal probability density
functions. 
The statistical uncertainties in the signal and background estimates are modeled with gamma functions. 
Finally, we present the cross section limit at
a particular mass point from the table scheme or the BDT training that gives the best expected limit. 
The details for each model are discussed in the subsequent paragraphs.

As a general remark, the model-independent advanced \ST table schemes are more sensitive than the lowest-mass BDT training process for all the models.
This is because at low signal masses, the training process is degraded by the low signal yield and the similar kinematic properties of signal and SM processes.

\begin{figure}[hbt!]
\centering
\includegraphics[width=0.49\textwidth]{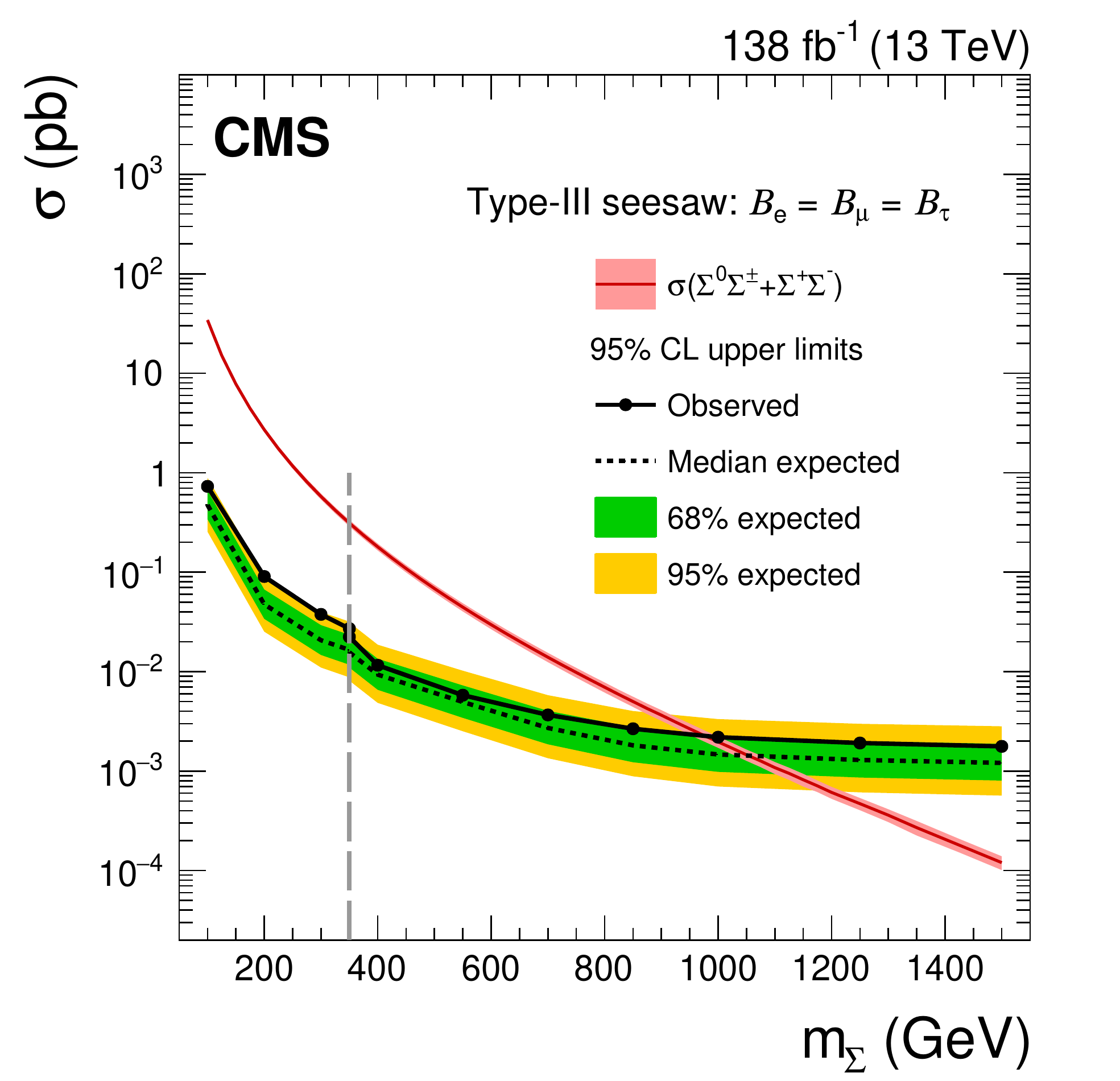}
\caption{\label{fig:limitsSeesaw} Observed and expected upper limits at 95\% \CL on the production cross section for the type-III seesaw fermions in the flavor-democratic scenario using the table schemes and the BDT regions of the \textit{SS-M} and the \textit{SS-H} $\mathcal{B}_{\Pe}=\mathcal{B}_{\PGm}=\mathcal{B}_{\PGt}$ BDTs. 
To the left of the vertical dashed gray line, the limits are shown from the advanced \ST table, and to the right the limits are shown from the BDT regions.
}
\end{figure}

\begin{figure}[hbt!]
\centering
\includegraphics[width=0.49\textwidth]{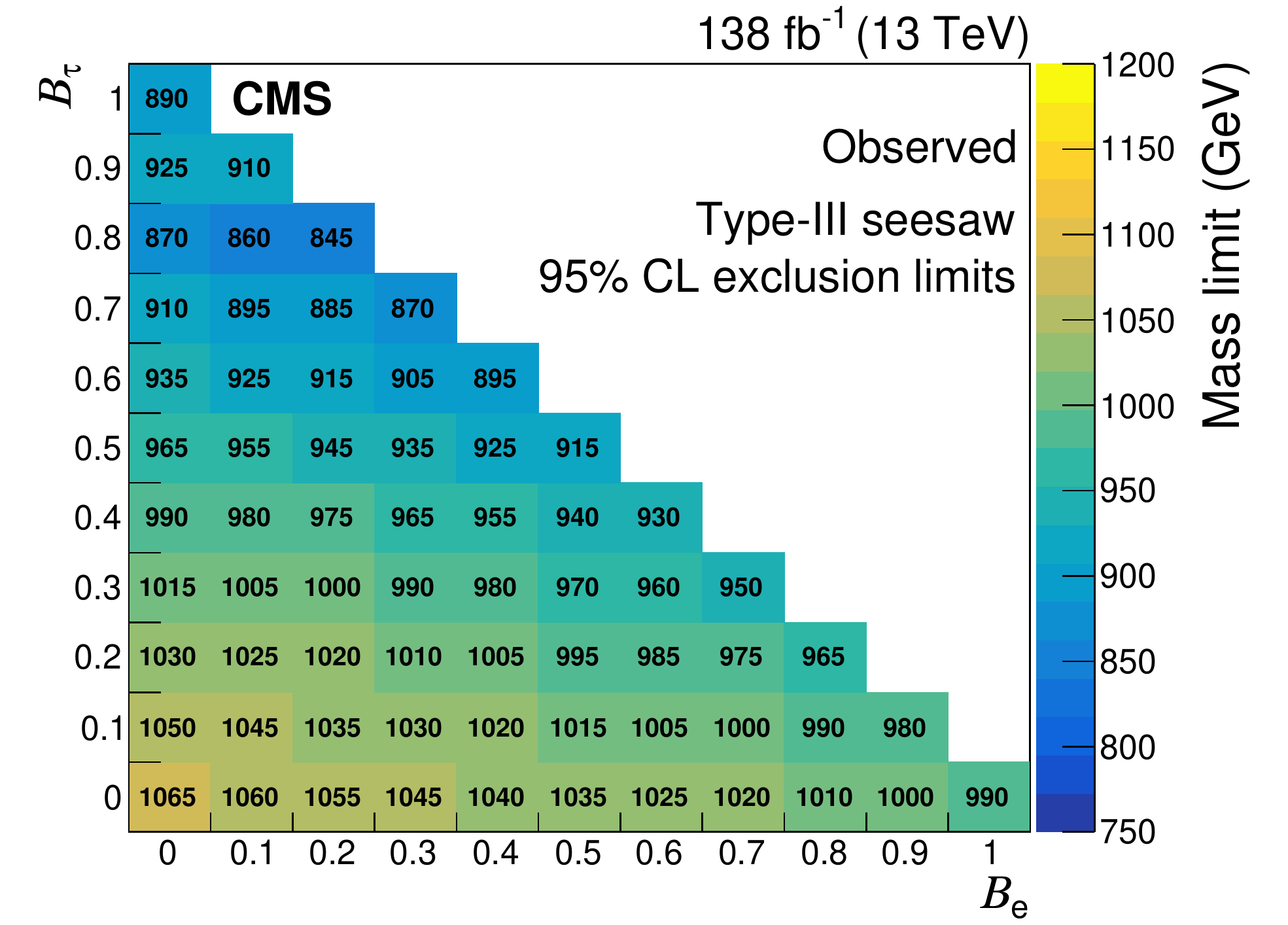}
\includegraphics[width=0.49\textwidth]{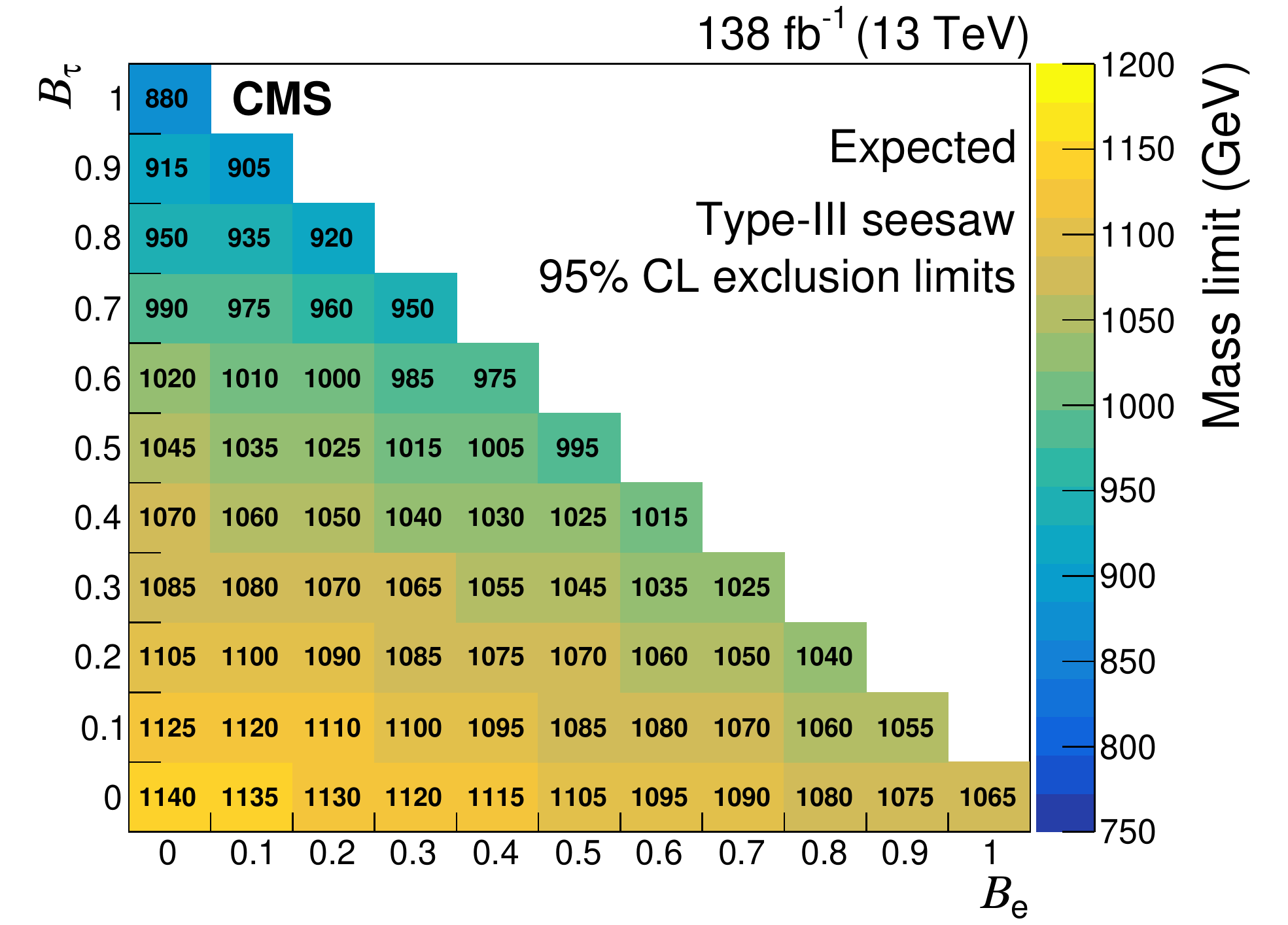}
\caption{\label{fig:limitsSeesawTri} Observed (\cmsLeft) and expected (\cmsRight) lower limits at 95\% \CL on the mass of the type-III seesaw fermions in the plane defined by $\mathcal{B}_{\Pe}$ and $\mathcal{B}_{\PGt}$, with the constraint that \mbox{$\mathcal{B}_{\Pe} + \mathcal{B}_{\Pgm} + \mathcal{B}_{\Pgt}=1$}. 
These limits arise from the \textit{SS-H} $\mathcal{B}_{\PGt}=1$ BDT when $\mathcal{B}_{\PGt} \ge 0.9$, and by the \textit{SS-H} $\mathcal{B}_{\Pe}=\mathcal{B}_{\PGm}=\mathcal{B}_{\PGt}$ BDT for the other decay branching fraction combinations.}
\end{figure}

Figure~\ref{fig:limitsSeesaw} shows the observed and expected cross section limits for the production of the type-III seesaw heavy fermions in the
flavor-democratic scenario. 
The observed (expected) lower limit on $m_{\Sigma}$ in this scenario is 980 (1060)\GeV.
The best expected limit is given by the advanced \ST table scheme for $m_{\Sigma}<350\GeV$, and by the BDT regions for higher signal mass values.
For arbitrary $\Sigma$ decay branching fractions to SM lepton flavors, subject to the constraint that $\mathcal{B}_{\Pe}+\mathcal{B}_{\PGm}+\mathcal{B}_{\PGt}=1$, the observed and expected lower limits on $m_{\Sigma}$ in the plane defined by $\mathcal{B}_{\Pe}$ and $\mathcal{B}_{\PGt}$ are shown in Fig.~\ref{fig:limitsSeesawTri}. 
These limits are given by the \textit{SS-H} $\mathcal{B}_{\PGt}=1$ BDT when $\mathcal{B}_{\PGt} \ge 0.9$, and by the \textit{SS-H} $\mathcal{B}_{\Pe}=\mathcal{B}_{\PGm}=\mathcal{B}_{\PGt}$ BDT for the other decay branching fraction combinations.
The strongest constraints are when $\mathcal{B}_{\PGm}=1$ ($m_{\Sigma}>1065\GeV$), while the weakest are when $\mathcal{B}_{\PGt}=0.8$, $\mathcal{B}_{\Pe}=0.2$ ($m_{\Sigma}>845\GeV$). This behavior is expected because of the greater efficiency of reconstructing and identifying muons versus \tauh candidates in the experiment.

\begin{figure}[hbt!]
\centering
\includegraphics[width=0.49\textwidth]{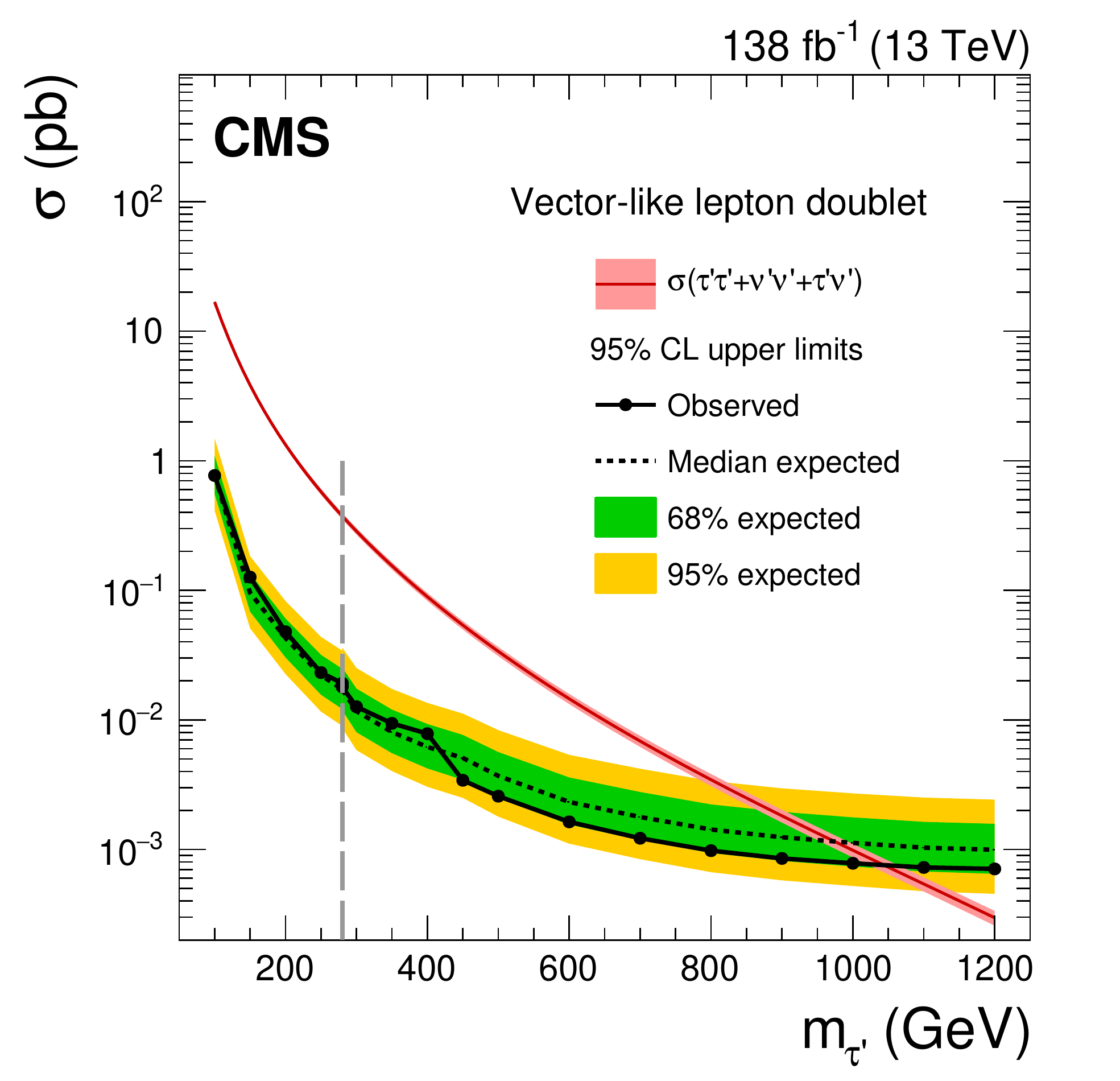}
\includegraphics[width=0.49\textwidth]{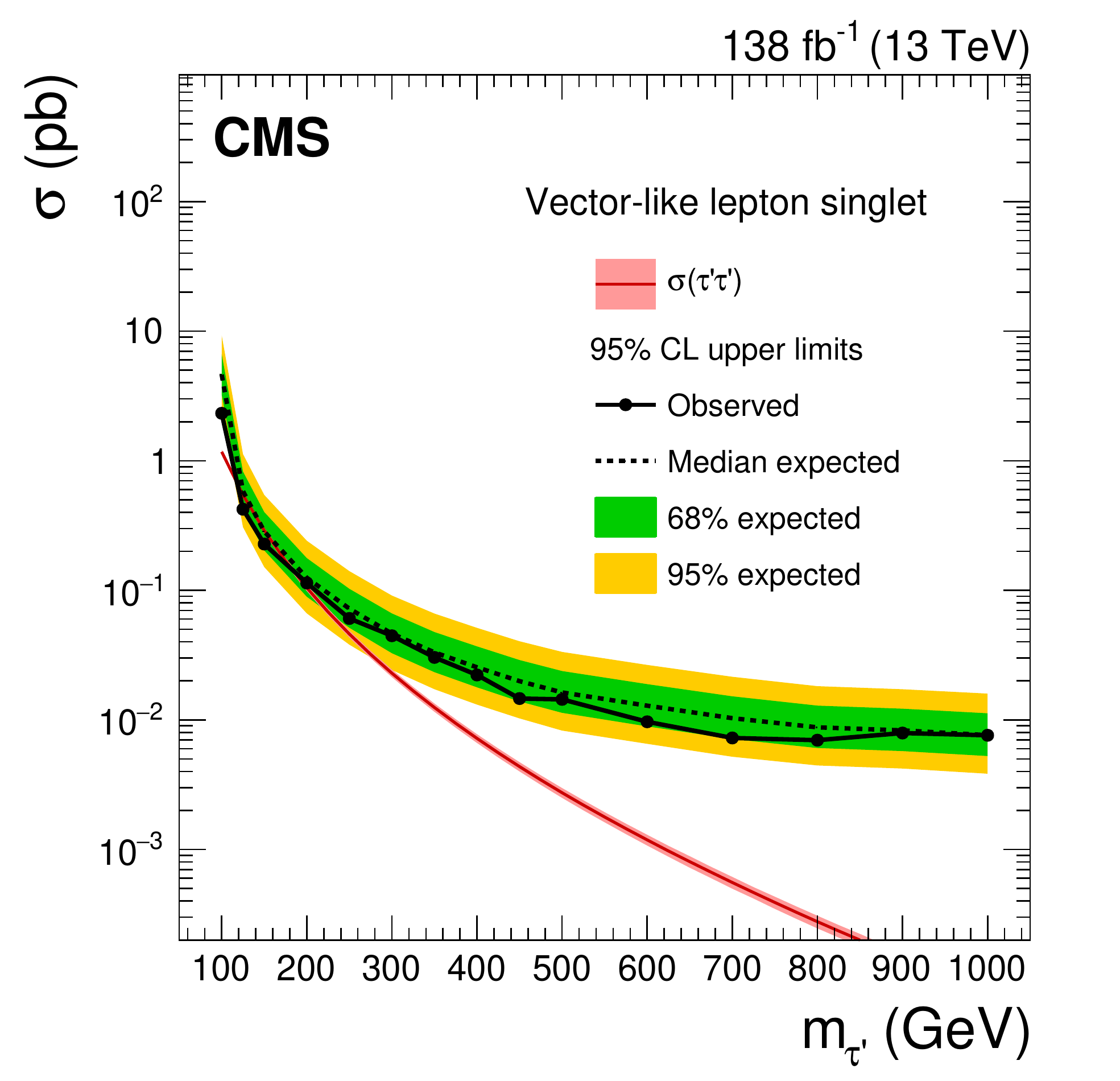}
\caption{\label{fig:limitsVLL} Observed and expected upper limits at 95\% \CL on the production cross section for the vector-like \PGt leptons: doublet model (\cmsLeft), and singlet model (\cmsRight).
  For the doublet vector-like lepton model, to the left of the vertical dashed gray line, the limits are shown from the advanced \ST table, while to the right
  the limits are shown from the BDT regions.
  For the singlet vector-like lepton model, the limit is shown from the advanced \ST table for all masses.}
\end{figure}

Figure~\ref{fig:limitsVLL} shows the observed and expected cross section limits for the doublet and singlet vector-like lepton models. 
For the doublet model, vector-like \PGt leptons are excluded up to a mass $m_{\vltau} $ of to 1045\GeV, where the expected mass exclusion is 975\GeV. 
The best expected limit for $m_{\vltau} < 280\GeV$ is given by the advanced \ST table scheme, and by the BDT regions for larger masses. 
For the singlet model, the best expected limits are given by the advanced \ST table over the entire mass range. 
Singlet vector-like \PGt leptons are excluded in the mass interval from 125 to 150\GeV, while the expected exclusion range is from 125 to 170\GeV.

\begin{figure*}[hbt!]
\centering
\includegraphics[width=0.49\textwidth]{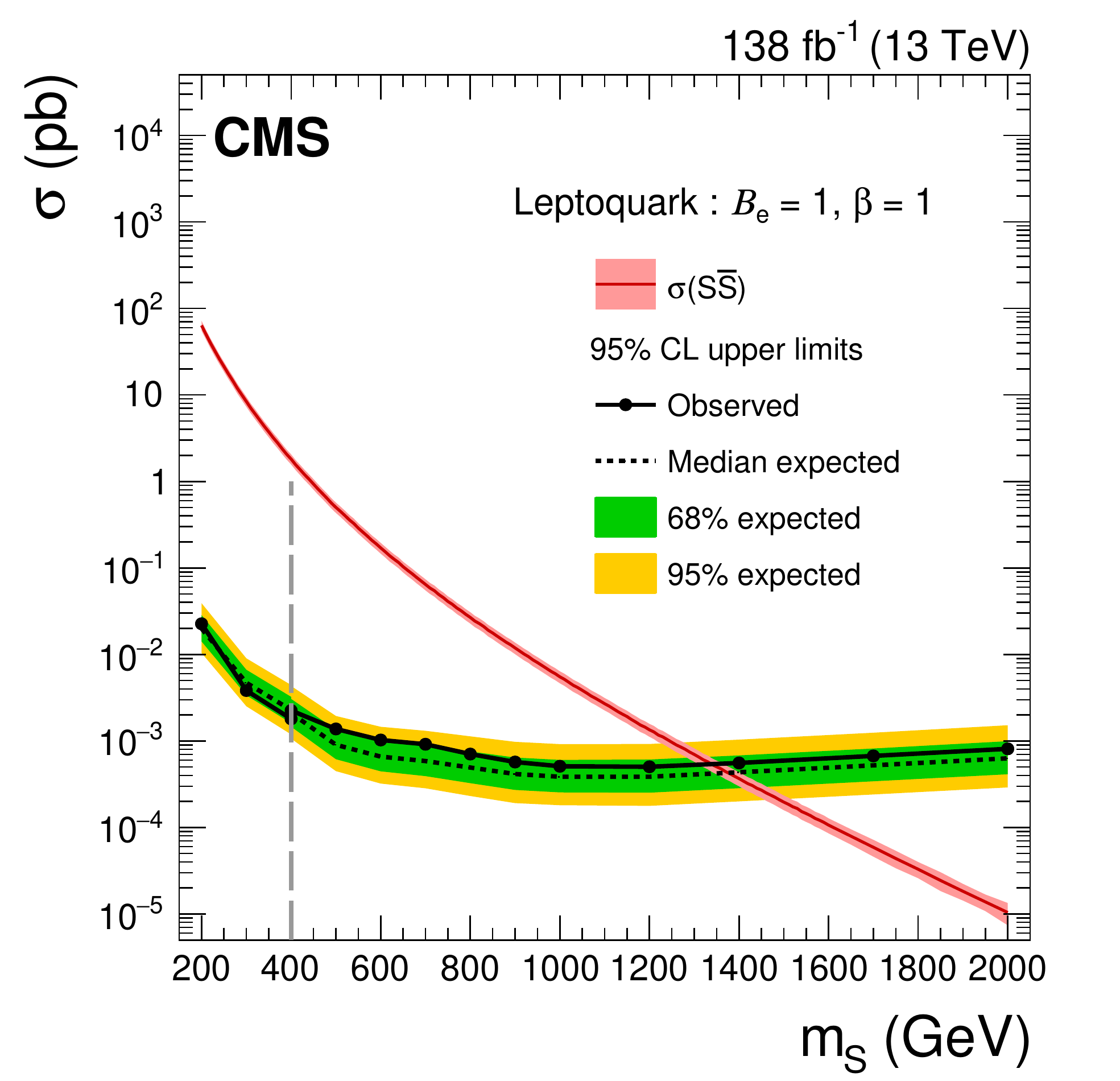}
\includegraphics[width=0.49\textwidth]{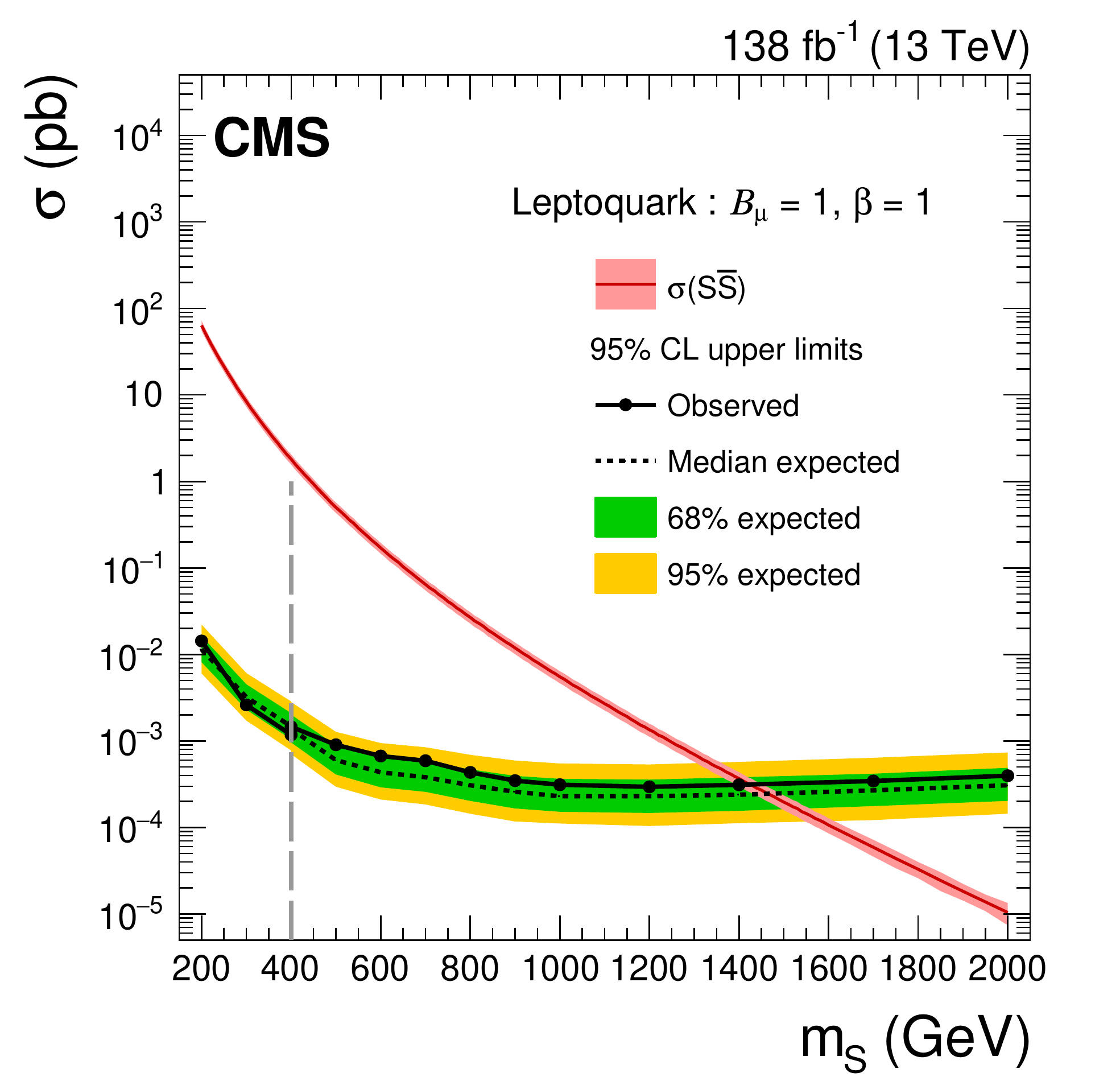}
\includegraphics[width=0.49\textwidth]{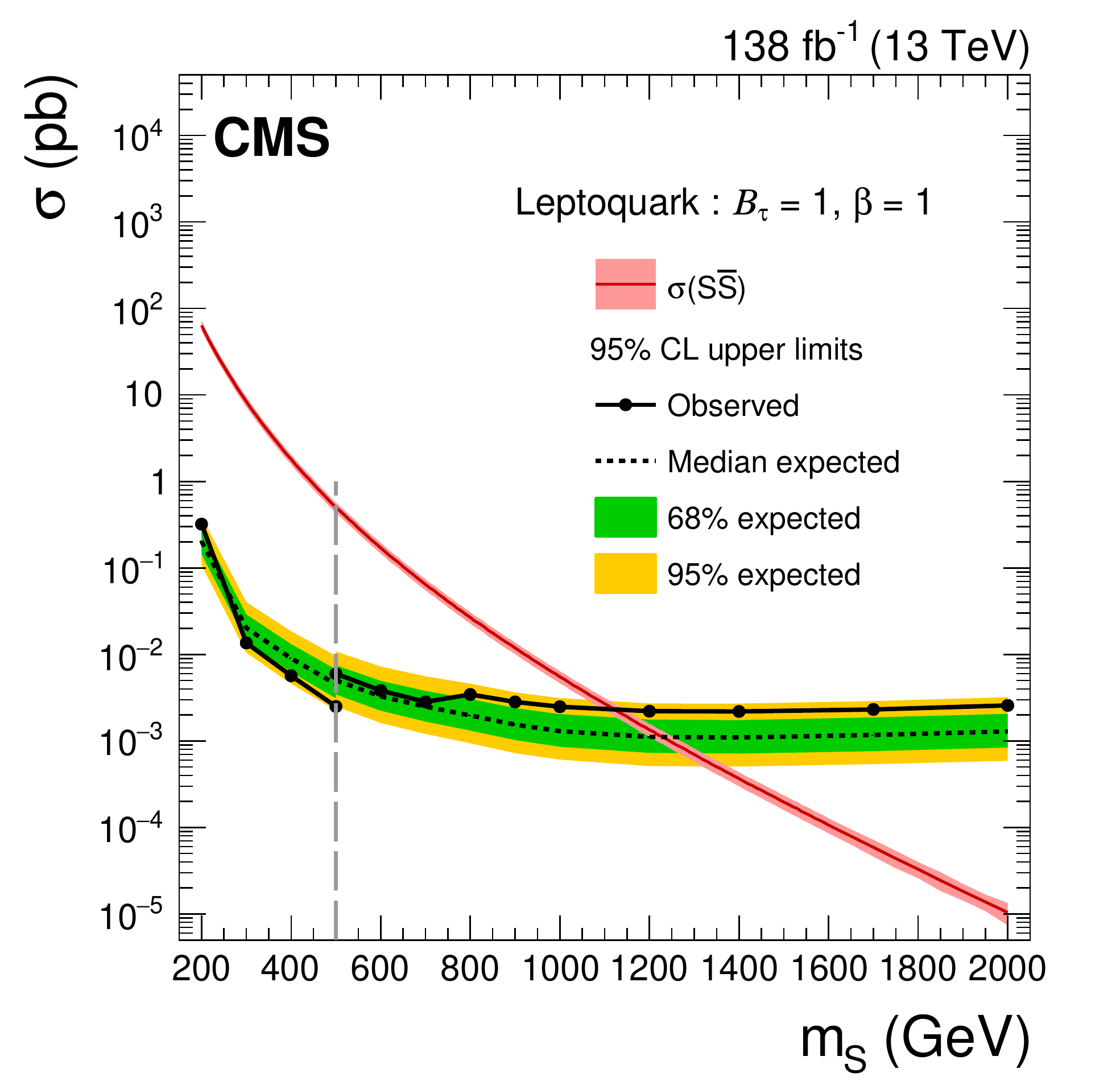}
\caption{\label{fig:limitsLQ} Observed and expected upper limits at 95\% \CL on the production cross section for the scalar leptoquarks: $\mathcal{B}_{\Pe}=1$ (upper left), $\mathcal{B}_{\PGm}=1$
  (upper right), and $\mathcal{B}_{\PGt}=1$ (lower). 
  In each figure, the limits to the left of the vertical dashed gray line are shown from the advanced \ST table,
  and to the right are shown from the BDT regions. }
\end{figure*}

The cross section limits for the leptoquark model are shown in Fig.~\ref{fig:limitsLQ}.
For a leptoquark \PS exclusively coupling to a top quark and a muon, the observed (expected) lower limit on the mass of pair produced leptoquarks is 1420 (1460)\GeV. 
For the top quark and electron decay scenario, the observed (expected) lower limit on $m_{\PS}$ is 1340 (1370)\GeV, while for the top quark and \PGt lepton decay scenario, the lower limit is 1120 (1235)\GeV.
The advanced \ST table gives the best expected limit for $m_{\PS}$ less than 400, 400, and 500\GeV for the $\mathcal{B}_{\Pe}=1$, $\mathcal{B}_{\PGm}=1$, and $\mathcal{B}_{\PGt}=1$ scenarios, respectively,
while the BDT regions give the best expected limits above these thresholds.

The results presented here can be reinterpreted to provide constraints on other BSM models. 
We provide the requisite information in a \textsc{HEPData} record~\cite{hepdata}, as detailed in the following description. 
Specifically, given a specific BSM model, a particular model-independent scheme should be selected. 
The fundamental $\LTmet$ table scheme will be sensitive to BSM models produced primarily via electroweak interactions, while the advanced \ST table scheme will be more sensitive for models which populate final states with several jets, which may or may not arise from \PQb quarks.
Following the choice of a scheme, the signal yield for the model should be obtained in the various categories of the scheme. 
This signal yield can be calculated using generator-level quantities. 
However, there will be a significant correction arising from detector effects, primarily the lepton reconstruction and identification efficiencies. 
We provide detailed efficiency maps for electrons, muons and \tauh, where the provided efficiency is that for a generator level lepton to be both reconstructed and identified as described in this analysis. 
In addition, we also provide the product of acceptance and efficiency for each probed signal model in this paper. 
We find that the yield calculated from generator-level quantities corrected by the efficiency maps agrees with the final analysis yields within 20\% for the channels with light leptons (4L, 3L) and within 30\% for channels that involve a \tauh (3L1T, 2L2T, 2L1T, 1L3T, 1L2T).

The obtained BSM model yields in the various categories can then be used along with the SM backgrounds, the background covariance matrix, and the observations to arrive at constraints for the model in the simplified likelihood framework~\cite{Collaboration:2242860}.

\clearpage

\clearpage
\section{Summary}\label{sec:summary}

A search has been performed for physics beyond the standard model (SM), using multilepton events in proton-proton collision data at $\sqrt{s} = 13\TeV$, collected in 2016--2018 by the CMS experiment at the LHC, corresponding to an integrated luminosity of 138\fbinv. 
The search is carried out in seven orthogonal channels based on the number of light leptons and hadronically decaying \PGt leptons. 
Three model-independent schemes are used to define signal regions for the search. 
In addition, for each model scenario considered, a boosted decision tree is used to define model-specific signal regions. 
In all cases, the observations are found to be consistent with the expectations from the SM processes.
Constraints are set on the production cross section of a number of beyond the SM signal models predicting a variety of multilepton final states.

Type-III seesaw heavy fermions are excluded at 95\% confidence level (\CL) with masses below 980\GeV (expected 1060\GeV), assuming flavor-democratic mixings with SM leptons, and below 990\GeV (expected 1065\GeV), 1065\GeV (expected 1140\GeV), and 890\GeV (expected 880\GeV), assuming mixings exclusively with electron, muon, and \PGt lepton flavors, respectively.
Lower limits on the masses of the heavy fermions are also presented for various decay branching fractions of the heavy fermions to the different SM lepton flavors. 
These are the most stringent constraints on the type-III seesaw heavy fermions to date.

In the vector-like lepton doublet model, vector-like \PGt leptons are excluded at 95\% \CL with masses below 1045\GeV, with an expected exclusion of 975\GeV. 
These are the most stringent constraints on the doublet model. 
For the singlet model, vector-like \PGt leptons are excluded in the mass range from 125 to 150\GeV, while the expected exclusion range is from 125 to 170\GeV.
These are the first constraints from the LHC on the singlet model.

Scalar leptoquarks coupled to top quarks and individual lepton flavors are also probed.
In the scenario with the leptoquark coupling to a top quark and a \PGt lepton, leptoquarks with masses below 1120\GeV are excluded at 95\% \CL (expected 1235\GeV). 
For the decay to a top quark and an electron, leptoquarks are excluded with masses below 1340\GeV (expected 1370\GeV), and for the decay into a top quark and a muon, masses below 1420\GeV (expected 1460\GeV) are excluded.

\begin{acknowledgments}
  We congratulate our colleagues in the CERN accelerator departments for the excellent performance of the LHC and thank the technical and administrative staffs at CERN and at other CMS institutes for their contributions to the success of the CMS effort. In addition, we gratefully acknowledge the computing centers and personnel of the Worldwide LHC Computing Grid and other centers for delivering so effectively the computing infrastructure essential to our analyses. Finally, we acknowledge the enduring support for the construction and operation of the LHC, the CMS detector, and the supporting computing infrastructure provided by the following funding agencies: BMBWF and FWF (Austria); FNRS and FWO (Belgium); CNPq, CAPES, FAPERJ, FAPERGS, and FAPESP (Brazil); MES and BNSF (Bulgaria); CERN; CAS, MoST, and NSFC (China); MINCIENCIAS (Colombia); MSES and CSF (Croatia); RIF (Cyprus); SENESCYT (Ecuador); MoER, ERC PUT and ERDF (Estonia); Academy of Finland, MEC, and HIP (Finland); CEA and CNRS/IN2P3 (France); BMBF, DFG, and HGF (Germany); GSRI (Greece); NKFIA (Hungary); DAE and DST (India); IPM (Iran); SFI (Ireland); INFN (Italy); MSIP and NRF (Republic of Korea); MES (Latvia); LAS (Lithuania); MOE and UM (Malaysia); BUAP, CINVESTAV, CONACYT, LNS, SEP, and UASLP-FAI (Mexico); MOS (Montenegro); MBIE (New Zealand); PAEC (Pakistan); MSHE and NSC (Poland); FCT (Portugal); JINR (Dubna); MON, RosAtom, RAS, RFBR, and NRC KI (Russia); MESTD (Serbia); MCIN/AEI and PCTI (Spain); MOSTR (Sri Lanka); Swiss Funding Agencies (Switzerland); MST (Taipei); ThEPCenter, IPST, STAR, and NSTDA (Thailand); TUBITAK and TAEK (Turkey); NASU (Ukraine); STFC (United Kingdom); DOE and NSF (USA).
 
  \hyphenation{Rachada-pisek} Individuals have received support from the Marie-Curie program and the European Research Council and Horizon 2020 Grant, contract Nos.\ 675440, 724704, 752730, 758316, 765710, 824093, 884104, and COST Action CA16108 (European Union); the Leventis Foundation; the Alfred P.\ Sloan Foundation; the Alexander von Humboldt Foundation; the Belgian Federal Science Policy Office; the Fonds pour la Formation \`a la Recherche dans l'Industrie et dans l'Agriculture (FRIA-Belgium); the Agentschap voor Innovatie door Wetenschap en Technologie (IWT-Belgium); the F.R.S.-FNRS and FWO (Belgium) under the ``Excellence of Science -- EOS" -- be.h project n.\ 30820817; the Beijing Municipal Science \& Technology Commission, No. Z191100007219010; the Ministry of Education, Youth and Sports (MEYS) of the Czech Republic; the Deutsche Forschungsgemeinschaft (DFG), under Germany's Excellence Strategy -- EXC 2121 ``Quantum Universe" -- 390833306, and under project number 400140256 - GRK2497; the Lend\"ulet (``Momentum") Program and the J\'anos Bolyai Research Scholarship of the Hungarian Academy of Sciences, the New National Excellence Program \'UNKP, the NKFIA research grants 123842, 123959, 124845, 124850, 125105, 128713, 128786, and 129058 (Hungary); the Council of Science and Industrial Research, India; the Latvian Council of Science; the Ministry of Science and Higher Education and the National Science Center, contracts Opus 2014/15/B/ST2/03998 and 2015/19/B/ST2/02861 (Poland); the Funda\c{c}\~ao para a Ci\^encia e a Tecnologia, grant CEECIND/01334/2018 (Portugal); the National Priorities Research Program by Qatar National Research Fund; the Ministry of Science and Higher Education, projects no. 0723-2020-0041 and no. FSWW-2020-0008 (Russia); MCIN/AEI/10.13039/501100011033, ERDF ``a way of making Europe", and the Programa Estatal de Fomento de la Investigaci{\'o}n Cient{\'i}fica y T{\'e}cnica de Excelencia Mar\'{\i}a de Maeztu, grant MDM-2017-0765 and Programa Severo Ochoa del Principado de Asturias (Spain); the Stavros Niarchos Foundation (Greece); the Rachadapisek Sompot Fund for Postdoctoral Fellowship, Chulalongkorn University and the Chulalongkorn Academic into Its 2nd Century Project Advancement Project (Thailand); the Kavli Foundation; the Nvidia Corporation; the SuperMicro Corporation; the Welch Foundation, contract C-1845; and the Weston Havens Foundation (USA).
\end{acknowledgments}

\bibliography{auto_generated}
\cleardoublepage \appendix\section{The CMS Collaboration \label{app:collab}}\begin{sloppypar}\hyphenpenalty=5000\widowpenalty=500\clubpenalty=5000\cmsinstitute{Yerevan~Physics~Institute, Yerevan, Armenia}
A.~Tumasyan
\cmsinstitute{Institut~f\"{u}r~Hochenergiephysik, Vienna, Austria}
W.~Adam\cmsorcid{0000-0001-9099-4341}, J.W.~Andrejkovic, T.~Bergauer\cmsorcid{0000-0002-5786-0293}, S.~Chatterjee\cmsorcid{0000-0003-2660-0349}, K.~Damanakis, M.~Dragicevic\cmsorcid{0000-0003-1967-6783}, A.~Escalante~Del~Valle\cmsorcid{0000-0002-9702-6359}, R.~Fr\"{u}hwirth\cmsAuthorMark{1}, M.~Jeitler\cmsAuthorMark{1}\cmsorcid{0000-0002-5141-9560}, N.~Krammer, L.~Lechner\cmsorcid{0000-0002-3065-1141}, D.~Liko, I.~Mikulec, P.~Paulitsch, F.M.~Pitters, J.~Schieck\cmsAuthorMark{1}\cmsorcid{0000-0002-1058-8093}, R.~Sch\"{o}fbeck\cmsorcid{0000-0002-2332-8784}, D.~Schwarz, S.~Templ\cmsorcid{0000-0003-3137-5692}, W.~Waltenberger\cmsorcid{0000-0002-6215-7228}, C.-E.~Wulz\cmsAuthorMark{1}\cmsorcid{0000-0001-9226-5812}
% ...
\vskip\cmsinstskip
% ...
1:~Also at TU Wien, Wien, Austria\\
% ...
\end{sloppypar}
\end{document}